\documentclass[runningheads]{llncs}
\usepackage[T1]{fontenc}
\usepackage{graphicx}
\usepackage{booktabs}
\usepackage[misc]{ifsym}
\newcommand{\corr}{(\Letter)}
\usepackage{mwe}

\let\Letter\relax  
\usepackage{marvosym}  
\usepackage{color}
\usepackage[show]{chato-notes}

\usepackage{tocloft} 

\usepackage{pgfmath} 

\usepackage{multirow}%
\usepackage{amsmath,amssymb,amsfonts}%
\usepackage{mathrsfs}%
\usepackage[toc, page, title]{appendix}%
\usepackage{xcolor}%
\usepackage{textcomp}%
\usepackage{manyfoot}%
\usepackage{booktabs}%
\usepackage{listings}%
\usepackage{algorithm}
\usepackage{algorithmic}

\usepackage{bibentry}                
\nobibliography*                     

\usepackage{pgfplots}
\pgfplotsset{compat=1.17}
\usepgfplotslibrary{colormaps, groupplots}
\usepackage{tikz}
\usepackage{pgfmath} 
\usepackage{xcolor}
\usetikzlibrary{matrix, positioning, shapes.geometric}

\definecolor{slateblue}{RGB}{106, 90, 205}       
\definecolor{darkolivegreen}{RGB}{85, 107, 47}   
\definecolor{crimson}{RGB}{220, 20, 60}     

\usepackage{adjustbox} 
\usepackage{array}     

\definecolor{darkmagenta}{rgb}{0.55, 0.0, 0.55} 

\usepackage[utf8]{inputenc} 
\usepackage[T1]{fontenc}    
\usepackage{hyperref}       
\usepackage{url}            
\usepackage{booktabs}       
\usepackage{amsfonts}       
\usepackage{nicefrac}       
\usepackage{microtype}      
\usepackage{amsmath}
\usepackage{amssymb}
\usepackage{mathtools}
\usepackage{graphicx}
\usepackage{xspace}
\usepackage{float}
\usepackage{varwidth}
\usepackage{caption}
\usepackage{enumitem}
\usepackage{pgfplots}
\pgfplotsset{compat=newest}
\usepgfplotslibrary{groupplots}
\usepgfplotslibrary{dateplot}

\makeatletter
\newcommand\footnoteref[1]{\protected@xdef\@thefnmark{\ref{#1}}\@footnotemark}
\makeatother

\usepackage[show]{chato-notes}

\DeclareMathOperator*{\argmax}{\arg\max}

\newcommand{\journal}[1]{\textcolor{red}{#1}}

\newcommand{\para}[1]{\noindent{\bf{#1}}}
\newcommand{\spara}[1]{\smallskip\noindent{\bf{#1}}}

\newcommand{\abs}[1]{{\left|#1\right|}}

\newcommand{\real}{\ensuremath{\mathbb{R}}\xspace}

\newcommand{\np}{\ensuremath{\mathbf{NP}}\xspace}
\newcommand{\bigO}{\ensuremath{\mathcal{O}}\xspace}

\newcommand{\inputmat}{\ensuremath{\mathbf{D}}\xspace}
\newcommand{\outputmat}{\ensuremath{\mathbf{X}}\xspace}

\newcommand{\outputmatestimate}{\ensuremath{\hat{\mathbf{X}}\xspace}}

\newcommand{\distributionz}{\ensuremath{\mathcal{Z}}\xspace}

\newcommand{\bfA}{\ensuremath{\mathbf{A}}\xspace}
\newcommand{\bfB}{\ensuremath{\mathbf{B}}\xspace}

\newcommand{\bfU}{\mathbf{U}\xspace}
\newcommand{\bfSigma}{\mathbf{\Sigma}\xspace}
\newcommand{\bfV}{\mathbf{V}\xspace}

\newcommand{\estimate}[1]{\hat{#1}}

\newcommand{\frobeniusnorm}[1]{\left\| {#1} \right\|_F}
\newcommand{\frobeniusnormshort}[1]{\| {#1} \|_F}

\newcommand{\spectralnorm}[1]{\left\| {#1} \right\|_2}
\newcommand{\spectralnormshort}[1]{\| {#1} \|_2}

\newcommand{\infinitynorm}[1]{\left\| {#1} \right\|_{\max}}
\newcommand{\infinitynormshort}[1]{\| {#1} \|_{\max}}

\newcommand{\generalnormshort}[1]{\| {#1} \|_{*}}

\newcommand{\lowrankness}[1]{\ensuremath{\mathit{{\ell}{r}}({#1})}\xspace}

\newcommand{\nphard}{\ensuremath{\mathbf{NP}}-hard\xspace}

\newcommand{\noisescale}{\ensuremath{\epsilon_{\text{noise}}}\xspace}

\newcommand{\initialmat}{\ensuremath{\mathbf{P}}\xspace}
\newcommand{\ind}{\ensuremath{\mathbf{I}}\xspace}
\newcommand{\indrows}{\ensuremath{\mathbf{I}^r}\xspace}
\newcommand{\indcols}{\ensuremath{\mathbf{I}^c}\xspace}
\newcommand{\generalrank}{\ensuremath{k}\xspace}

\newcommand{\initialmatnospace}{\ensuremath{\mathbf{P}}}

\newcommand{\initialmats}{\ensuremath{\initialmatnospace_{k}}\xspace}
\newcommand{\initialerror}{\ensuremath{\mathbf{E}_{\initialmatnospace}}\xspace}
\newcommand{\identitymat}{\ensuremath{\mathbf{I}}\xspace}

\newcommand{\outputerror}{\ensuremath{\mathbf{E}_{\outputmat, \estimate{\outputmat}}}\xspace}
\newcommand{\rank}{\ensuremath{\mathrm{rank}}\xspace}

\newcommand{\initialmatcols}{\ensuremath{\mathcal{C}_{\initialmatnospace}}\xspace}

\newcommand{\gmax}[1]{g_{\max}(#1)}
\newcommand{\gF}[1]{g_{F}(#1)}

\newcommand{\rowset}{\ensuremath{\mathcal{R}}\xspace}
\newcommand{\colset}{\ensuremath{\mathcal{C}}\xspace}
\newcommand{\sinematrix}{\ensuremath{\mathbf{S}_{\mathcal{C}}}\xspace}

\newcommand{\nearrankone}{{LNROS}\xspace}
\newcommand{\nearrankk}{{LNR\ensuremath{k}S}\xspace}

\newcommand{\nrows}{\ensuremath{n}\xspace}
\newcommand{\ncols}{\ensuremath{m}\xspace}
\newcommand{\nrowssubmatrix}{\ensuremath{n'}\xspace}
\newcommand{\ncolssubmatrix}{\ensuremath{m'}\xspace}

\newcommand{\toleranceinit}{\ensuremath{\delta_{\mathit{init}}}\xspace}
\newcommand{\tol}{\ensuremath{\delta}\xspace}

\newcommand{\errorbound}{\ensuremath{\epsilon}\xspace}

\newcommand{\simplifiedproblem}{LNROSR\xspace}

\newcommand{\f}[1]{f(#1)}

\newcommand{\errors}
{\ensuremath{\mathbf{E}^{abs}}\xspace}
\newcommand{\svd}{{SVD}\xspace}
\newcommand{\svp}{\textsc{SVP}\xspace}
\newcommand{\pca}{{PCA}\xspace}
\newcommand{\sparsePCA}{\textsc{Sparse\-PCA}\xspace}
\newcommand{\cvx}{\textsc{CVX}\xspace}
\newcommand{\rpsp}{\textsc{RPSP}\xspace}
\newcommand{\ourmethod}{\textsc{Sample\-And\-Expand}\xspace}

\newcommand{\gram}
{\ensuremath{\mathbf{G}}\xspace}
\newcommand{\anchorrow}
{\ensuremath{\mathbf{x}^r}\xspace}
\newcommand{\anchorrowentryi}
{\ensuremath{x^r_i}\xspace}
\newcommand{\anchorrowentryis}
{\ensuremath{{x^r_i}}\xspace}
\newcommand{\anchorrowentryj}
{\ensuremath{x^r_j}\xspace}
\newcommand{\anchorrowentryjs}
{\ensuremath{{x^r_j}}\xspace}

\newcommand{\anchorindexrow}
{\ensuremath{i_a}\xspace}
\newcommand{\anchorindexcol}
{\ensuremath{j_a}\xspace}

\newcommand{\anchorcolumn}
{\ensuremath{\mathbf{x}^c}\xspace}

\newcommand{\nrep}{\ensuremath{N_{\mathit{init}}}\xspace}
\newcommand{\p}{\ensuremath{|\outputmat|}\xspace}

\newcommand{\pfirststar}{\ensuremath{p}\xspace}
\newcommand{\pN}{\ensuremath{\mathit{Pr}(\mathit{itr}=N)}\xspace}
\newcommand{\generalentry}{\ensuremath{(i,j)}\xspace}

\newcommand{\rowratios}{\ensuremath{\mathbf{R}^r}\xspace}
\newcommand{\columnratios}{\ensuremath{\mathbf{R}^c}\xspace}

\newcommand{\rowratiosij}{\ensuremath{\mathbf{R}^r_{i,j}}\xspace}
\newcommand{\columnratiosij}{\ensuremath{\mathbf{R}^c_{i,j}}\xspace}

\newcommand{\Hyperspectral}{\textsc{Hyperspectral}\xspace}
\newcommand{\Isolet}{\textsc{Isolet}\xspace}
\newcommand{\Olivetti}{\textsc{Olivetti}\xspace}
\newcommand{\MoiveLens}{\textsc{MovieLens}\xspace}
\newcommand{\MovieLens}{\textsc{MovieLens}\xspace}
\newcommand{\OrlRnSp}{\textsc{ORL}\xspace}
\newcommand{\Jester}{\textsc{Jester}\xspace}
\newcommand{\Golub}{\textsc{AL-Genes}\xspace}
\newcommand{\Mandrill}{\textsc{Mandrill}\xspace}
\newcommand{\Ozone}{\textsc{Ozone}\xspace}
\newcommand{\SKCM}{\textsc{SKCM1}\xspace}
\newcommand{\BRCA}{\textsc{BRCA-Genes}\xspace}
\newcommand{\PRAD}{\textsc{PRAD1}\xspace}
\newcommand{\Google}{\textsc{Google}\xspace}
\newcommand{\NPAS}{\textsc{NPAS}\xspace}
\newcommand{\Cameraman}{\textsc{Cameraman}\xspace}
\newcommand{\MovieTrust}{\textsc{MovieTrust}\xspace}
\newcommand{\Hearth}{\textsc{Hearth}\xspace}
\newcommand{\Imagenet}{\textsc{Imagenet}\xspace}

\newcommand{\maxnorm}{\ensuremath{\gamma_{max}}\xspace}

\newcommand{\errorvec}{\ensuremath{\boldsymbol{\epsilon}}\xspace}

\newcommand{\vecu}{\ensuremath{\mathbf{u}}\xspace}
\newcommand{\vecv}{\ensuremath{\mathbf{v}}\xspace}
\newcommand{\budget}{\ensuremath{\mathcal{B}}\xspace}

\newcommand{\yvec}{\ensuremath{\mathbf{y}}\xspace}
\newcommand{\xvec}{\ensuremath{\mathbf{x}}\xspace}
\newcommand{\proj}[2]{\operatorname{Proj}_{{#1}} #2}

\newcommand{\var}{\ensuremath{\mathit{Var}}\xspace}

\tikzset{%
  point/.style={circle, inner sep=2pt}, 
  other point/.style={fill=black, point}  
}

\makeatletter
\providecommand*{\cupdot}{%
  \mathbin{%
    \mathpalette\@cupdot{}%
  }%
}
\newcommand*{\@cupdot}[2]{%
  \ooalign{%
    $\m@th#1\cup$\cr
    \sbox0{$#1\cup$}%
    \dimen@=\ht0 %
    \sbox0{$\m@th#1\cdot$}%
    \advance\dimen@ by -\ht0 %
    \dimen@=.5\dimen@
    \hidewidth\raise\dimen@\box0\hidewidth
  }%
}

\providecommand*{\bigcupdot}{%
  \mathop{%
    \vphantom{\bigcup}%
    \mathpalette\@bigcupdot{}%
  }%
}
\newcommand*{\@bigcupdot}[2]{%
  \ooalign{%
    $\m@th#1\bigcup$\cr
    \sbox0{$#1\bigcup$}%
    \dimen@=\ht0 %
    \advance\dimen@ by -\dp0 %
    \sbox0{\scalebox{2}{$\m@th#1\cdot$}}%
    \advance\dimen@ by -\ht0 %
    \dimen@=.5\dimen@
    \hidewidth\raise\dimen@\box0\hidewidth
  }%
}
\makeatother

\begin{document}


\title{Sample and Expand: Discovering Low-rank Submatrices With Quality Guarantees}

\titlerunning{Sample and Expand: Discovering Low-rank Submatrices}


\author{Martino Ciaperoni \inst{1}\thanks{The work was done while the author was at Aalto University.} \corr  \and
Aristides Gionis \inst{2}  \and
Heikki Mannila \inst{3}}


\authorrunning{M. Ciaperoni et al.}

\institute{
Scuola Normale Superiore, Italy
\\
\email{martino.ciaperoni@sns.it}
\and
KTH Royal Institute of Technology, Sweden 
\\
\email{argioni@kth.se}
\and
Aalto University, Finland 
\\
\email{heikki.mannila@aalto.fi}
}

\maketitle              

\toctitle{Sample and Expand: Discovering Low-rank Submatrices}

\tocauthor{Martino Ciaperoni, Aristides Gionis, and Heikki Mannila}

\begin{abstract}
The problem of approximating a matrix by a low-rank one has been extensively studied. This problem assumes, however, that the whole matrix has a low-rank structure. This assumption is often false for real-world matrices. We consider the problem of discovering sub\-matrices from the given matrix with bounded deviations from their low-rank approximations. We introduce an effective two-phase method for this task: first, we use sampling to discover small nearly low-rank sub\-matrices, and then they are expanded while preserving proximity to a low-rank approximation. An extensive experimental evaluation confirms that the method we introduce compares favorably to existing approaches.

\end{abstract}



\section{Introduction}\label{sec:introduction}


Low-rank approximation has emerged as a fundamental task in many data-analysis applications, including machine-learning pipelines~\cite{wang2017research}, large language models~\cite{hu2021lora}, recommender systems~\cite{lee2016llorma}, 
image compression and de\-noising~\cite{guo2015efficient}.
The goal of low-rank approximation is to represent an input matrix as accurately as possible using a small number of row and column vectors. 
%

For decades, the \emph{singular value decomposition} (\svd), 
with the closely related principal component analysis (PCA), 
has remained the gold standard for low-rank approximation~\cite{golub2013matrix}.
%
Despite its 
success, \svd has certain limitations. 
Among others, when applying \svd we aim to find a low-rank approximation for the entire input matrix. 
This assumption can be rather restrictive, as in real-world data
it might be that only certain sub\-matrices are well approximated by low-rank structures.
For instance, in ratings data originating in movie recommender systems, 
low-rank submatrices occur because subsets of users may share a similar taste only for a subset of movies. 
Similar local patterns could be observed in data coming from other domains, 
such as market-basket analysis, image processing, and~biology~\cite{dang2023generalized}. 
The \svd can fail to identify  local low-rank submatrices.

\spara{Existing approaches to find local low-rank sub\-matrices.}
The task of identifying sub\-matrices that are well described by low-rank structures has been largely overlooked until recently~\cite{dang2023generalized}. 
Existing work in this direction is based primarily on 
the \svd, and 
does not provide any guarantee on the quality of the approximation for the identified sub\-matrices.  

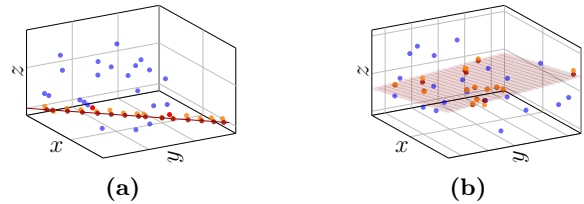
\begin{figure}[t]
    \centering
    \begin{tabular}{c@{\hspace{1.5cm}}c}  
         \scalebox{0.375}{  \definecolor{darkred}{rgb}{0.6,0,0}  
\begin{tikzpicture}
    \begin{axis}[
        view={60}{30},              
        axis lines=box,             
        xlabel={$x$}, ylabel={$y$}, zlabel={$z$},
        xlabel style={font=\Huge}, 
        ylabel style={font=\Huge}, 
        zlabel style={font=\Huge}, 
        ticks=none,                 
        grid=major,                 
        width=9cm,                  
        height=7cm,                 
        scatter/classes={%
            blue={mark=*,blue,opacity=0.6},  
            red={mark=*,red,thick},          
            orange={mark=*,orange,thick,opacity=0.8},    
           darkred={mark=*,darkred,thick,opacity=0.8}} 
    ]

    \addplot3[
        only marks,
        scatter,
        scatter src=explicit symbolic
    ] table[meta=class] {
        x       y       z       class
        2.3     1.5     3.1     blue
        0.5     2.2     0.0     blue
        1.9     1.3     2.5     blue
        3.5     2.8     1.7     blue
        0.7     0.9     2.1     blue
        1.2     2.8     3.3     blue
        2.6     0.4     1.5     red
        3.0     1.9     3.5     blue
        2.7     2.2     0.1     blue
        1.4     3.1     1.8     blue
        0.3     0.5     0.8     blue
        2.9     3.3     2.5     blue
        1.7     1.5     3.0     blue
        2.5     3.0     0.1     blue
        0.6     0.4     0.9     blue
        3.2     1.8     2.4     blue
        1.9     0.7     1.3     blue
        0.5     1.0     2.7     blue
        3.3     2.4     1.6     blue
        2.8     2.7     3.0     blue
        2.1     1.3     0.0     blue
        1.6     0.8     1.2     blue
        3.4     3.0     1.1     red
        2.2     2.9     3.1     blue
    };

    \addplot3[
        only marks,
        scatter,
        scatter src=explicit symbolic
    ] table[meta=class] {
        x       y       z       class
        0.1     0.7     0.0     orange
        0.5     0.7     0.0     orange
        0.6     1.2     0.1     orange
        0.7     1.7     0.0     orange
        1.0     2.1     0.0     orange
        1.2     2.5     0.1     orange
        1.4     2.9     0.0     orange
        1.6     3.3     0.0     orange
        1.7     3.5     0.0     orange
        2.1     3.9     0.0     orange
        2.3     4.2     0.0     orange
        2.4     4.7     0.1     orange
        2.6     5.1     0.0     orange
        2.7     5.5     0.0     orange
        3.0     5.7     0.1     orange
    };

    \addplot3[
        thick, darkred
    ] coordinates {
        (0,0,0) (3.2,5.8,0)
    };


    \addplot3[
        only marks,
        scatter,
        scatter src=explicit symbolic
    ] table[meta=class] {
    x       y       z       class
    0.32075 	 0.57736 	 0.0 	 darkred
    0.41509 	 0.74717 	 0.0 	 darkred
    0.65094 	 1.1717 	 0.0 	 darkred
    0.88679 	 1.59623 	 0.0 	 darkred
    1.12736 	 2.02925 	 0.0 	 darkred
    1.34434 	 2.41981 	 0.0 	 darkred
    1.56132 	 2.81038 	 0.0 	 darkred
    1.7783 	 3.20094 	 0.0 	 darkred
    1.88679 	 3.39623 	 0.0 	 darkred
    2.15094 	 3.8717 	 0.0 	 darkred
    2.32547 	 4.18585 	 0.0 	 darkred
    2.56132 	 4.61038 	 0.0 	 darkred
    2.7783 	 5.00094 	 0.0 	 darkred
    2.9717 	 5.34906 	 0.0 	 darkred
    3.12736 	 5.62925 	 0.0 	 darkred
    };
    \end{axis}
\end{tikzpicture}} & 
        \scalebox{0.375}{ \definecolor{darkred}{rgb}{0.6,0,0}  

\begin{tikzpicture}
    \begin{axis}[
        view={60}{30},              
        axis lines=box,             
        xlabel={$x$}, ylabel={$y$}, zlabel={$z$},
        xlabel style={font=\Huge}, 
        ylabel style={font=\Huge}, 
        zlabel style={font=\Huge}, 
        ticks=none,                 
        grid=major,                 
        width=9cm,                  
        height=7cm,                 
        scatter/classes={%
            blue={mark=*,blue,opacity=0.6},  
            orange={mark=*,orange,thick,opacity=0.8},    
            darkred={mark=*,darkred,thick,opacity=0.8}}  
    ]

\addplot3 [
    surf,
    shader=flat,
    opacity=0.1, 
    colormap={darkredmap}{rgb255(0cm)=(153,0,0) rgb255(1cm)=(102,0,0)}
]
{0.5*x + 0.3*y};

    \addplot3[
        only marks,
        scatter,
        scatter src=explicit symbolic
    ] table[meta=class] {
        x       y       z       class
-2.7     0.1     8.2     blue
-4.2     -0.8     -5.2     blue
-2.1     -2.8     -7.1     blue
-3.4     -3.8     -0.2     blue
4.3     -1.6     9.7     blue
3.1     4.4     -5.2     blue
1.3     -1.8     3.4     blue
3.7     0.2     5.2     blue
3.0     2.0     -5.2     blue
-3.1     -1.4     4.6     blue
3.9     4.7     -2.6     blue
0.4     4.6     2.6     blue
3.1     -2.5     2.7     blue
4.0     -0.0     0.7     blue
-1.8     -2.0     -8.2     blue
-3.9     -2.2     6.7     blue
-2.7     -4.6     -3.6     blue
-0.7     1.1     -6.3     blue
3.2     0.0     -9.2     blue
3.6     -4.5     1.8     blue
-4.9     -2.2     3.6     blue

    };

    \addplot3[
        only marks,
        scatter,
        scatter src=explicit symbolic
    ] table[meta=class] {
        x       y       z        class
        -1.3     -3.2     -1.61     darkred
        4.5     -2.0     1.65     darkred
        2.3     0.2     1.21     darkred
        1.0     -0.7     0.29     darkred
        -3.4     -2.1     -2.33     darkred
        -3.4     1.1     -1.37     darkred
        -4.4     -3.6     -3.28     darkred
        3.7     -2.1     1.22     darkred
        1.0     -1.3     0.11     darkred
        2.1     -0.4     0.93     darkred
        -4.8     2.9     -1.53     darkred
        4.7     -3.0     1.45     darkred
        3.3     0.1     1.68     darkred
        -2.9     0.9     -1.18     darkred
        4.0     5.0     3.50     darkred
    };

\addplot3[
    only marks,
    scatter,
    scatter src=explicit symbolic
] table[meta=class] {
x       y       z        class
-1.3     -3.2     -1.81     orange
4.5     -2.0     0.48    orange
2.3     0.2     1.39     orange
1.0     -0.7     0.49     orange
-3.4     -2.1     -1.3     orange
-3.4     1.1     -0.7     orange
-4.4     -3.6     -3.1     orange
3.7     -2.1     1.0     orange
1.0     -1.3     0     orange
2.1     -0.4     0.9     orange
-4.8     2.9     -1.0     orange
4.7     -3.0     1.75     orange
3.3     0.1     2     orange
-2.9     0.9     -0.24     orange
4     5     4.5     orange
};

 \end{axis}
\end{tikzpicture}} \\
        \textbf{(a)} & \textbf{(b)} \\ 
    \end{tabular}
    \caption{\label{fig:line_three_d} Example. A subset of data points (in orange) in the $3$-dimensional space are close to their projection (in red) onto a line in the $xy$-plane (a) or to a plane in the $3$-dimensional space (b), while other points (in blue) can be further away. }
\end{figure}

\spara{Our approach.}
In this work, we adopt a different perspective on discovering local low-rank patterns, and we address the problem of identifying sub\-matrices 
that are guaranteed to be close to a low-rank approximation. 
Our quality guarantees hold with respect to an approximation that can be easily interpreted 
in terms of the original data, which can be particularly valuable for applications in different domains. 
For example, near-rank-$1$ sub\-matrices can be accurately approximated 
by each row in the submatrix being colinear with a single row.  
Unlike previous work, our work does not directly rely on the \svd. 
Nearly-low-rank sub\-matrices correspond to subsets of points (matrix rows) 
that approximately lie on a low-dimensional sub\-space, 
for a subset of dimensions (matrix columns).
For rank equal to~$1$, which is a particularly interesting case, the points approximately lie on a line through the origin, and for rank equal to~$2$ the points are close to a plane through the origin, as in the example of Figure~\ref{fig:line_three_d}, which shows data points 
identifying a $15 \times 2$ near-rank-$1$ sub\-matrix and a $15 \times 3$ near-rank-$2$ sub\-matrix. 

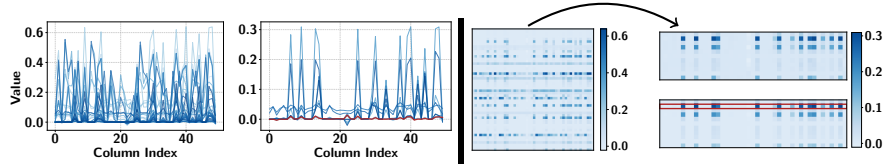
\begin{figure}[t]
    \centering
    \begin{tabular}{c c !{\vrule width 2pt} c c}  
        \adjustbox{valign=m}{\includegraphics[width=0.24\textwidth]{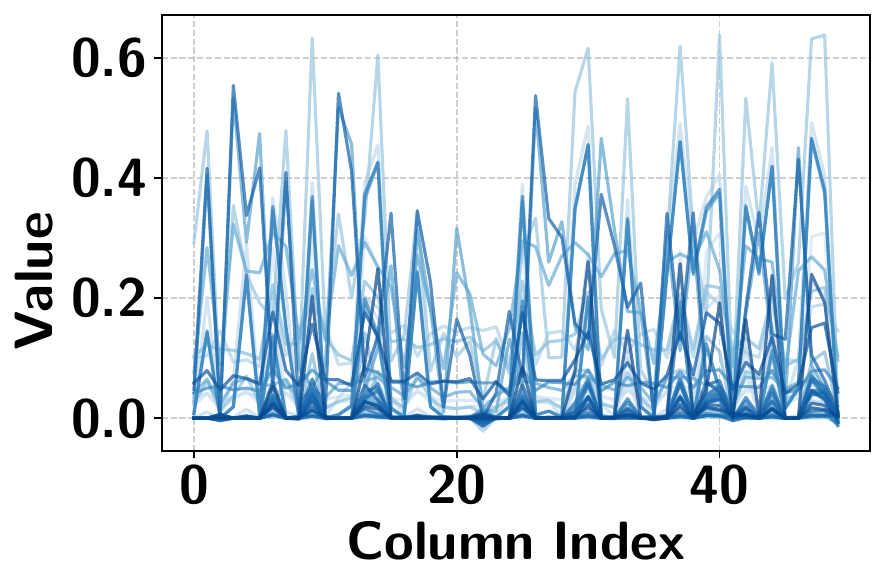}} &
        \adjustbox{valign=m}{\includegraphics[width=0.24\textwidth]{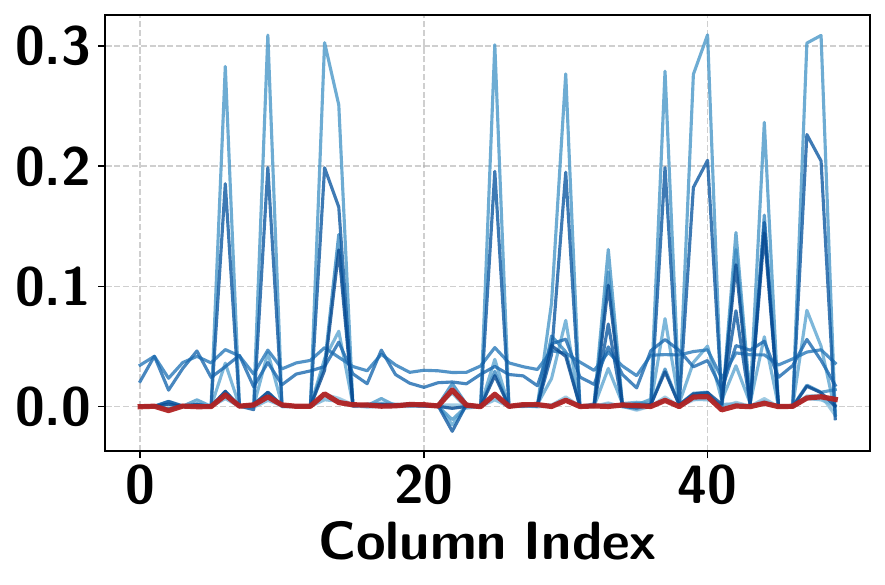}} &
        \adjustbox{valign=m}{\includegraphics[width=0.18\textwidth]{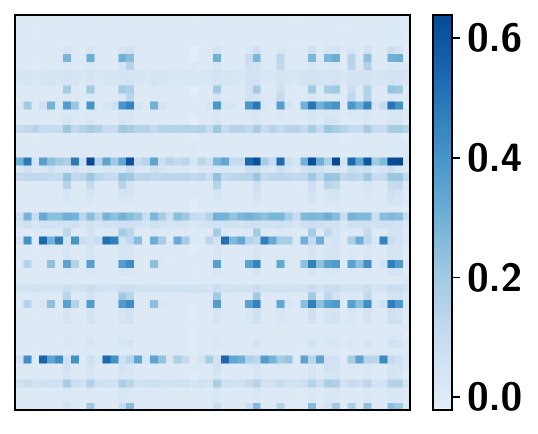} 
            \begin{tikzpicture}[overlay, remember picture]
                \draw[thick, ->, black] (-1.5,1.75) to[out=30, in=150] (0.5,1.75);
            \end{tikzpicture}
        } &
        \adjustbox{valign=m}{\includegraphics[width=0.25\textwidth]{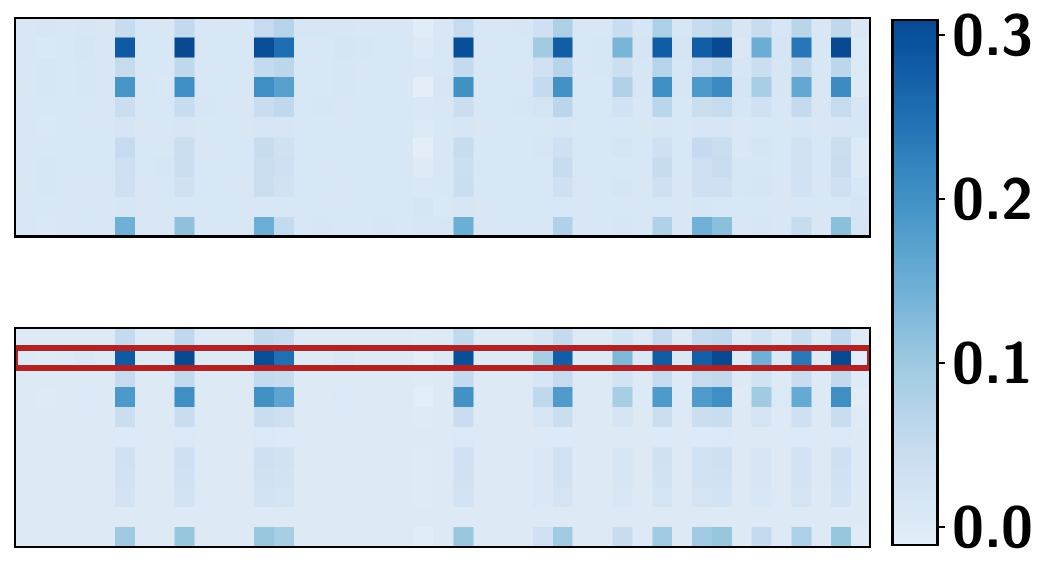}} 
    \end{tabular}
    \caption{\label{fig:first_figure} \Hyperspectral dataset. On the left, we show the values of the rows in a $50 \times 50$ matrix and the nearly-proportional values of the rows in a near-rank-$1$ $11 \times 44$ sub\-matrix discovered by our method. On the right we show the matrix and, next to it, the discovered sub\-matrix (top) and its accurate approximation expressing each row as collinear with the row highlighted in red (bottom). 
    }
\end{figure}

While \svd may fail to reveal dense lines in the data,
it is possible to find such structures by sampling. 
A na\"ive approach would be to sample subsets of points and dimensions 
until a large set of nearly-collinear points is found. 
However, this procedure quickly becomes inefficient. 
Instead, to identify points approximately distributed along a line, we introduce a method that only relies on sampling in an initialization phase to find a minimal structure that can exhibit this property, i.e., two points in two-dimensions. 
In a subsequent phase, the $2 \times 2$ sub\-matrix is expanded deterministically to obtain the entire subset of points and dimensions associated with the target line. 
Based on such a two-phase method, we discover arbitrary sub\-matrices that admit low-rank approximations that can be easily interpreted in terms of the original sub\-matrix rows or columns, and are supported by quality guarantees. A real-world example is given in Figure~\ref{fig:first_figure}. 


\spara{Our contributions.}
Our main contributions can be summarized as follows. 

\begin{itemize}
    \item We formalize the problem of finding sub\-matrices that are provably close to a low-rank approximation. 
    
    \item We introduce an effective method for finding sub\-matrices that are provably close to rank $1$. Then, we generalize this method to the 
    case of rank-\generalrank sub\-matrices. 

    \item We analyze the theoretical properties of the method we introduce. 

    \item We demonstrate the advantages of our method over previous work. 
\end{itemize}

\spara{Roadmap.} The rest of the paper is organized as follows. 
In Section~\ref{sec:related} we discuss related work. 
Section~\ref{sec:preliminaries} introduces the notation used throughout the paper as well as key preliminary concepts. 
In Section~\ref{sec:prob} we present the problem we study and in Section~\ref{sec:algorithms} we illustrate our method to address it. In 
Section~\ref{sec:analysis} we analyze the properties of the method, and in Section~\ref{sec:experiments} we assess its empirical performance.
Finally, Section~\ref{sec:conclusion} provides conclusions.


\section{Related Work}
\label{sec:related}

\para{Low-rank matrix approximation.}
Low-rank approximation techniques are widely used 
to decompose a matrix into simpler components, 
capturing essential patterns while reducing noise and dimensionality. 
The \svd and the related \pca are among the most popular 
techniques~\cite{golub2013matrix}. 
Nonnegative matrix factorization techniques~\cite{gillis2010using} 
have become popular in applications where the goal is to decompose data in nonnegative components. 
Boolean matrix decomposition relies on boolean algebra instead of linear algebra~\cite{miettinen2020recent}.
Column subset selection~\cite{boutsidis2009improved} and the CUR decomposition~\cite{mahoney2009cur}   
have emerged as more interpretable alternatives to the \svd. 
In 2019, Gillis and Shitov~\cite{gillis2019low}  studied the problem of low-rank approximation 
to minimize the maximum entry-wise deviation. 
More recently, an approach to low-rank approximation that accounts for multiplicative effects was introduced~\cite{ciaperoni2024hadamard}. 

\spara{Local low-rank matrix approximation.} 
Relatively less research has been conducted for finding decompositions
that do not assume a \emph{global} low-rank structure, 
which is the focus of our paper.
The goal here is to find sub\-matrices that are \emph{locally} well-approximated by a low-rank structure.
%
A simple heuristic to local low-rank approximation 
is obtained by imposing a sparsity constraint on matrix decomposition, 
and sparse PCA~\cite{mairal2009online} is a prominent example of such~methods.

Doan and Vavasis proposed the problem of recovering near-rank-$1$ sub\-matrices 
by framing it as a convex-optimization problem  
\cite{doan2013finding}. 
Lee et al.~\cite{lee2016llorma} introduced the LLORMA method to address matrix-completion tasks while relaxing the assumption that the entire matrix has low rank. 
LLORMA approximates the entire input matrix, 
and thus, it is fundamentally different from our work, 
which focuses on detecting local low-rank patterns. 
On the other hand, the problem we study finds application in matrix completion, 
as shown by the work of Ruchansky et al.~\cite{ruchansky2017targeted}, which introduces the \svp method to quickly detect low-rank sub\-matrices with the ultimate goal of improving the accuracy in matrix completion.  
While \svp cannot discover arbitrary low-rank sub\-matrices, Dang et al.~\cite{dang2023generalized} introduced the \rpsp method, which addresses this lack of generality. 
The core idea behind \rpsp is to sample sub\-matrices and count
the number of times that each entry belongs to a low-rank sub\-matrix. 
%
Like \rpsp, our method targets arbitrary near-low-rank sub\-matrices. Unlike previous methods, our method can in principle identify  sub\-matrices that are close to a specific target rank.

\spara{Co-clustering, projective clustering, and subspace clustering.}
Co-cluster\-ing  algorithms~\cite{dhillon2001co}
simultaneously cluster the rows and columns of a matrix. 
Although co-clustering algorithms can be used 
for detecting low-rank submatrices,  
they cannot generally discover such structures, 
except in specific cases where the values of the low-rank submatrices deviate significantly from the background. 
Projective clustering and subspace clustering are also related problems. 
In projective clustering~\cite{agarwal2003approximation}, 
the goal is to partition the data into subsets such that the points in each subset are 
close to each other in some subspace. 
In subspace clustering~\cite{vidal2011subspace} the goal is to find a representation of the input data as 
a union of different subspaces. 
In general, clustering problems are fundamentally different from the problem we study, 
as they seek a partitioning of the entire data matrix.

\section{Preliminaries}\label{sec:preliminaries}


\para{Notation and basic definitions.}
Matrices are denoted by upper-case boldface letters, and we use $\inputmat$ to denote the input data matrix.  
$\inputmat_{i,j}$ indicates the entry of \inputmat in row $i$ and column $j$, while the $i$-th row and $j$-th column of \inputmat are denoted by $\inputmat_{i,:}$ and $\inputmat_{:,j}$, respectively. 
Sets are denoted by upper-case letters and scalars by lower-case letters. 
Vectors are denoted by lower-case boldface letters, e.g., $\xvec = (x_1, \dots, x_d)$.
We denote the $L_2$ (or Euclidean) norm of a vector $\xvec$ as $\| \xvec \|_2$ and 
the inner product between two vectors \xvec and \yvec as $\xvec^T \yvec$. 
We consider different matrix norms: 
the Frobenius norm  $\frobeniusnormshort{\inputmat} = (\sum_{i,j} |\inputmat_{i,j}|^2)^{1/2}$, 
the spectral norm $\spectralnormshort{\inputmat} = \sup_{||\xvec||_2\le 1} ||\inputmat\xvec||_2$, and 
the max norm 
$\infinitynormshort{\inputmat} = \max_{i,j} |\inputmat_{i,j}|$.
We use $\generalnormshort{\inputmat}$ with $*= \{ 2, F, \text{max}\}$ to indicate the above norms. 
We refer to the total number of entries of a (sub)matrix  \inputmat\ as its \emph{size}, which is also simply  denoted by~$\abs{\inputmat}$, if there is no risk of confusion with the entry-wise absolute value. 
For a matrix \outputmat, we denote by $\estimate{\outputmat}$ a low-rank approximation of \outputmat and by 
$\outputerror$
the difference
$\outputerror = \outputmat - \estimate{\outputmat}$.



\spara{Orthogonal projections.}
Given a nonzero vector $\xvec$ and a  vector $\yvec$, the \emph{orthogonal projection} of $\yvec$ onto $\xvec$ 
is given by
$\proj{\xvec}{\yvec} = ({\yvec^T \xvec})/({\xvec^T \xvec}) \xvec$.
Similarly, given a matrix~$\bfB$ and a matrix $\bfA$ with linearly independent columns, 
the \emph{orthogonal projection} of~$\bfB$ onto the column space of $\bfA$ is given by
$\proj{\bfA}{\bfB} =  \bfA \bfA^+ \bfB$, where $\bfA^+ = (\bfA^T \bfA)^{-1} \bfA^T$ is the 
%
\emph{Moore-Penrose pseudoinverse} of $\bfA$.  
The orthogonal projection
$\proj{\bfA}{\bfB}$ is the closest matrix to \bfB in the column space of \bfA under the Frobenius~norm.



\spara{Low-rank approximation and \svd.}
The \emph{singular value decomposition} (\svd) of a matrix \(\inputmat \in \mathbb{R}^{\nrows \times \ncols}\) 
is given by $\inputmat = \bfU \bfSigma \bfV^T$,
where
\(\bfU \in \mathbb{R}^{\nrows \times \nrows}\) and  \(\bfV \in \mathbb{R}^{ \ncols  \times  \ncols  }\) 
are unitary matrices, and
\(\bfSigma \in \mathbb{R}^{ \nrows \times \ncols }\) is a diagonal matrix 
with singular values $\{ \sigma_1, \sigma_2 \dots \sigma_{\min\{\nrows,\ncols\}} \}$  as diagonal entries, conventionally sorted in decreasing order.
If the matrix is not clear from the context, we denote as $\sigma_i(\outputmat)$ the $i$-th singular value of $\outputmat$. 

It is known that 
the optimal rank-\generalrank approximation of \inputmat for the Frobenius and the spectral norm (but not for the max norm) is obtained from the \svd
by retaining the first 
\generalrank~singular values, along with the associated \generalrank~columns  of $\bfU$ and \generalrank~rows of $\bfV^T$\,\cite{golub2013matrix}. 
The largest singular value of a matrix equals its spectral norm, and the number of non-zero singular values indicates the rank of the matrix. 
As real-world data are often noisy, the singular values are seldom exactly zero.  
Accordingly, to measure the proximity of a matrix to rank $1$, in this work, 
we use the \emph{low-rankness score}~\cite{dang2023generalized}, which is given by 
\( \lowrankness{\outputmat} = \frac{\sigma_1(\outputmat)^2}{  \sum_{i = 1}^{\min(\nrows, \ncols)} \sigma_i(\outputmat)^2} \). 
A matrix whose singular values after the $\generalrank$-th one are close to zero
can be accurately approximated by a rank-\generalrank matrix, and is loosely referred to as \emph{near-rank-\generalrank} matrix. 
\smallskip

\section{Problem Formulation}\label{sec:prob}

Next, we formalize the problems we study in this paper. 
To provide better insight, we first present a special case,
and then introduce the more general problems.


\spara{Searching for a near-rank-$\mathbf{1}$ subset of rows or columns.}
As a warm-up, 
we introduce a simpler problem that fixes the matrix columns or rows.
In other words, in this simplified scenario, we are not looking for 
 arbitrary sub\-matrices, but for sub\-matrices that  include all rows or all columns of the input matrix. 

\begin{problem}[Largest near-rank-$1$ subset of rows (\simplifiedproblem)]
\label{prob:simplified}
    Given a matrix \inputmat $\in \real^{\nrows \times \ncols}$  with set of rows \rowset and a threshold $\errorbound \in \real^+$, 
    find the largest subset of rows $\rowset' \subseteq \rowset$ such that there exist a 
    rank-$1$ matrix $\estimate{\outputmat} \propto \xvec \yvec^T$, where 
    $\yvec^T \in \real^\ncols$ is a row of \inputmat, 
    satisfying
    \begin{equation}
    \label{equation:problem-rows-only}
        \|\inputmat_{i,:} - \estimate{\outputmat}_{i,:}\|_2 \leq \errorbound, \quad ~~\mbox{for all}~~ i \in \rowset'. 
    \end{equation}
\end{problem}

Problem~\ref{prob:simplified} asks to find the largest near-rank-$1$ sub\-matrix
defined over a subset of rows of \inputmat and all columns.
This problem 
is computationally tractable. 


\begin{proposition}
The \simplifiedproblem 
problem can be solved in polynomial time.
\end{proposition}
The proof, via a simple algorithm, 
is presented in~\ref{app:specialcase}.

While Problem~\ref{prob:simplified} asks for a subset of rows, the symmetric problem asking for a subset of columns can be solved simply by considering~$\inputmat^T$ in place of $\inputmat$.

\spara{Searching for a near-rank-$\mathbf{1}$ sub\-matrix.}
Next, we discuss the more challenging problem of finding a near-rank-$1$ sub\-matrix,
without fixing neither the rows nor the columns of the input matrix. 

\begin{problem}[Largest near-rank-$1$ sub\-matrix (\nearrankone)]
\label{prob:rank-1}
    Given a matrix \inputmat $\in \real^{\nrows \times \ncols}$, and a threshold $\errorbound \in \real^+$, 
    find a sub\-matrix \outputmat $\in \real^{\nrowssubmatrix \times \ncolssubmatrix}$ of maximum size 
    such that there exist a rank-$1$ matrix $\estimate{\outputmat}$ 
    satisfying
    \begin{equation}
    \label{problem:near-rank-1-submatrix}
       \generalnormshort{\outputerror} =  \generalnormshort{\outputmat - \hat{\outputmat}} \leq \errorbound, ~
            \text{ where } * \text{ can be any of the norms } \{F, 2, \text{max} \}. 
    \end{equation}
\end{problem}

Unfortunately, due to the interaction between rows and columns,
the \nearrankone problem is computationally intractable.

\begin{proposition}
\label{proposition:np-hardness}
The \nearrankone problem is \np-hard. 
\end{proposition}

The $\mathbf{NP}$-hardness of \nearrankone follows from that of the largest rank-$1$ sub\-matrix problem~\cite{doan2013finding}
by setting $\errorbound=0$, and 
highlights the connection with the \emph{maximum-edge biclique} problem~\cite{lyu2020maximum}, 
which is made evident in Section~\ref{sec:algorithms}.

\spara{Searching for a near-rank-$\mathbf{k}$ submatrix.}
We generalize the \nearrankone problem to the case of near-rank-$k$ sub\-matrices. 

\begin{problem}[Largest near-rank-$k$ submatrix (\nearrankk)]
\label{prob:rank-k}
    Given a matrix \inputmat $\in \real^{\nrows \times \ncols}$, and a threshold $\errorbound \in \real^+$, 
    find a sub\-matrix \outputmat $\in \real^{\nrowssubmatrix \times \ncolssubmatrix}$ of maximum size 
    such that there exist a rank-$k$ matrix $\estimate{\outputmat}$ 
    satisfying
    \begin{equation}
        \generalnormshort{\outputmat - \hat{\outputmat}} \leq \errorbound, ~
         \text{ where } * \text{ can be any of the norms } \{F, 2, \text{max} \}. 
    \end{equation}
\end{problem}


As \nearrankk is a generalization of \nearrankone, the \nearrankk problem is also \nphard. 




\spara{Extensions.}
The problem formulations presented above focus on extracting a single sub\-matrix. 
In practice, one may wish to find a representation of the input matrix
as a sum of $N$ local low-rank patterns.
Such a problem is a generalization of both \nearrankone and \nearrankk, and hence, 
inherits their hardness. 

Additionally, it may be of interest to identify sub\-matrices that define affine subspaces. 
Extending our problem formulations and method to the case of affine subspaces (or  
near-low-rank sub\-matrices up to a particular translation) is straightforward. 
The details are deferred to an extended version of this work.

\section{Algorithms}\label{sec:algorithms}
In this section, we present \ourmethod, our method for discovering near-low-rank sub\-matrices. We first give an overview of the method, and then we present the algorithms to detect near-rank-$1$ and near-rank-\generalrank sub\-matrices.


\subsection{High-level Overview of the Method}\label{sec:algo_overview}
\ourmethod is based on a simple two-phase procedure.
The first phase \emph{samples} small {seed} sub\-matrices, 
and the second phase \emph{expands} those seed sub\-matrices into larger near-low-rank sub\-matrices.

The main idea relies on the simple principle that any sub\-matrix of a rank-$k$ matrix 
must also have rank at most $k$. 
Thereby, a near-rank-$1$ sub\-matrix \outputmat of size $\nrowssubmatrix \times \ncolssubmatrix$ 
contains a large number of $2 \times 2$ near-rank-$1$ sub\-matrices. 
Thus, if we are looking for a rank-1 sub\-matrix, 
in the first phase (initialization or sampling phase) we identify a \emph{seed}, 
which is a $2 \times 2$ submatrix 
that can expanded into a larger sub\-matrix that is still close to a rank-1 approximation. 
The goal of the second phase (or expansion phase) is to expand the seed into a large near-rank-1 sub\-matrix. 
Similarly, if we are looking for a near-rank-$\generalrank$ sub\-matrix, in the first phase we identify a seed sub\-matrix of minimal size that is close to rank $\generalrank$. 
In the second phase, the seed is expanded as much as possible while preserving the proximity to rank $\generalrank$. 

The two-phase procedure is repeated~\nrep\ times, to explore different random initializations. 
Each repetition outputs a near-low-rank sub\-matrix \outputmat.
\ourmethod accepts a parameter $\tol$ that controls the trade-off between proximity to a low-rank approximation and size of the output matrices.  
Higher values of $\tol$ tend to yield sub\-matrices that are larger but deviate more from a low-rank structure. 


After the last repetition, we return the sub\-matrix \outputmat that maximizes the objective function 
$\f{\outputmat} = \abs{\outputmat} - \frac{\lambda}{\abs{\outputmat}} \frobeniusnormshort{\outputerror}^2$. 
By default, in the absence of prior information, we standardize the error term and the size, and set $\lambda=1$. 
However, \ourmethod is flexible and supports any 
objective function. 

The high-level pseudocode of the \ourmethod method is given in~Algorithm~\ref{alg:general}.
The details of the initialization and expansion phase of the method for the algorithm specialized to near-rank-$1$ sub\-matrix detection and for the algorithm for general near-rank-$\generalrank$ sub\-matrix detection, described later, are different.

\makeatletter
\renewcommand{\algorithmiccomment}[1]{\hfill {\color{black!80} \textit{// #1}}}
\makeatother

\begin{algorithm}[t]
\begin{algorithmic}[1]
\STATE \textbf{Input:} Matrix \inputmat, target rank $\generalrank$, number of initializations $\nrep$, initial tolerance $\toleranceinit$, tolerance $\tol$. 
\STATE \textbf{Output:} Near-rank-$\generalrank$ sub\-matrix $\outputmat^*$. 
\STATE $\outputmat^* \gets \mathbf{0}$
\FOR{$i = 1$ to $\nrep$}
    \STATE $\initialmat \gets $ \textbf{Initialization}(\inputmat, \generalrank, \toleranceinit)\COMMENT{first phase: initialization  (sampling)} 
    \STATE $\outputmat \gets $ \textbf{Expansion}(\initialmat, \generalrank, \tol)\COMMENT{second phase: expansion} 
    \IF{$\f{\outputmat} \geq \f{\outputmat^*}$}
    \STATE $\outputmat^* \gets \outputmat$\COMMENT{select the best sub\-matrix across different initializations}   
\ENDIF
\ENDFOR
\STATE \textbf{Return} $\outputmat^*$
\end{algorithmic}
\caption{Overview of \ourmethod.} 
\label{alg:general} 
\end{algorithm}

\spara{Approximating the discovered sub\-matrices.}
The discovered sub\-matrices can be approximated via \svd. 
Further, 
\ourmethod also naturally leads to a low-rank approximation that is more interpretable than the \svd since it is based on the rows (or columns) of \outputmat. 
For the rank-$1$ case, this approximation is given by $\estimate{\outputmat} = \xvec \yvec^T$, where, either $\yvec^T$ is a row or $\xvec$ a column of $\outputmat$. If, e.g., $\yvec^T$ is a row of \outputmat, $\xvec$  can be chosen to minimize \( \frobeniusnormshort{\outputmat - \xvec \yvec^T}^2 \). The resulting optimal $\xvec$ is the vector of coefficients that describe the orthogonal projections of the rows of \outputmat onto $\yvec^T$. An analogous argument also applies to the columns. 
As discussed in Section~\ref{sec:analysis}, this approximation is supported by quality guarantees. 

Although the rank-$1$ \svd may be more accurate than the interpretable alternative, if a matrix is sufficiently close to rank $1$, the difference is typically negligible. To gain some intuition for this claim, note that a matrix that has exactly rank $1$ can be represented with no error not only by the discussed interpretable approximation, but also by the rescaled outer product \( \alpha \outputmat_{:,j}  \outputmat_{i,:}^T \) of any of its rows and columns, for some $\alpha \in \real$.  If instead the matrix deviates significantly from  rank $1$, the rank-$\generalrank$ interpretable approximation based on orthogonal projections is often not as accurate as the rank-$\generalrank$ \svd. 


\spara{Discovering multiple sub\-matrices.}
In practice, we may wish to discover multiple sub\-matrices within a single matrix \inputmat and eventually obtain an approximation $\estimate{\inputmat}$ of the matrix as sum of local low-rank patterns. To achieve this, we run \ourmethod iteratively. In each iteration, the method finds a single near-rank-$\generalrank$ sub\-matrix, and then updates $\estimate{\inputmat}$ and the input matrix for the next iteration. This simple procedure is summarized in~Algorithm~\ref{alg:iterated}. 

\begin{algorithm}[t]
\begin{algorithmic}[1]
\STATE \textbf{Input:} Matrix \inputmat, vector of target ranks $\mathbf{\generalrank} \in \real^{\nrows \times \ncols}$, number of initializations $\nrep$, number of near-low-rank sub\-matrices $N_{patterns}$, initial tolerance $\toleranceinit$, tolerance $\tol$. 
\STATE \textbf{Output:} Estimate $\estimate{\inputmat}$ of \inputmat. 

\STATE $\estimate{\inputmat} \gets \mathbf{0}$
\FOR{$h = 1$ to $N_{patterns}$}

    \STATE $\estimate{\outputmat}_h \gets \mathbf{0}, \rowset_h \gets \emptyset, \colset_h \gets \emptyset$ \COMMENT{initialize best sub\-matrix across target ranks}
    
    \FOR{$k \in \mathbf{\generalrank}$}

    \STATE $\outputmat, \estimate{\outputmat}, \rowset , \colset \gets  
    \textsc{FindNearLowRankSubmatrix}(\inputmat, \generalrank,  \nrep, \toleranceinit, \tol)$ \COMMENT{return a near-rank-$k$ sub\-matrix with its rank-$k$ estimate, row and column indices} 

    \IF{$f(\estimate{\outputmat}) > f(\estimate{\outputmat}_h)$}
    \STATE $\estimate{\outputmat}_h \gets \estimate{\outputmat},\rowset_h \gets \rowset, \colset_h \gets \colset$\COMMENT{update best sub\-matrix across target ranks}
    \ENDIF
    \ENDFOR
    \STATE $\estimate{\inputmat}_{\rowset_h , \colset_h} = \estimate{\inputmat}_{\rowset_h , \colset_h}  + \estimate{\outputmat}_h $\COMMENT{update current estimate}

    \STATE $ \inputmat_{\rowset_h , \colset_h} \gets \inputmat_{\rowset_h , \colset_h} - \estimate{\outputmat}_h $\COMMENT{update input for the next iteration}
    \ENDFOR
\STATE \textbf{Return} $\estimate{\inputmat}$
\end{algorithmic}
\caption{\label{alg:iterated} Iterative algorithm to find multiple near-rank-$k$ sub\-matrices.}
\end{algorithm}

\subsection{Recovering a Near-rank-1 Sub\-matrix}
\label{sec:algo_rank1}

Here, we present the initialization (sampling) and expansion phases of the algorithm to discover near-rank-$1$ sub\-matrices. Algorithm~\ref{alg:rankone} presents the pseudo\-code.

\begin{algorithm}[t!]
\begin{algorithmic}[1]
\STATE \textbf{Input:} Matrix $\inputmat \in \real^{\nrows \times \ncols}$, number of initializations $\nrep$, initial tolerance $\toleranceinit$, tolerance $\tol$. 
\STATE \textbf{Output:} Near-rank-$1$ sub\-matrix $\outputmat^*$. 
\STATE $\outputmat^* \gets \mathbf{0}$\COMMENT{start initialization phase}
\STATE $\Omega \gets \{ (i,j) \}, \quad \forall i \in \{1, \dots, \nrows \}, \forall j \in \{1, \dots, \ncols\}$
\FOR{$t = 1$ to $\nrep$}
    \STATE \( \Gamma' \gets \{ (i_1, j_1), (i_2, j_2)  \sim Uniform(\Omega) \} \)\COMMENT{sample two entries at random}
    \STATE \( \Gamma \gets \ \Gamma' \cup \{ (i_2, j_1), (i_1, j_2) \} \)\COMMENT{complete $2 \times 2$ sub\-matrix} 
    \STATE \( \initialmat'_{i_h, j_h} \gets \inputmat_{i_h, j_h}, \quad \forall (i_h , j_h) \in \Gamma \) 
    \WHILE{$ |\det(\initialmat')| > \toleranceinit$}
    \STATE  \( \Gamma' \gets \{ (i_1, j_1), (i_2, j_2)  \sim Uniform(\Omega) \} \)\COMMENT{repeat until $\initialmat'$ is close to rank $1$} 
    \STATE \( \Gamma \gets \ \Gamma' \cup \{ (i_2, j_1), (i_1, j_2) \} \) 
    \STATE \( \initialmat'_{i_h, j_h} \gets \inputmat_{i_h, j_h}, \quad \forall (i_h , j_h) \in \Gamma \) 
    \ENDWHILE
    \STATE $\initialmat \gets \initialmat'$\COMMENT{start expansion phase}  
    \STATE  $(i_a, j_a) \sim Uniform(\Gamma)$ \COMMENT{select anchor} 
    \STATE $(\anchorrow, \anchorcolumn) \gets (\inputmat_{i_a,:}, \inputmat_{:,j_a})$
    \STATE \( \rowratiosij \gets \frac{\inputmat_{i,j}}{\anchorrow_j}, \quad \forall i \in \{1, \dots, \nrows \}, \forall j \in \{1, \dots, \ncols \} \)\COMMENT{compute row-wise ratios} 
    \STATE \( \columnratiosij \gets \frac{\inputmat_{i,j}}{\anchorcolumn_i}, \quad \forall i \in \{1, \dots, \nrows \}, \forall j \in \{1, \dots, \ncols \} \)\COMMENT{compute column-wise ratios}
    %
    \STATE \( \indrows \gets \mathbf{0}\)\COMMENT{compute indicator matrix for the rows}
    \FOR{$i=1$ to $\nrows$}
    \STATE $\Psi_i \gets \argmax_{\{ \Psi \subseteq \{ 1, \dots , \ncols \} \mid \anchorindexcol \in \Psi, \,  | \rowratios_{i,j_1}- \rowratios_{i,j_2} | \leq \tol \,  \forall j_1, j_2 \in \Psi  \}} |\Psi|$ 
    \STATE \( \indrows_{i, \Psi_i} \gets 1 \)
    \ENDFOR
    \STATE \( \indcols \gets \mathbf{0} \)\COMMENT{compute indicator matrix for the columns}
    \FOR{$j=1$ to $\ncols$}
    \STATE $\Psi_j \gets \argmax_{\{ \Psi \subseteq \{ 1, \dots , \nrows \} \mid \anchorindexrow \in \Psi, \,  | \columnratios_{i_{1},j}- \columnratios_{i_{2},j} | \leq \tol \,  \forall i_1, i_2 \in \Psi  \}} |\Psi|$ 
    \STATE \( \indcols_{  \Psi_j  , j  } \gets 1 \)
    \ENDFOR 
    \STATE \( \ind \gets \indrows \cap \indcols \) \COMMENT{compute indicator matrix} 
    \STATE \( \outputmat \gets \textsc{ExtractMaximumEdgeBiclique}(\ind) \) 
    \IF{$f(\outputmat) > f(\outputmat^*)$}
    \STATE $\outputmat^* \gets \outputmat$  
    \ENDIF
    \ENDFOR
\STATE \textbf{Return} $\outputmat^*$
\end{algorithmic}
\caption{\label{alg:rankone} Algorithm to find a near rank-$1$ sub\-matrix.}
\end{algorithm}


\spara{Initialization.}
To find the initial $2 \times 2$ near-rank-$1$ sub\-matrix $\initialmat$, we sample
two distinct row indices $\{i_1,i_2\}$ and  column indices $\{j_1,j_2\}$ of the input matrix \inputmat,
and then we compute the determinant of the associated $2 \times 2$ sub\-matrix $\initialmat'$:  
\begin{equation*}
   \abs{\det(\initialmat')} = \abs{\inputmat_{i_1, j_1} \inputmat_{i_2,j_2} -  \inputmat_{i_1, j_2} \inputmat_{i_2,j_1} }. 
\end{equation*}
If $\abs{\det(\initialmat')} \leq \toleranceinit$,
for some input $\toleranceinit \in \real^+$,  $\initialmat'$
is close to rank $1$, and hence it may be contained into a larger near-rank-$1$ sub\-matrix.
Therefore, $\initialmat'$ is the seed $\initialmat$ that will be expanded. 
If instead $\abs{\det(\initialmat')} > \toleranceinit$, we sample different $2 \times 2$ sub\-matrices $\initialmat'$ until we find a seed to expand.  
In practice, \toleranceinit is initialized to a small value ($10^{-11}$ by default) and progressively increased until the seed is found. 

\spara{Expansion.}
To extend \initialmat into a larger sub\-matrix, 
we consider one of the entries $(\anchorindexrow, \anchorindexcol)$ in \initialmat, which we call \emph{anchor}. 
Then, we divide all rows in \inputmat by the $\anchorindexrow$-th row, obtaining the row-wise ratio matrix 
$\rowratios$ and 
all columns by the $\anchorindexcol$-th column, obtaining the column-wise ratio matrix 
$\columnratios$. 
If an entry in the $\anchorindexrow$-th row or $\anchorindexcol$-th column of \inputmat is zero, we add a small positive constant to prevent division by zero. 

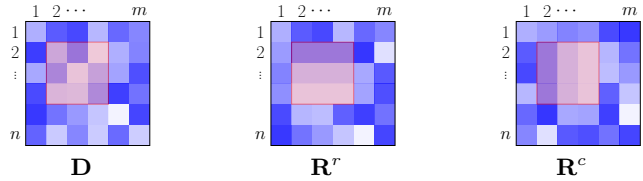
\begin{figure}[t]
\centering
\begin{tabular}{c@{\hspace{1.25cm}}c@{\hspace{1.25cm}}c} 
    \scalebox{0.55}{\scalebox{0.5}{
\begin{tikzpicture}

\pgfmathsetseed{555}

\foreach \i in {0, 1, 2, 3, 4, 5} {
    \foreach \j in {0, 1, 2, 3, 4, 5} {
        \pgfmathsetmacro{\randcolor}{rnd * 100} 
        \fill[blue!\randcolor, opacity=0.8] (\j, 6-\i) rectangle (\j+1, 6-\i-1);
    }
}

\draw[thick] (0, 0) rectangle (6, 6);

\node at (-0.5, 5.5) {\Huge 1};
\node at (-0.5, 4.5) {\Huge 2};
\node at (-0.5, 3.5) {\Huge $\vdots$};
\node at (-0.5, 0.5) {\Huge $\nrows$};
\node at (0.5, 6.5) {\Huge 1};
\node at (1.5, 6.5) {\Huge 2};
\node at (2.5, 6.5) {\Huge $\cdots$};
\node at (5.5, 6.5) {\Huge $\ncols$};

\draw[thick, red, fill=red!30, opacity=0.5] (1, 2) rectangle (4, 5);

\end{tikzpicture}
}} &
    \scalebox{0.55}{\scalebox{0.5}{
\begin{tikzpicture}
\pgfmathsetseed{555}

\foreach \i in {0, 1, 2, 3, 4, 5} {
    \ifnum\i>0\relax 
        \ifnum\i<4\relax
            \pgfmathsetmacro{\rowcolor}{rnd * 100} 
        \fi
    \fi

    \foreach \j in {0, 1, 2, 3, 4, 5} {
        \ifnum\i>0\relax 
            \ifnum\i<4\relax
                \ifnum\j>0\relax 
                    \ifnum\j<4\relax
                        \fill[blue!\rowcolor, opacity=0.8] (\j, 6-\i) rectangle (\j+1, 6-\i-1);
                    \else
                        \pgfmathsetmacro{\randcolor}{rnd * 100}
                        \fill[blue!\randcolor, opacity=0.8] (\j, 6-\i) rectangle (\j+1, 6-\i-1);
                    \fi
                \else
                    \pgfmathsetmacro{\randcolor}{rnd * 100}
                    \fill[blue!\randcolor, opacity=0.8] (\j, 6-\i) rectangle (\j+1, 6-\i-1);
                \fi
            \else
                \pgfmathsetmacro{\randcolor}{rnd * 100}
                \fill[blue!\randcolor, opacity=0.8] (\j, 6-\i) rectangle (\j+1, 6-\i-1);
            \fi
        \else
            \pgfmathsetmacro{\randcolor}{rnd * 100}
            \fill[blue!\randcolor, opacity=0.8] (\j, 6-\i) rectangle (\j+1, 6-\i-1);
        \fi
    }
}

\draw[thick] (0, 0) rectangle (6, 6);

\node at (-0.5, 5.5) {\Huge 1};
\node at (-0.5, 4.5) {\Huge 2};
\node at (-0.5, 3.5) {\Huge $\vdots$};
\node at (-0.5, 0.5) {\Huge $\nrows$};
\node at (0.5, 6.5) {\Huge 1};
\node at (1.5, 6.5) {\Huge 2};
\node at (2.5, 6.5) {\Huge $\cdots$};
\node at (5.5, 6.5) {\Huge $\ncols$};

\draw[thick, red, fill=red!30, opacity=0.5] (1, 2) rectangle (4, 5);
\end{tikzpicture}
}} &
    \scalebox{0.55}{\scalebox{0.5}{
\begin{tikzpicture}
\pgfmathsetseed{555}

\foreach \j in {0, 1, 2, 3, 4, 5} {  
    \ifnum\j>0\relax 
        \ifnum\j<4\relax
            \pgfmathsetmacro{\colcolor}{rnd * 100} 
        \fi
    \fi

    \foreach \i in {0, 1, 2, 3, 4, 5} {  
        \ifnum\i>0\relax 
            \ifnum\i<4\relax
                \ifnum\j>0\relax 
                    \ifnum\j<4\relax
                        \fill[blue!\colcolor, opacity=0.8] (\j, 6-\i) rectangle (\j+1, 6-\i-1);
                    \else
                        \pgfmathsetmacro{\randcolor}{rnd * 100}
                        \fill[blue!\randcolor, opacity=0.8] (\j, 6-\i) rectangle (\j+1, 6-\i-1);
                    \fi
                \else
                    \pgfmathsetmacro{\randcolor}{rnd * 100}
                    \fill[blue!\randcolor, opacity=0.8] (\j, 6-\i) rectangle (\j+1, 6-\i-1);
                \fi
            \else
                \pgfmathsetmacro{\randcolor}{rnd * 100}
                \fill[blue!\randcolor, opacity=0.8] (\j, 6-\i) rectangle (\j+1, 6-\i-1);
            \fi
        \else
            \pgfmathsetmacro{\randcolor}{rnd * 100}
            \fill[blue!\randcolor, opacity=0.8] (\j, 6-\i) rectangle (\j+1, 6-\i-1);
        \fi
    }
}

\draw[thick] (0, 0) rectangle (6, 6);

\node at (-0.5, 5.5) {\Huge 1};
\node at (-0.5, 4.5) {\Huge 2};
\node at (-0.5, 3.5) {\Huge $\vdots$};
\node at (-0.5, 0.5) {\Huge $\nrows$};
\node at (0.5, 6.5) {\Huge 1};
\node at (1.5, 6.5) {\Huge 2};
\node at (2.5, 6.5) {\Huge $\cdots$};
\node at (5.5, 6.5) {\Huge $\ncols$};

\draw[thick, red, fill=red!30, opacity=0.5] (1, 2) rectangle (4, 5);
\end{tikzpicture}
}} \\
    \inputmat & \rowratios & \columnratios
\end{tabular}
\caption{\label{fig:ratios_diagram}
Example of row-wise (\rowratios) and column-wise (\columnratios) ratio matrices associated with an input matrix (\inputmat) containing a rank-$1$ sub\-matrix (highlighted in red). 
Within this rank-$1$ sub\-matrix, the entries of $\rowratios$ are constant across rows, and the entries of $\columnratios$ are constant across columns.
}
\end{figure}

As illustrated in Figure~\ref{fig:ratios_diagram}, if the matrix \inputmat contains a sub\-matrix \outputmat of rank $1$, 
the entries corresponding to \outputmat in \rowratios and \columnratios will be row-wise and column-wise constant, respectively.
More generally, as we explain in in Section~\ref{sec:analysis}, 
bounding the variation in all the row-wise and column-wise ratios in a sub\-matrix leads to quality guarantees for its rank-$1$ approximation. 
Therefore, the goal of the expansion stage is to identify a sub\-matrix of maximum size with bounded variation in the row-wise and column-wise ratios. 

To this end, our algorithm examines the rows of \rowratios to find subsets of near-constant entries including column~$\anchorindexcol$ and 
the columns of \columnratios to find subset of near-constant entries including row $\anchorindexrow$. 
More specifically, for an input parameter $\tol \in \real^+$, 
we select, for each row of \rowratios, the $\anchorindexcol$-th entry and all other entries such that the maximum variation is less than $\tol$ in absolute value. Similarly, for each column of \columnratios, we select  the $\anchorindexrow$-th entry and all other entries such that the maximum variation is less than $\tol$ in absolute value. 
Subsets within each row can be efficiently retrieved by sorting the row elements by their absolute deviation from the $\anchorindexcol$-th element and analogously for the columns. 

Given the identified subsets, we construct two indicator matrices: $\indrows \in {0,1}^{\nrows \times \ncols}$, where the entries with value $1$ correspond to subsets of near-constant row-wise ratios; and $\indcols \in {0,1}^{\nrows \times \ncols}$, where the entries with value $1$ correspond to subsets of near-constant column-wise ratios.
%
We can then compute the intersection of the two matrices $\indrows$ and $\indcols$ to obtain the intersection indicator matrix $\ind$ of the same dimensions. 
The problem of extracting a sub\-matrix  of maximum size  with all row-wise and column-wise ratios with bounded variations can then be framed as the problem of finding the largest possible all-ones sub\-matrix within \ind.
This problem is equivalent to the extraction of a  maximum-edge bi\-clique~\cite{lyu2020maximum} from the bipartite graph $\mathcal{G}_\ind$ that has \ind as adjacency matrix. 
Although this is an \np-hard problem~\cite{lyu2020maximum}, so that it cannot be solved in polynomial time, we can leverage recent algorithmic advances that solve the problem quickly in considerably dense and large bipartite graphs~\cite{lyu2020maximum}. 
In addition, to avoid possible scalability issues that may still arise, we also rely on effective heuristics, as discussed in Section~\ref{sec:scalability}. 

\subsection{Recovering a Near-rank-$\generalrank$ Sub\-matrix}
\label{sec:algo_general_susbpace}
Next, we illustrate the adaptation of the initialization (sampling) and expansion phases of \ourmethod to the general case of recovery of near-rank-$k$ sub\-matrices. 
The pseudo\-code 
is given in Algorithm~\ref{alg:rankk}. 
\begin{algorithm}[t]
\begin{algorithmic}[1]
\STATE \textbf{Input:} Matrix $\inputmat \in \real^{\nrows \times \ncols}$, target rank \generalrank, number of initializations $\nrep$, initial tolerance $\toleranceinit$, tolerance $\tol$.
\STATE \textbf{Output:} Near-rank-$\generalrank$ sub\-matrix $\outputmat^*$.
\STATE $\outputmat^* \gets \textbf{0}$\COMMENT{start initialization phase} 
\STATE $\Omega \gets \{ (i,j) \}, \quad \forall i \in \{1, \dots, \nrows \}, \forall j \in \{1, \dots, \ncols\}$
\FOR{$t = 1$ to $\nrep$}
    \STATE \( \Gamma \gets \{(i_h, j_h) \sim Uniform(\Omega) \mid h = 1, \dots,  \generalrank + 1 \} \)
    \STATE \( \initialmat'_{i_h, j_h} \gets \inputmat_{i_h, j_h}, \quad \forall (i_h , j_h) \in \Gamma \) 
    \WHILE{$ | \det(\initialmat') | > \toleranceinit$}
    \STATE \( \Gamma \gets \{(i_h, j_h) \sim Uniform(\Omega) \mid h = 1, \dots, \generalrank + 1 \} \)
    \STATE \( \initialmat'_{i_h, j_h} \gets \inputmat_{i_h, j_h}, \quad \forall (i_h , j_h) \in \Gamma \)\COMMENT{repeat until $\initialmat'$ is rank-deficient} 
    \ENDWHILE
    \STATE $\initialmat \gets \initialmat'$\COMMENT{start expansion phase}
    \STATE \( \indrows \gets \textsc{ExtractIndicatorMatrix}( \inputmat, \nrows,  \initialmat, \Gamma, \generalrank, \tol) \) 
    \STATE \(  \Gamma_{\mathcal{C}} \gets \{ (j,i)  \}, \quad \forall (i,j) \in \Gamma \) 
    \STATE      \( \indcols \gets \textsc{ExtractIndicatorMatrix}( \inputmat^T,  \ncols,   \initialmat^T, \Gamma_{\mathcal{C}} , \generalrank, \tol) \) 
    \STATE \( \ind \gets \indrows \cap {\indcols}^T \) \COMMENT{compute indicator matrix} 
    \STATE \( \outputmat \gets \textsc{ExtractMaximumEdgeBiclique}(\ind) \) 
    \IF{$f(\outputmat) > f(\outputmat^*)$}
    \STATE $\outputmat^* \gets \outputmat$  
    \ENDIF
    \ENDFOR
\STATE \textbf{Return} $\outputmat^*$
\STATE \textbf{Procedure }{\textsc{ExtractIndicatorMatrix}}
    \STATE \textbf{Input:} Input matrix \( \inputmat^{\Lambda}  \), number of rows \( 
    \nrows^{\Lambda} \),  initial submatrix $\initialmat^{\Lambda} $, set of indices $\Gamma^{\Lambda}$, target rank 
    $\generalrank$, tolerance $\tol$.  
    \STATE \textbf{Output:} indicator matrix \( {\ind}^{\Lambda} \). 
    \STATE \( \Phi \gets \{ i_h \sim Uniform(\{1, \dots, \nrows^{\Lambda} \}) \mid h = 1, \dots, \generalrank \} \)\COMMENT{sample \generalrank indices}
    \STATE \( \mathbf{C} \gets \inputmat_{:, j; j \in \Gamma_{\Lambda}} \initialmat_{\Phi, :}^+ \) \COMMENT{compute orthogonal-projection coefficients}
    \STATE \( \estimate{\inputmat}^\perp \gets \mathbf{C} \inputmat_{\Phi,:}  \) \COMMENT{compute orthogonal projections}
    \STATE \( \errors \gets | \estimate{\inputmat}^\perp - \inputmat |  \)  \COMMENT{compute matrix of absolute errors}
    \STATE \( {\ind^{\Lambda}}_{i,j} \gets  \mathbf{1}( \errors_{i,j} \leq \tol )  \) 
    \STATE \textbf{Return} \( \ind^{\Lambda} \)
\end{algorithmic}
\caption{\label{alg:rankk} Algorithm to find a near-rank-$k$ sub\-matrix.}
\end{algorithm}

\spara{Initialization.}
To find a seed, that is, a minimal near-rank-$k$ sub\-matrix, 
we sample matrices with $k+1$ rows and columns 
until we find a matrix $\initialmat$ that, as indicated by a near-zero determinant, is close to rank $k$ or lower, and thus represents the seed to expand in the second phase. 
In particular, to determine whether a matrix $\initialmat'$ has rank $\generalrank$ or lower, we check whether $|\det(\initialmat')| \leq \toleranceinit$, for a small $\toleranceinit \in \real^+$, which, as for the algorithm tailored to near-rank-$1$ sub\-matrices, is first initialized to a to a small value and then increased until a seed \initialmat is found.   

\spara{Expansion.}
Given the seed matrix $\initialmat \in \real^{\generalrank+1, \generalrank+1}$, of rank $\generalrank' \leq \generalrank$, we 
sample $\generalrank'$ anchor rows from the rows of $\initialmat$. Let $\initialmatcols$ denote the set of the $\generalrank + 1$ indices of the columns in $\initialmat$.  
Considering only the columns in $\initialmatcols$, we compute the coefficients of the orthogonal projection of each row of \inputmat onto the subspace spanned by the $\generalrank'$ anchor rows. 

We then compute the matrix of orthogonal projections $\estimate{\inputmat}^{\perp} \in \real^{\nrows \times \ncols}$ expressing each row as a linear combination of the anchor rows with weights given by the orthogonal-projection coefficients. 
The coefficients are obtained by considering only the columns in $\initialmatcols$, identified in the initialization phase. Nevertheless, the matrix $\estimate{\inputmat}^{\perp} \in \real^{\nrows \times \ncols}$, similarly to the ratio matrices in the rank-$1$ case, indicates which additional columns and rows are close to a rank-$\generalrank$ approximation. More specifically,
all entries $\inputmat_{i,j}$ that are closely approximated by $\estimate{\inputmat}^{\perp}_{i,j}$ lie close to the $\generalrank$-dimensional subspace identified in the initialization phase.

\begin{figure}[t]
\centering
\begin{tabular}{c@{\hspace{1.25cm}}c@{\hspace{1.25cm}}c}
    \scalebox{0.55}{\scalebox{0.5}{
\begin{tikzpicture}

\newcommand{\values}[2]{%
    \ifcase#1 \ifcase#2 42\or 87\or 23\or 91\or 56\or 12\fi
    \or \ifcase#2 74\or 19\or 88\or 33\or 67\or 25\fi
    \or \ifcase#2 31\or 76\or 55\or 92\or 14\or 60\fi
    \or \ifcase#2 85\or 47\or 29\or 73\or 38\or 94\fi
    \or \ifcase#2 58\or 99\or 41\or 16\or 82\or 37\fi
    \or \ifcase#2 69\or 21\or 50\or 95\or 13\or 77\fi
    \fi
}

\foreach \i in {0, 1, 2, 3, 4, 5} {
    \foreach \j in {0, 1, 2, 3, 4, 5} {
        \pgfmathsetmacro{\colorval}{\values{\i}{\j}}
        \fill[blue!\colorval, opacity=0.8] (\j, 6-\i) rectangle (\j+1, 6-\i-1);
    }
}

\draw[thick] (0, 0) rectangle (6, 6);

\node at (-0.5, 5.5) {\Huge 1};
\node at (-0.5, 4.5) {\Huge 2};
\node at (-0.5, 3.5) {\Huge $\vdots$};
\node at (-0.5, 0.5) {\Huge $\nrows$};
\node at (0.5, 6.5) {\Huge 1};
\node at (1.5, 6.5) {\Huge 2};
\node at (2.5, 6.5) {\Huge $\cdots$};
\node at (5.5, 6.5) {\Huge $\ncols$};

\draw[thick, red, fill=red!10, opacity=0.5] (1, 1) rectangle (5, 5);

\end{tikzpicture}
}} &
    \scalebox{0.55}{ \scalebox{0.5}{
\begin{tikzpicture}

\newcommand{\values}[2]{%
    \ifcase#1 \ifcase#2 93\or 54\or 23\or 78\or 11\or 99\fi  
    \or \ifcase#2 74\or 19\or 88\or 33\or 67\or 35\fi
    \or \ifcase#2 22\or 76\or 55\or 92\or 14\or 66\fi
    \or \ifcase#2 35\or 47\or 29\or 73\or 38\or 44\fi
    \or \ifcase#2 28\or 99\or 41\or 16\or 82\or 27\fi
    \or \ifcase#2 11\or 62\or 45\or 27\or 90\or 22\fi  
    \fi
}

\foreach \i in {0, 1, 2, 3, 4, 5} {
    \foreach \j in {0, 1, 2, 3, 4, 5} {
        \pgfmathsetmacro{\colorval}{\values{\i}{\j}}
        \fill[blue!\colorval, opacity=0.8] (\j, 6-\i) rectangle (\j+1, 6-\i-1);
    }
}

\draw[thick] (0, 0) rectangle (6, 6);

\node at (-0.5, 5.5) {\Huge 1};
\node at (-0.5, 4.5) {\Huge 2};
\node at (-0.5, 3.5) {\Huge $\vdots$};
\node at (-0.5, 0.5) {\Huge $\nrows$};
\node at (0.5, 6.5) {\Huge 1};
\node at (1.5, 6.5) {\Huge 2};
\node at (2.5, 6.5) {\Huge $\cdots$};
\node at (5.5, 6.5) {\Huge $\ncols$};

\draw[thick, red, fill=red!10, opacity=0.5] (1, 1) rectangle (5, 5);
\end{tikzpicture}
}} &
     \scalebox{0.55}{\scalebox{0.5}{
\begin{tikzpicture}

\newcommand{\values}[2]{%
    \ifcase#1 \ifcase#2 51\or 33\or 0\or 13\or 45\or 87\fi  
    \or \ifcase#2 0\or 0\or 0\or 0\or 0\or 10\fi
    \or \ifcase#2 9\or 0\or 0\or 0\or 0\or 6\fi
    \or \ifcase#2 50\or 0\or 0\or 0\or 0\or 50\fi
    \or \ifcase#2 30\or 0\or 0\or 0\or 0\or 10\fi
    \or \ifcase#2 58\or 41\or 5\or 68\or  77\or 55\fi  
    \fi
}

\foreach \i in {0, 1, 2, 3, 4, 5} {
    \foreach \j in {0, 1, 2, 3, 4, 5} {
        \pgfmathsetmacro{\colorval}{\values{\i}{\j}}
        \fill[blue!\colorval, opacity=0.8] (\j, 6-\i) rectangle (\j+1, 6-\i-1);
    }
}

\draw[thick] (0, 0) rectangle (6, 6);

\node at (-0.5, 5.5) {\Huge 1};
\node at (-0.5, 4.5) {\Huge 2};
\node at (-0.5, 3.5) {\Huge $\vdots$};
\node at (-0.5, 0.5) {\Huge $\nrows$};
\node at (0.5, 6.5) {\Huge 1};
\node at (1.5, 6.5) {\Huge 2};
\node at (2.5, 6.5) {\Huge $\cdots$};
\node at (5.5, 6.5) {\Huge $\ncols$};

\draw[thick, red, fill=red!10, opacity=0.5] (1, 1) rectangle (5, 5);

\begin{scope}[shift={(7,-0.5)}] 
    \foreach \v [count=\i from 0] in {0, 20, 40, 60, 80} {
        \pgfmathsetmacro{\colorval}{\v}
        \fill[blue!\colorval, opacity=0.8] (0, 5-\i) rectangle (1, 6-\i);
        \node[right] at (1.2, 5.5-\i) {\Huge \v};
    }
    \draw[thick] (0, 1) rectangle (1, 6);
    \node[above] at (0.5, 6.5) {\Huge Value};
\end{scope}

\end{tikzpicture}
}} \\
     \hspace*{0.2cm}\inputmat & \hspace*{0.2cm}$\estimate{\inputmat}^\perp$ & \hspace*{-0.4cm}\errors \\
\end{tabular}
\caption{\label{fig:linear_combination_diagram}
Example of matrices of projections ($\estimate{\inputmat}^\perp$) and absolute errors (\errors) associated with an input matrix (\inputmat) containing a rank-$\generalrank$ sub\-matrix (highlighted in red). Within this rank-$\generalrank$ sub\-matrix, the entries of $\estimate{\inputmat}^\perp$ are equal to those of \inputmat, and the entries of $\errors$ are all identically zero. 
}
\end{figure}
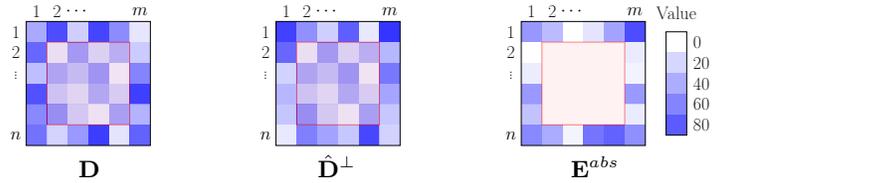

Therefore, to find a near-rank-\generalrank sub\-matrix of maximum size, we need to identify the largest sub\-matrix of $\estimate{\inputmat}^{\perp}_{i,j}$ where all entries nearly match the corresponding entries of \inputmat. 
To obtain such a sub\-matrix, we calculate the matrix of absolute errors $\errors = | \inputmat - \estimate{\inputmat}^{\perp} |$, and from it, the indicator matrix $\indrows$, which takes value $1$ for entry $\generalentry$ if $\errors_{i,j} \leq \tol$ and $0$ otherwise, for some input $\tol \in \real^+$.
Figure~\ref{fig:linear_combination_diagram} presents an example of the matrices $\estimate{\inputmat}^{\perp}$ and $\errors$. 

The same procedure followed to determine $\indrows$, 
but on input $\inputmat^T$ and $\initialmat^T$ yields~${\indcols}$ . 
Similarly to the case of near-rank-$1$ sub\-matrix discovery, the intersection of $\indrows$ and~${\indcols}^T$ gives the matrix $\ind$ and the associated bipartite graph $\mathcal{G}_{\ind}$. 
The desired output near-rank-$\generalrank$ sub\-matrix is then given by a sub\-matrix of $\ind$ consisting of all ones, or, equivalently, by a maximum-edge bi\-clique of $\mathcal{G}_{\ind}$.

\subsection{Scalability Considerations}\label{sec:scalability}
One limitation of \ourmethod is its reliance on solving the maximum-edge biclique problem, 
which is  \nphard. 
While the algorithm we use to extract these bicliques is often efficient in practice~\cite{lyu2020maximum}, scalability issues may still arise.
%
To address such issues, 
the algorithm for finding maximum-edge bicliques can be replaced with a more scalable heuristic.  
Among many possible different heuristic approaches, by default,
we rely on spectral bi\-clustering~\cite{kluger2003spectral}, which is empirically found to be particularly effective in quickly identifying a dense sub\-matrix of \ind.
Even more efficient and scalable approaches include
algorithms to extract dense bipartite subgraphs, 
a greedy algorithm removing rows and columns from \ind, e.g., based on the  amount of ones, and a randomized algorithm sampling sub\-matrices from \ind according to the amount of ones they contain~\cite{boley2011direct}. 
A comprehensive evaluation of the performance of various heuristics for approximating maximum-edge bicliques is left to future work.

\section{Analysis}\label{sec:analysis}
In this section, we explain how the proposed methods yield sub\-matrices 
with bounded approximation error.
We also provide a brief discussion on the probabilistic aspects and on the computational complexity of the methods.

\subsection{Approximation Error Guarantees}

In global low-rank approximation, 
the presence of outliers in the data may lead to situations where
the whole matrix cannot be approximated with a low-rank structure 
without compromising the overall approximation quality. 
However, as our problem definition lifts the requirement that the whole matrix must be approximated, 
it is interesting to control the \emph{entry-wise} maximum approximation error in the discovered sub\-matrices. 
We thus provide
approximation-error guarantees in terms of the max norm.
A bound on the max norm yields  bounds on the Frobenius and spectral norms, albeit loose.
In the case of near-rank-$1$ sub\-matrix discovery, 
we also provide interesting bounds on the spectral and Frobenius norms  that are not a direct consequence of the bound on the max~norm. 
\spara{Near-rank-$\mathbf{1}$ submatrices.}
As mentioned in Section~\ref{sec:algo_overview}, $\nrowssubmatrix \times \ncolssubmatrix$ near-rank-$1$ sub\-matrices contain many near-rank-$1$  $2 \times 2$ sub\-matrices. Building on this intuition, \ourmethod starts by locating a $2 \times 2$ sub\-matrix \initialmat with bounded determinant, and hence close to rank $1$. 
Then, it computes row-wise and column-wise ratios dividing all rows (columns) by a single anchor row (column) with index sampled from those of \initialmat, 
and finds sub\-matrices with rows (columns) of nearly-constant ratios. 
Nearly-constant ratios correspond to bounded $2 \times 2$ determinants. For instance, if  
\( \left| \frac{\inputmat_{i,j_{1}}}{x^r_{j_1}} - \frac{\inputmat_{i,j_{2}}}{x^r_{j_2}} \right| \leq \tol\), where the left-hand side is a difference of ratios, 
then 
\( 
| \inputmat_{i,j_{1}} x^r_{j_2}  - \inputmat_{i,j_{2}}  x^r_{j_1} | \leq \tol  | x^r_{j_1} | | x^r_{j_2} | \), 
where the left-hand side is a $2 \times 2$ determinant.
Bounding the variation of all the ratios within each row and column of a sub\-matrix, and thus the corresponding $2 \times 2$ determinants, \ourmethod yields sub\-matrices composed of $2 \times 2$ near-rank-$1$ sub\-matrices, which, as formalized in Theorem~\ref{th:expansion},  results in approximation guarantees.

\begin{theorem}\label{th:expansion}
Let $\outputmat \in \real^{\nrowssubmatrix \times \ncolssubmatrix}$ be a near-rank-$1$ sub\-matrix output by \ourmethod with anchor row $\anchorrow$, anchor column $\anchorcolumn$ and input tolerance $\tol$. 
There exists a rank-$1$ approximation $\estimate{\outputmat}$ of \outputmat such that for $\outputerror = \outputmat - \estimate{\outputmat}$ it~holds:
\begin{equation}
\label{eq:initialization_guarantee_infinity}
\infinitynormshort{\outputerror} \leq \min\left\{
\tol \gmax{\anchorrow}, 
\tol \gmax{\anchorcolumn}
\right\},
\end{equation}
and 
\begin{equation}\label{eq:expansion_guarantee}
  \spectralnormshort{\outputerror}  \leq 
\frobeniusnormshort{\outputerror}  \leq \min\left\{ \tol  \sqrt{ (\nrows-1) \gF{\anchorrow} } , 
\tol  \sqrt{ (\ncols-1) \gF{\anchorcolumn} } 
\right\},
\end{equation}
where 
$\gmax{\xvec} = \frac{\max_{i} |{x}_i|^3 }{2 \min_{i} {x}_i^2}$ and $\gF{\xvec} = \frac{\sum_{i < j} {x_i}^2 {x_j}^2 }{\|\xvec\|_2^2}$.  

\end{theorem}

Theorem~\ref{th:expansion} suggests that the low-rank-approximation error incurred 
by the near-rank-$1$ sub\-matrices discovered by \ourmethod 
can be bounded by a function of the input parameter $\tol$ and of the scale of $\anchorrow$ or $\anchorcolumn$. 
Therefore, given the anchor row and column, 
one can set the value of \tol to guarantee that the maximum or the total approximation error 
is bounded by a user-specified threshold $\errorbound \in \real^+$, as requested by Problem~\ref{prob:rank-1}. 
However, the approximation-error guarantees given in Theorem~\ref{th:expansion} only hold if \ourmethod extracts a biclique in the last step. Alternative heuristic approaches that do not extract a biclique can be effective in practice, but they are not supported by approximation-error guarantees. 

Notably, the approximation-error guarantees are achieved by the interpretable rank-$1$ approximation discussed in Section~\ref{sec:algo_overview}. 
In addition, for the rank-1 \svd approximation $\estimate{\outputmat}$,  
Theorem~\ref{th:expansion_appendix} in~\ref{ap:additional_analysis} bounds the spectral norm of the error \( \outputerror = \outputmat - \estimate{\outputmat} \).

\spara{Near-rank-$\mathbf{k}$ sub\-matrices.}
The algorithm for the more general task of identifying near-rank-$\generalrank$ sub\-matrices does not admit the same analysis as the algorithm for identifying near-rank-$1$ sub\-matrices.  
However, the algorithm for the rank-$\generalrank$ case, by design, discovers sub\-matrices \outputmat such that
$\outputerror = \outputmat - \estimate{\outputmat}$ satisfies 
\( \infinitynormshort{\outputerror} \leq \delta \). 
As mentioned, the bound on the max norm leads to a straightforward bound on the Frobenius and spectral norms, namely \( \spectralnormshort{\outputerror} \leq \frobeniusnormshort{\outputerror} \leq \tol \sqrt{  (\nrows-\generalrank) (\ncols-\generalrank) }  \), which can also be used to set the value of $\tol$ based on a user-specified error threshold $\errorbound$ on the Frobenius or spectral norm.

\subsection{Probabilistic Analysis}
In this section, we discuss simple probabilistic aspects of our method. 

\spara{Probability of discovering a near-rank-$\mathbf{1}$ sub\-matrix.} 
Let \outputmat be a target near-rank-$1$ sub\-matrix of size $| \outputmat |$ within $\inputmat \in \real^{\nrows \times \ncols}$. 
The probability that \ourmethod 
 discovers \outputmat by one sample is 
\(\pfirststar = \frac{\p}{\nrows \ncols}\frac{\p-1}{\nrows \ncols-1}\). 
Hence, the probability of discovering \outputmat in \nrep iterations is \( 1 -  (1 - \pfirststar)^{\nrep} \), and
therefore the number of iterations required to discover $\outputmat$ with probability at least $\alpha_p$ is 
\(\nrep \ge \frac{\ln(1 - \alpha_p)}{\ln(1 - \pfirststar)}.\)
For instance, if \( \pfirststar = 0.1 \) and \( \alpha_p = 0.9 \), we need \(
\nrep > \frac{\ln(1 - 0.9)}{\ln(1 - 0.1)} \approx 22 
\) iterations. 

Basic probability theory implies that, 
in expectation, 
the number of iterations necessary to discover \outputmat is $\frac{1}{\pfirststar}$, 
and we discover it $\pfirststar\nrep$ times in \nrep iterations. 

\spara{Probability of discovering a near-rank-$\mathbf{\generalrank}$ sub\-matrix.} 
The simple probabilistic analysis presented above for near-rank-$1$ sub\-matrices also applies
to near-rank-$\generalrank$ sub\-matrices. 
The only difference is that, in this case, we have 
\(
\pfirststar = \frac{\p}{\nrows \ncols} \frac{\p-1}{\nrows \ncols} \dots  \frac{\p-\generalrank}{\nrows \ncols-\generalrank},  
\)
which can become small as $\generalrank$ grows. Yet, larger values of \generalrank tend to be associated with larger values of  $\p$ and, in practice, we are interested in small values of \generalrank. 

\spara{Probability of occurrence of a $\mathbf{2 \times 2}$ near-rank-$\mathbf{1}$ matrix.} 
We conclude the section by investigating the probability with which 
\ourmethod identifies a seed $2 \times 2$ sub\-matrix with near-zero determinant in random matrices.
Let \inputmat be a random matrix with i.i.d. entries distributed according to \distributionz, and let $E(\distributionz) = \mu$ and  
$\var(\distributionz) = \sigma^2$ be the expectation and variance of \distributionz. 
To study the probability of occurrence of $2 \times 2$ sub\-matrices with near-zero determinant, we consider the random variable $\mathcal{W} = x_1 y_2 - x_2 y_1$, where $x_1$, $x_2$, $y_1$ and $y_2$ are the entries of a $2 \times 2$ sub\-matrix.

By independence, 
\(
E(x_1 y_2)    = E(x_1) E(y_2)
\) and 
\(
\var(x_1 y_2) = \var(x_1) \var(y_2) + E(y_2)^2 \var(x_1) + E(x_1)^2 \var(y_2) = \sigma^4 + 2 \mu^2 \sigma^2, 
\)
and similarly for $x_2 y_1$.

Further, since $x_2 y_1$ and $x_2 y_1$ are independent, 
\begin{align*}
E(x_1 y_2 - x_2 y_1) & = E(x_1 y_2) - E(x_2 y_1) = 0 \text{ and} \\ 
\var(x_1 y_2 - x_2 y_1) & = \var(x_1 y_2) + \var(x_2 y_1)=2 \sigma^4 + 4 \mu^2\sigma^2.
\end{align*}
Chebyshev’s inequality~\cite{saw1984chebyshev} then implies that: 
\begin{equation*}
    P(|\mathcal{W}| \geq \toleranceinit) \leq \frac{2 \sigma^4 + 4 \mu^2\sigma^2}{\toleranceinit^2},
\end{equation*}
giving a bound on the probability that a $2\times2$-submatrix deviates significantly from rank $1$. 
The preliminary experiments presented in~\ref{ap:additional_experiments} additionally provide an empirical investigation of this probability.
Assumptions on \distributionz may lead to tighter bounds, a question that we leave open for future work.

\subsection{Computational Complexity}
Finally, we discuss the computational complexity of \ourmethod. 

Consider a single iteration of the method.
The runtime bottleneck is due to finding a maximum-edge bi\-clique, which, in the worst case can take exponential time. However, the algorithm introduced by Lyu et al.~\cite{lyu2020maximum} prunes large portions of the search space and can be very efficient in practice. 
As discussed in Section~\ref{sec:scalability}, to improve scalability, we can use a more scalable heuristic for finding an approximate maximum-edge biclique.
The spectral bi\-clustering algorithm, which is the heuristic we rely on by default, has computational complexity determined by the computation of the (truncated) \svd, 
which is $\bigO \left( \min(\nrows^2 \ncols, \ncols^2 \nrows) \right)$. 
If an even more scalable heuristic is leveraged, 
such as a basic linear-time algorithm removing rows and columns of \ind with less than a given proportion of ones, 
\ourmethod for the rank-$1$ case and for the general rank-\generalrank case incurs computational complexity $\bigO( \nrows +  \ncols)$ and $\bigO( \nrows \generalrank \ncols)$, respectively. 

As \ourmethod generally explores different initializations, if $\tau$ is the complexity of a single iteration, then $\bigO(\nrep \tau)$ is the overall complexity.

\section{Experiments}\label{sec:experiments}

In this section, we evaluate the performance of \ourmethod against existing approaches.  
We consider both synthetic data and real-world data. 
More details on the experimental setup are provided in~\ref{ap:additional_experimental_setup}, and
additional experimental results are presented in~\ref{ap:additional_experiments}. 


\subsection{Experimental Setup}\label{sec:exp_setup}

\para{Datasets.}
We conduct experiments on both synthetic and real-world datasets.

The synthetic data are generated by planting near-rank-$1$ sub\-matrices into larger matrices. 
To make the 
discovery task as challenging as possible, the entries of the planted sub\-matrices and of the background 
are generated 
from the same distributions. 
We consider six different distributions. 
The details of the data-generating mechanisms are given in~\ref{ap:additional_experimental_setup}. 

Additionally, we consider $15$ real-world datasets from different applications, 
including user ratings, images, and gene-expression data. 
We report summary characteristics for the real-world datasets in Table~\ref{tab:real_world_datasets}, 

\begin{table}[t]
\centering
\caption{\label{tab:real_world_datasets} Summary characteristics for real-world datasets. We report the number of rows, columns,  the low-rankness score, the entry-wise maximum squared deviation from the rank-1 \svd (Max rank-$1$ deviation) and a reference.}
    \label{tab:datasets}
\begin{tabular}{lrrrrr}  
    \toprule
    {{{Dataset}}} & 
    {{~\# Rows}} & 
    {{~\# Columns}} & 
    {{~Low-rankness}} & 
    {{~Max rank-${1}$}} & 
    {{~Reference}} \\
    & & & {{score}} & {{deviation}} & \\
    \midrule
    \Hyperspectral & 5\,554 & 2\,151 & 0.89 & 0.23 & \cite{leone2022hyperspectral} \\
    \Isolet & 7797 & 617 & 0.88  & 0.98  & \cite{asuncion2007uci} \\
    \Olivetti & 400  & 4096 & 0.95  & 0.53  & \cite{asuncion2007uci} \\
    \MoiveLens & 943 & 1682 & 0.30  & 1.00 & \cite{harper2015movielens} \\ 
    \OrlRnSp & 400 & 1024 & 0.95  & 0.55  & \cite{orl_faces} \\
     \Golub & 7129 & 38 & 1.00  & 0.28  & \cite{esposito2021review} \\
    \Mandrill & 512 & 512 & 0.92  & 0.36  & \cite{usc-sipi} \\
    \Ozone & 2536 & 72 & 1.00  & 0.08  & \cite{asuncion2007uci} \\
    \BRCA & 317 & 496 & 0.72  & 0.89  & \cite{tomczak2015review} \\
    \Google & 5456 & 24 & 0.76  & 0.77  & \cite{asuncion2007uci} \\
    \NPAS & 1418 & 78 & 0.85  & 0.15  & \cite{kaggle} \\
    \Cameraman & 256 & 256 & 0.86  & 0.56  & \cite{usc-sipi} \\
    \MovieTrust & 200 & 200 & 0.77  & 1  & \cite{kaggle} \\
    \Hearth & 920 & 15 & 0.94  & 0.08  & \cite{asuncion2007uci} \\
    \Imagenet & 64 & 27 & 0.92  & 0.22  & \cite{deng2009imagenet} \\
    \bottomrule
\end{tabular}
\end{table}

\spara{Baselines.}
We compare \ourmethod against baselines discussed in Section~\ref{sec:related}. 
Specifically, we consider 
a method (\cvx) based on convex optimization~\cite{doan2013finding}, 
PCA with sparsity constraints (\sparsePCA)~\cite{mairal2009online}, 
\svp~\cite{ruchansky2017targeted}, and \rpsp~\cite{dang2023generalized}.
In the experiments with real data, we restrict the comparison to  the most recently introduced methods, \svp and \rpsp, which  specifically aim at discovering (possibly multiple) near-low-rank sub\-matrices.  

\spara{Metrics.}
In experiments with synthetic data, 
all methods output matrices $\estimate{\inputmat}$ that contain low-rank approximations of the identified sub\-matrices and zero entries for all indices that are not part of such sub\-matrices.
To measure the ability of a method in recovering the indices of the planted ground-truth sub\-matrices,
we report the $F_1$ score.
Based on the same output,  we also report the error (squared Frobenius norm averaged over the entries) incurred in approximating the ground-truth sub\-matrices. 
\ourmethod approximates sub\-matrices through the interpretable approach discussed in Section~\ref{sec:algo_overview} for $\generalrank=1$ and via \svd for $\generalrank>1$. All baselines approximate sub\-matrices via \svd.   



In real-world datasets, where no ground truth is available, 
we report the size and low-rankness score (introduced in Section~\ref{sec:preliminaries})
of the returned sub\-matrices.  

In all cases, we measure run\-times in seconds.

\spara{Parameters.} 
The important parameter to set for our method 
is the tolerance $\tol$ controlling the trade-off between low-rankness and size. 
As explained in Section~\ref{sec:analysis}, 
one can set $\tol$ to match an input bound \errorbound on the allowed low-rank-approximation error. 
In our experiments, however, we explore few fixed values of $\tol$. 
Specifically, for experiments with synthetic data, we set $\tol$ to $0.05$ 
and the number of initializations $\nrep$ to $25$. 
For experiments with real-world data, we let $\tol$ vary in $\{ 10^{-1}, 10^{-2}, 10^{-3}, 10^{-4} \}$, 
and we consider $\nrep = 25$ initializations for each value of \tol. 
Finally, the initialization parameter $\toleranceinit$ is set to $10^{-11}$ and  
is increased by $10$ every $10000$ samples that do not result in a sub\-matrix to expand. 

\spara{Implementation.}
Our Python implementation is available online\footnote{\url{https://github.com/maciap/SaE}}.
Experiments are performed on a computer with
$2 \times 10$ core Xeon E5 processor and 256 GB memory. All reported results are averages over $10$ runs. 

\subsection{Experiment Results}\label{sec:exp_results}
We first present results for the experiments in synthetic data 
and then the results for the experiments in real-world datasets. 

\begin{figure}[p]
\begin{center}
\begin{tikzpicture}[scale=0.1]
\matrix [matrix of nodes, 
column sep=1.0pt, 
row sep=0pt, 
nodes={anchor=center}] (m) {
\tikz \draw[slateblue, thick] (-0.18,0) -- (0.18,0) 
node[fill=slateblue, circle, inner sep=2.1pt] {};
& {\small Sparse PCA}
& 
\tikz \draw[darkolivegreen, thick] (-0.18,0) -- (0.18,0) 
node[fill=darkolivegreen, regular polygon, regular polygon sides=3, inner sep=1.4pt] {};
& {\small \cvx}
& 
\tikz \draw[darkmagenta, thick] (-0.18,0) -- (0.18,0) 
node[fill=darkmagenta, regular polygon, regular polygon sides=3, inner sep=1.7pt, rotate=180] {};
& {\small \svp}
& 
\tikz \draw[crimson, thick] (-0.18,0) -- (0.18,0) 
node[fill=crimson, diamond, inner sep=1.7pt] {};
& {\small \rpsp} 
& 
\tikz \draw[orange, thick] (-0.18,0) -- (0.18,0) 
node[fill=orange, rectangle, inner sep=2.6pt] {};
& {\small \ourmethod} \\
};
\end{tikzpicture}
    \begin{tabular}{ccc}
        & \textsc{Normal} & \\
        \includegraphics[width=0.3\textwidth]{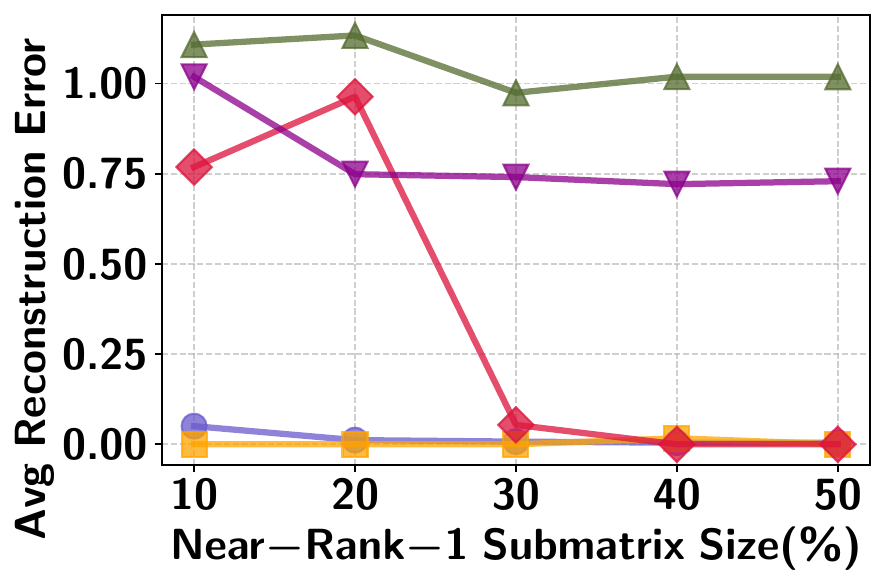} & 
        \includegraphics[width=0.3\textwidth]{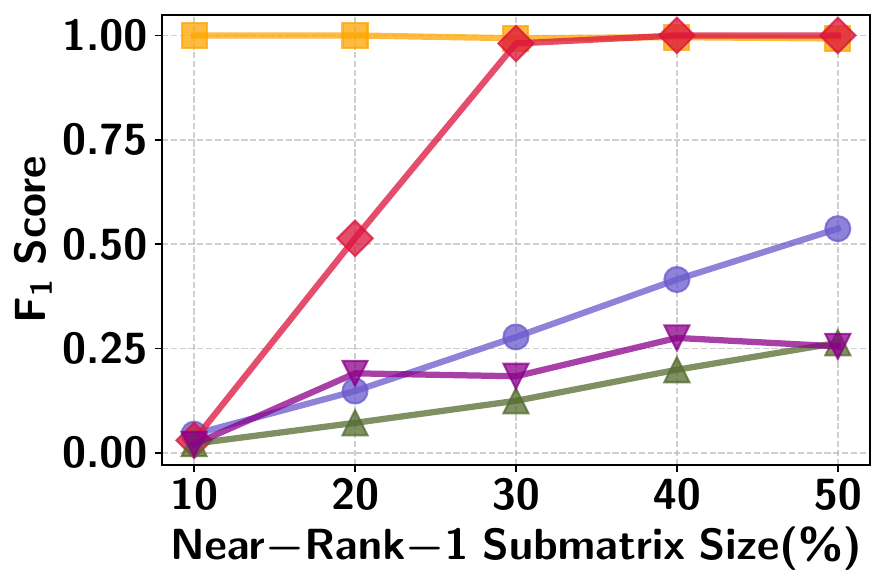} & 
        \includegraphics[width=0.3\textwidth]{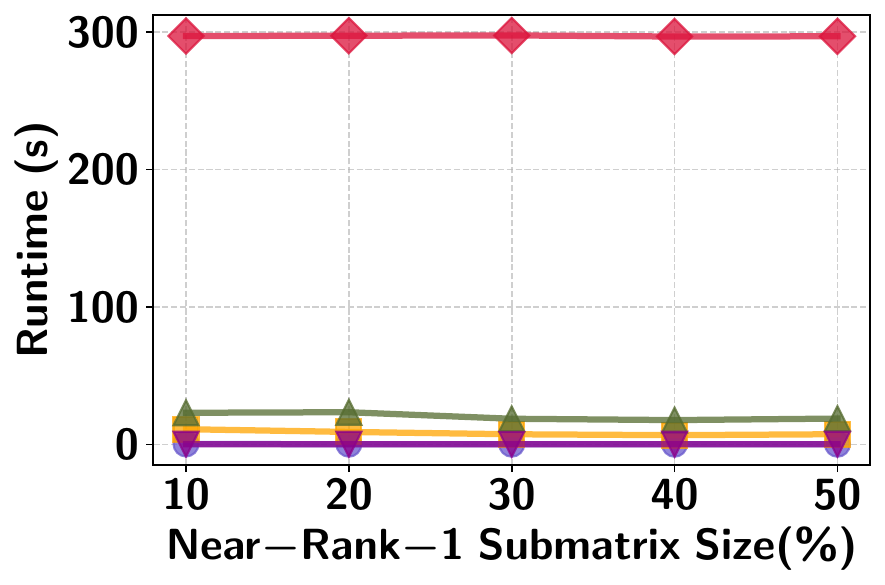} \\
         & \textsc{Uniform} & \\
        \includegraphics[width=0.3\textwidth]{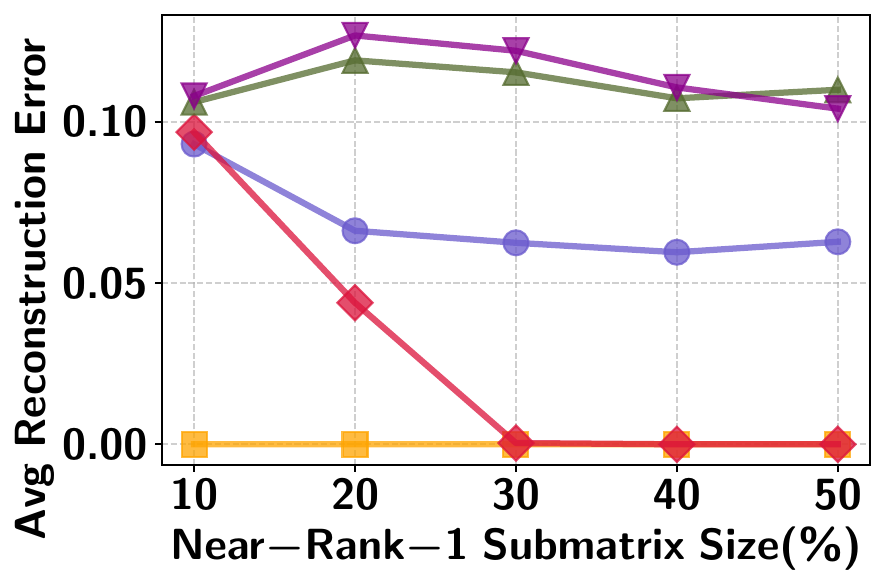} & 
        \includegraphics[width=0.3\textwidth]{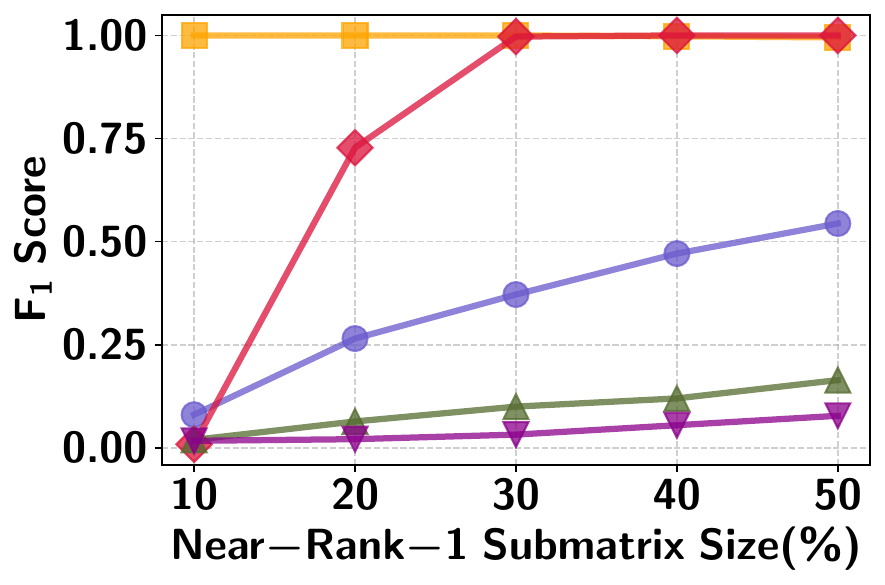} & 
        \includegraphics[width=0.3\textwidth]{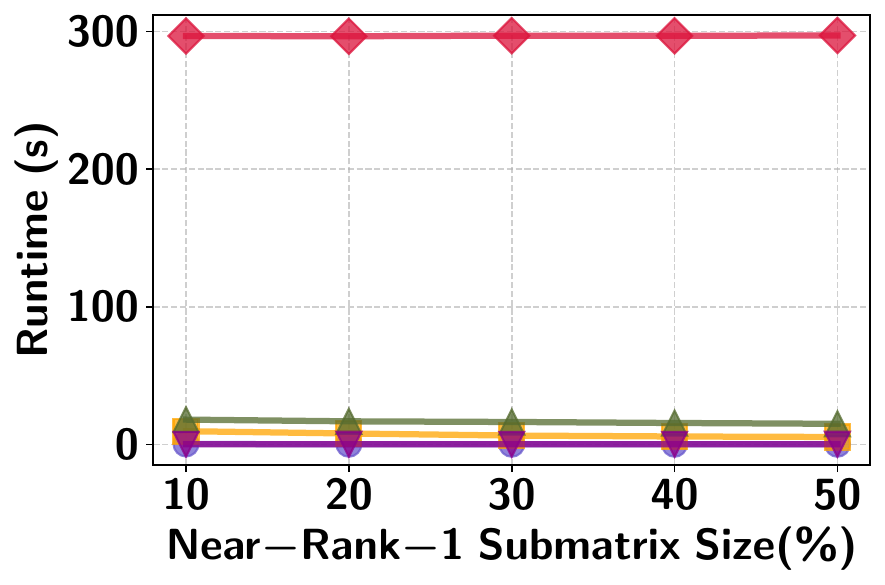} \\

        & \textsc{Exponential} & \\
        \includegraphics[width=0.3\textwidth]{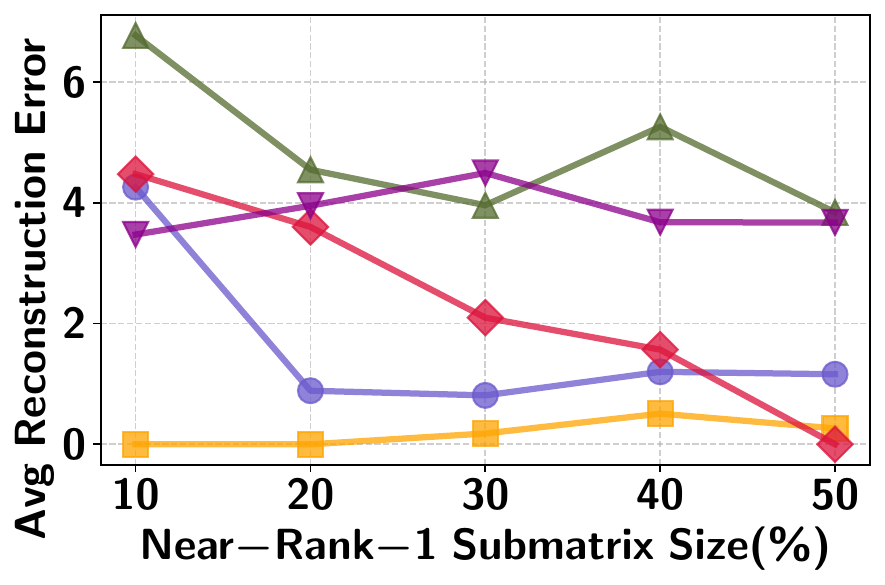} & 
        \includegraphics[width=0.3\textwidth]{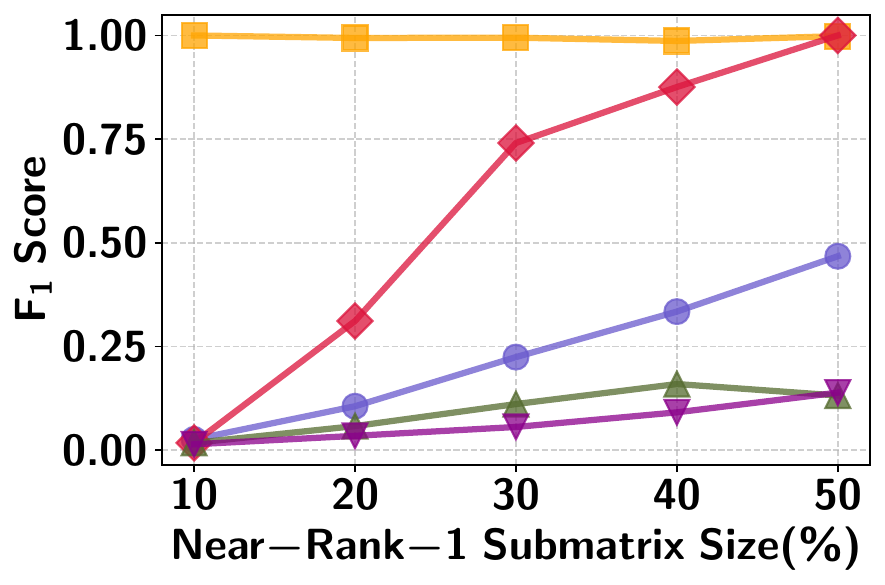} & 
        \includegraphics[width=0.3\textwidth]{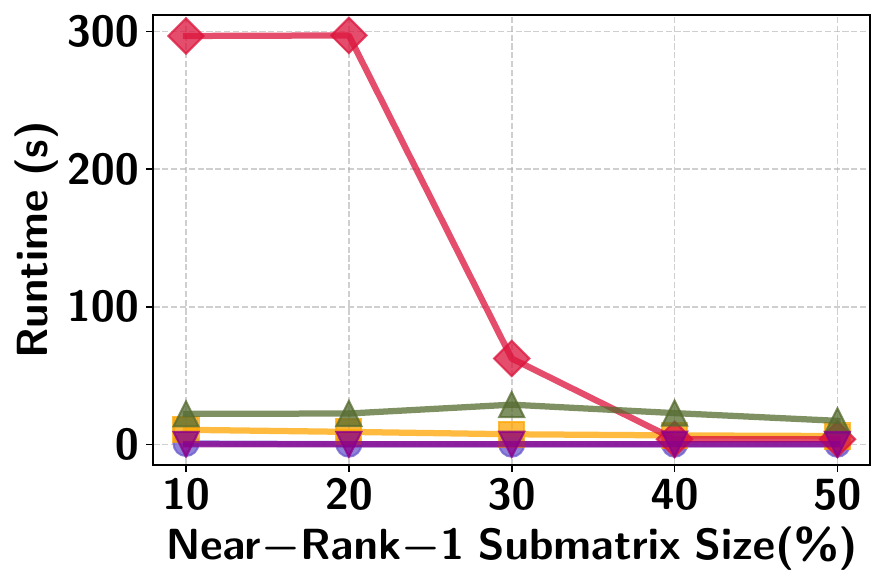} \\

        & \textsc{Beta} & \\
        \includegraphics[width=0.3\textwidth]{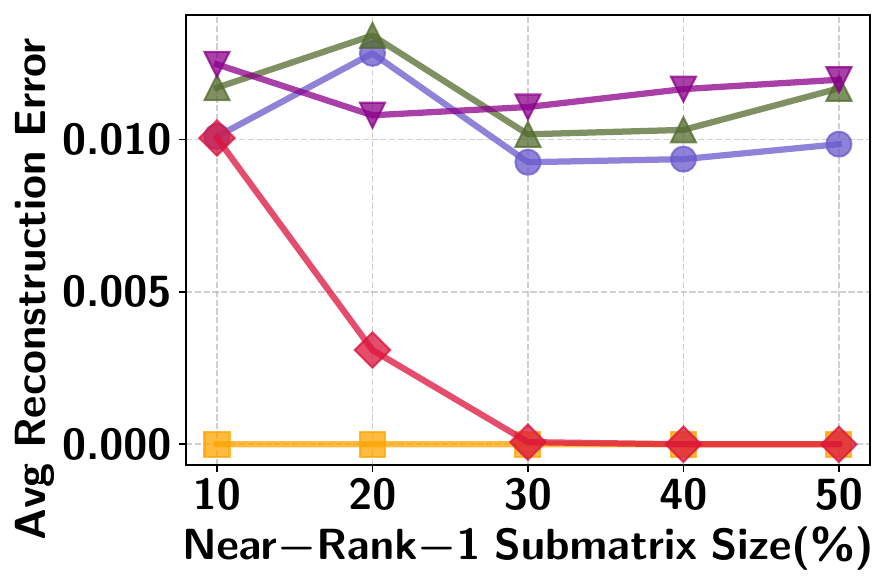} & 
        \includegraphics[width=0.3\textwidth]{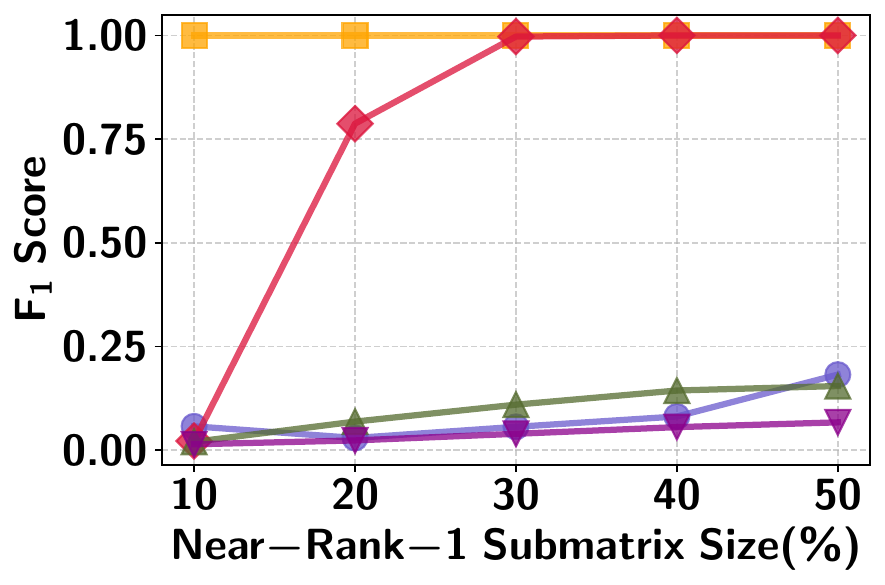} & 
        \includegraphics[width=0.3\textwidth]{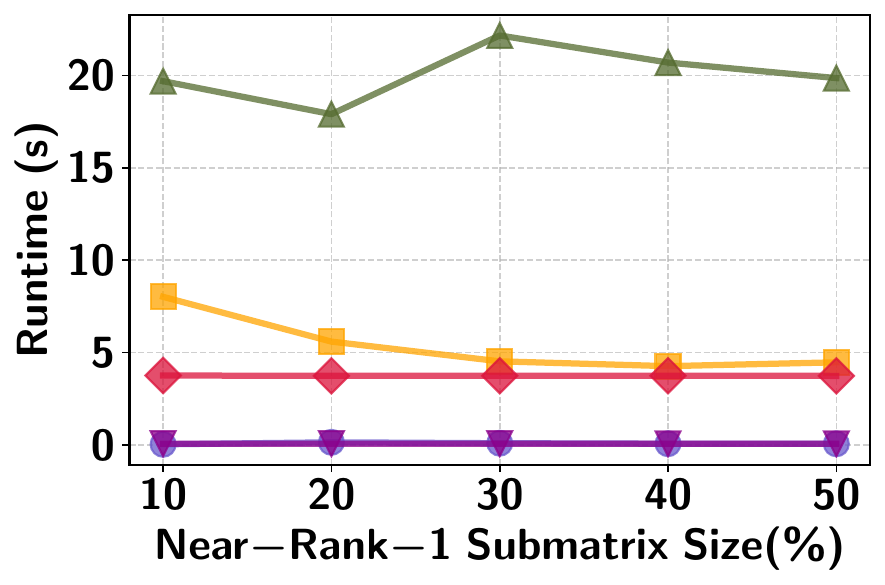} \\

        & \textsc{Gamma} & \\
        \includegraphics[width=0.3\textwidth]{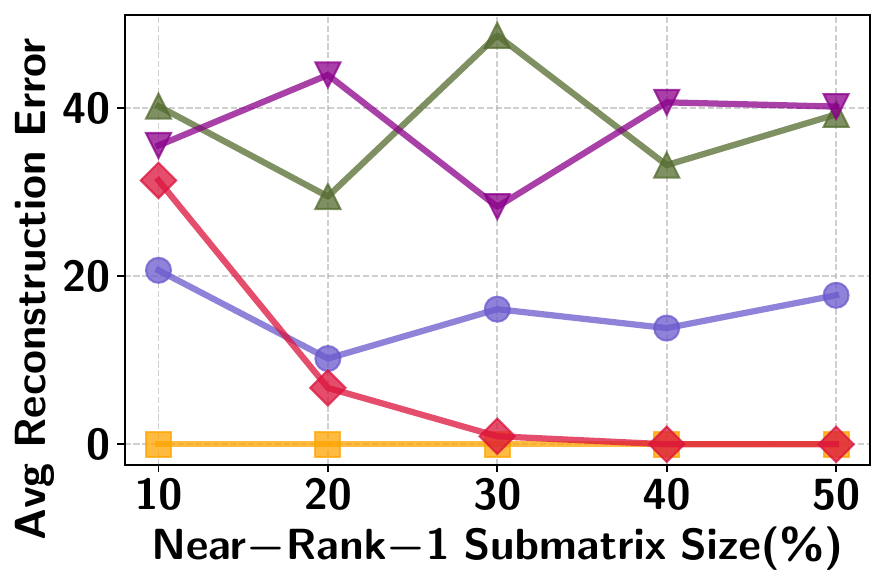} & 
        \includegraphics[width=0.3\textwidth]{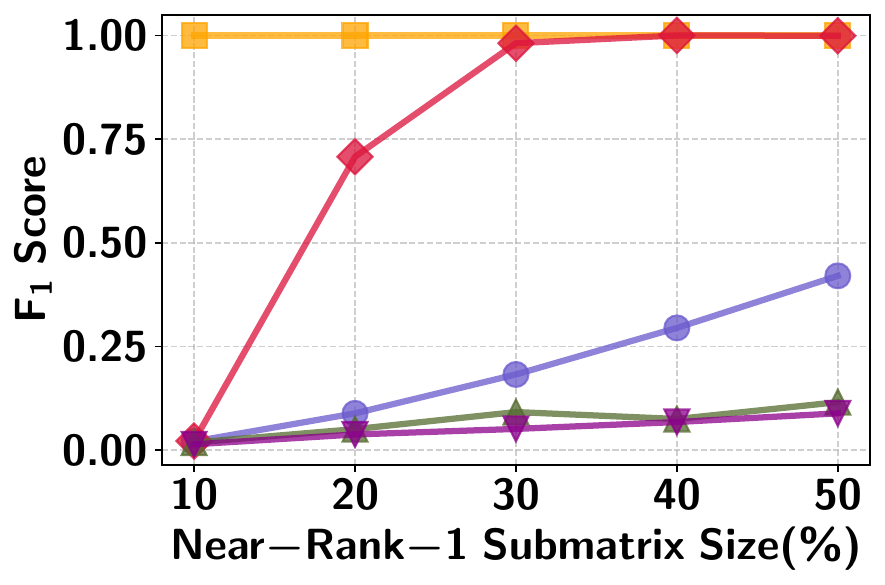} & 
        \includegraphics[width=0.3\textwidth]{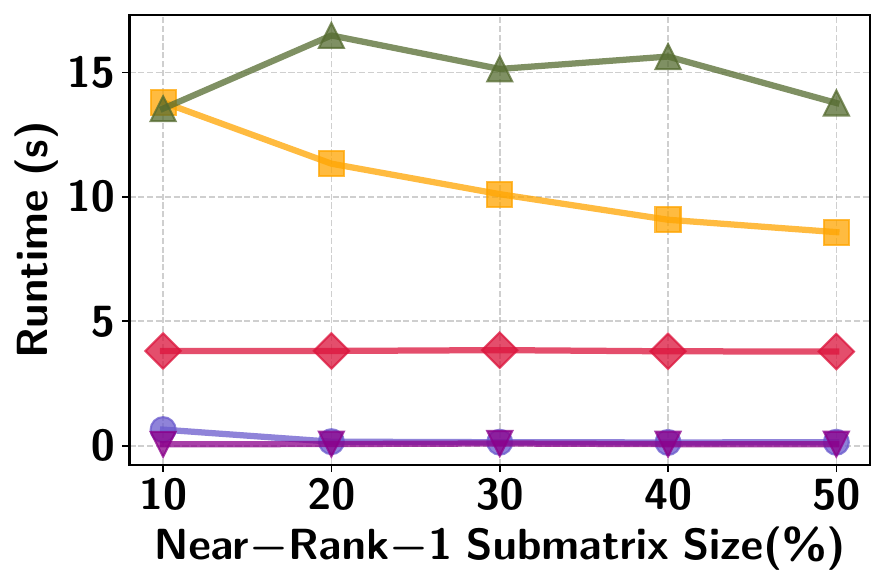} \\

        & \textsc{Poisson} & \\
         \includegraphics[width=0.3\textwidth]{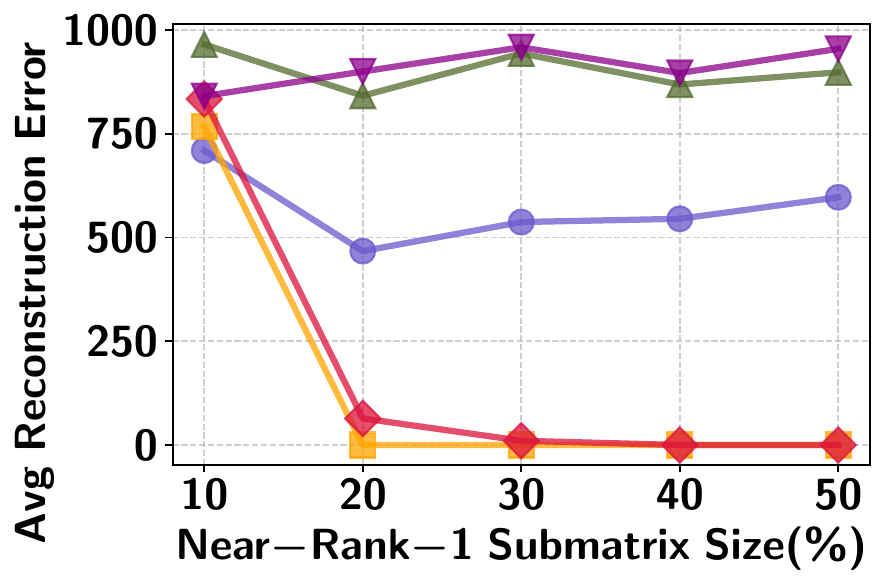} & 
        \includegraphics[width=0.3\textwidth]{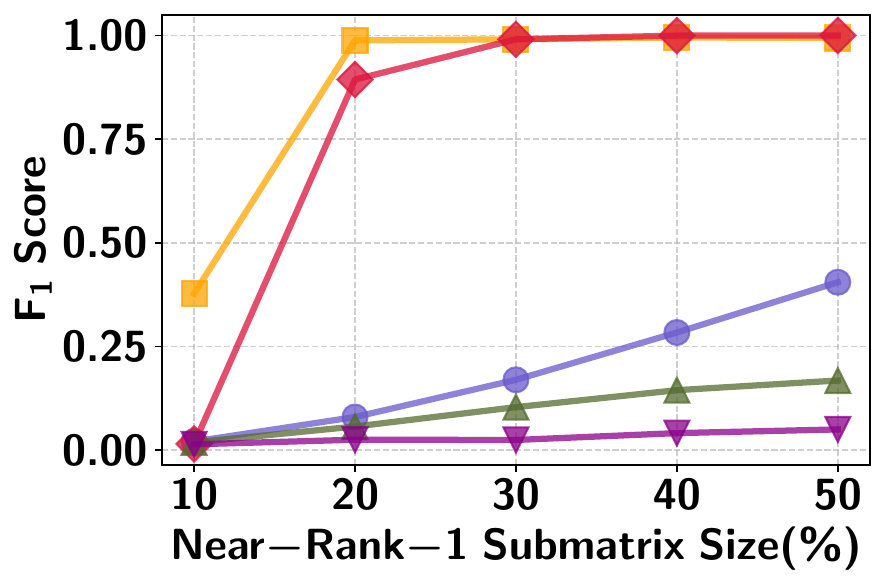} & 
        \includegraphics[width=0.3\textwidth]{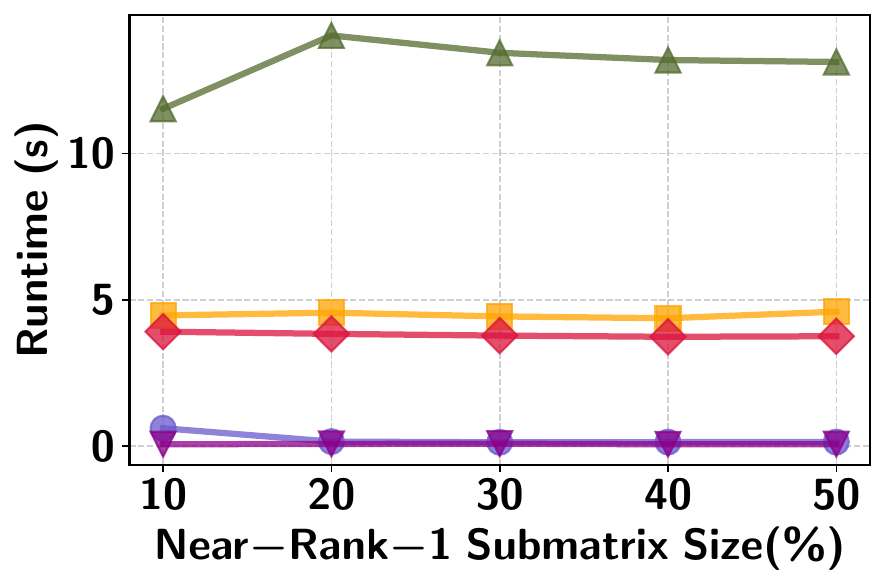} \\
        \vspace*{-0.65cm}
    \end{tabular}
\end{center}
\caption{
\label{fig:near_rank_one_recovery}
Full-rank synthetic $250 \times 250$ matrices generated from different probability distributions with a planted near-rank-$1$ submatrix.
Performance of different methods in the task of near-rank-$1$ submatrix discovery.  
We show the average (per-entry) reconstruction error (left), the $F_1$ score (center) and the runtime (right) of different methods as a function of planted submatrix size.
}
\end{figure}
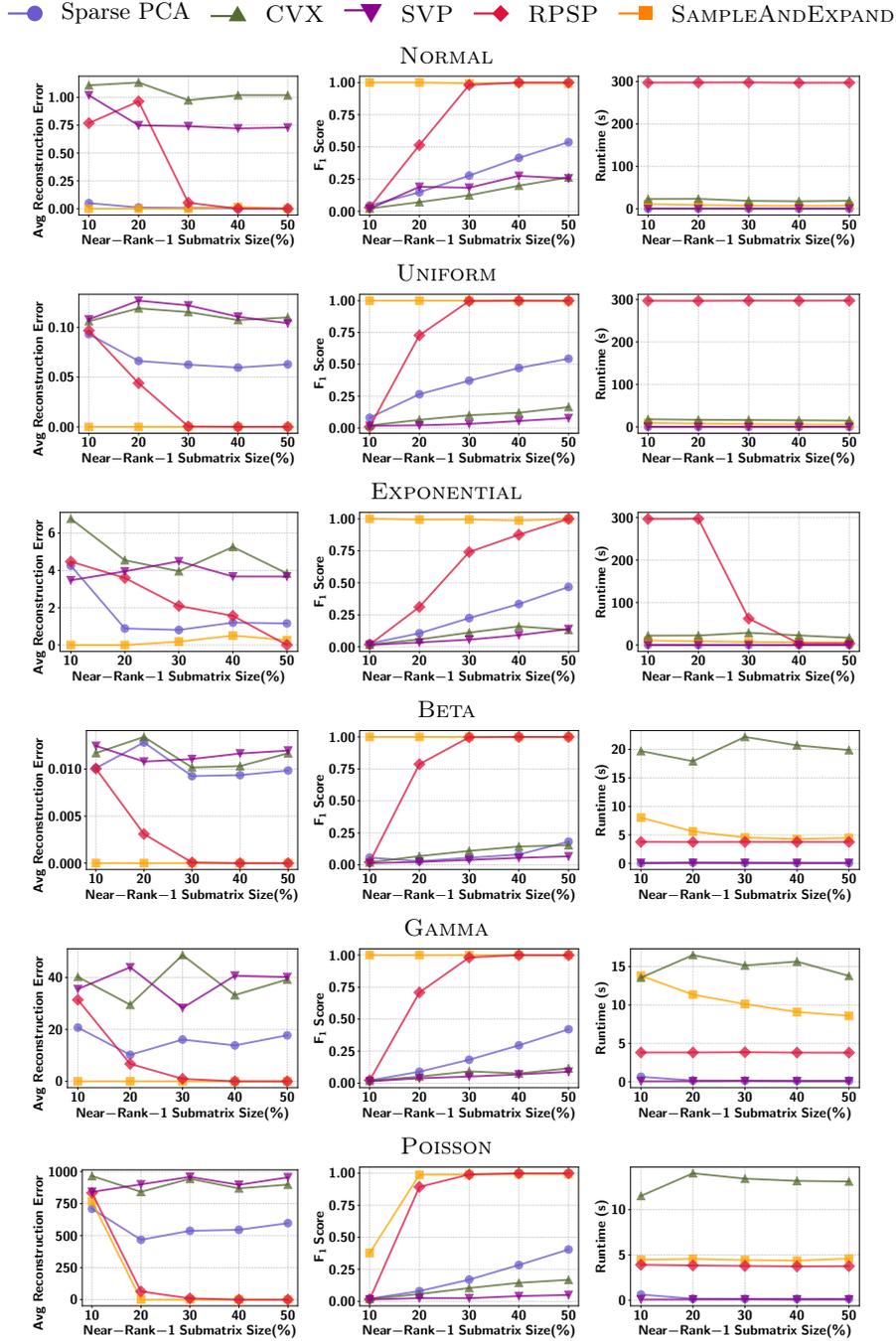

\spara{Results on synthetic datasets.}
Figure~\ref{fig:near_rank_one_recovery} presents results for the task of near-rank-$1$-submatrix discovery in $250 \times 250$ matrices of entries generated from six different probability distributions. 
The results show that our method  consistently recovers the ground truth (as indicated by $F_1$ score close to $1$ and reconstruction error close to $0$), which is not the case for the baseline approaches. 
More specifically, \rpsp tends to recover the ground-truth sub\-matrix as its size increases, but it is also considerably slower than the other methods. 
\sparsePCA and \svp are the fastest algorithms, but, like \cvx, they often fail in detecting the ground truth. 

We also observe significant variability in the results associated with different data-generating distributions. 
In particular, as suggested by Figure~\ref{fig:near_rank_one_recovery},  the only case where \ourmethod does not achieve an average $F_1$ score close to $1$ in near-rank-$1$ sub\-matrix recovery is for Poisson-distributed data with a planted near-rank-$1$ sub\-matrix accounting for the $10 \%$ percent of the total amount of entries.
The synthetic data are not scaled in pre\-processing. Thus, it is 
possible that the $F_1$ score would approach $1$ also in this case after suitably adjusting the parameter $\tol$ input to \ourmethod or normalizing the input matrix \inputmat.

\begin{figure}[p]
\begin{center}
\begin{tikzpicture}[scale=0.1]
\matrix [matrix of nodes, 
column sep=1.0pt, 
row sep=0pt, 
nodes={anchor=center}] (m) {
\tikz \draw[slateblue, thick] (-0.18,0) -- (0.18,0) 
node[fill=slateblue, circle, inner sep=2.1pt] {};
& {\small Sparse PCA}
& 
\tikz \draw[darkolivegreen, thick] (-0.18,0) -- (0.18,0) 
node[fill=darkolivegreen, regular polygon, regular polygon sides=3, inner sep=1.4pt] {};
& {\small \cvx}
& 
\tikz \draw[darkmagenta, thick] (-0.18,0) -- (0.18,0) 
node[fill=darkmagenta, regular polygon, regular polygon sides=3, inner sep=1.7pt, rotate=180] {};
& {\small \svp}
& 
\tikz \draw[crimson, thick] (-0.18,0) -- (0.18,0) 
node[fill=crimson, diamond, inner sep=1.7pt] {};
& {\small \rpsp} 
& 
\tikz \draw[orange, thick] (-0.18,0) -- (0.18,0) 
node[fill=orange, rectangle, inner sep=2.6pt] {};
& {\small \ourmethod} \\
};
\end{tikzpicture}
    \begin{tabular}{ccc}
         & \textsc{Normal} & \\
        \includegraphics[width=0.3\textwidth]{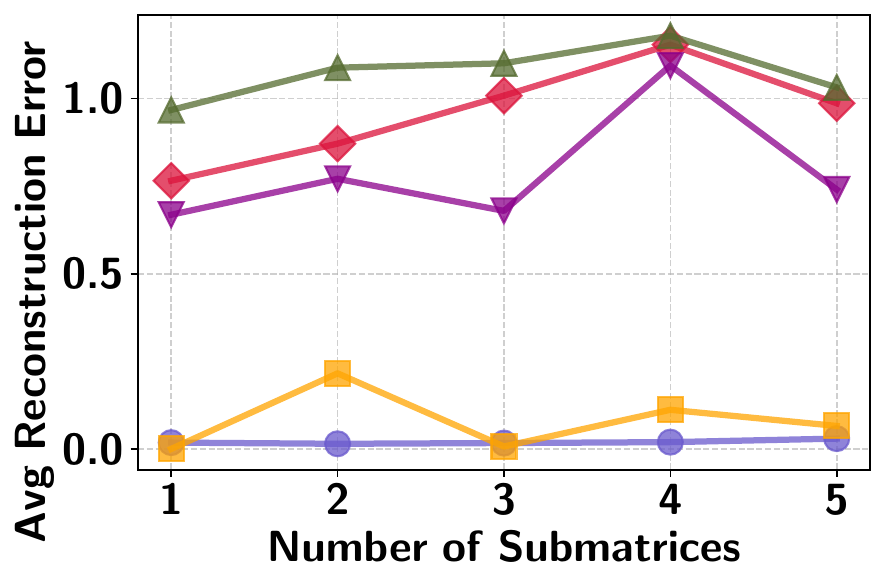} & 
        \includegraphics[width=0.3\textwidth]{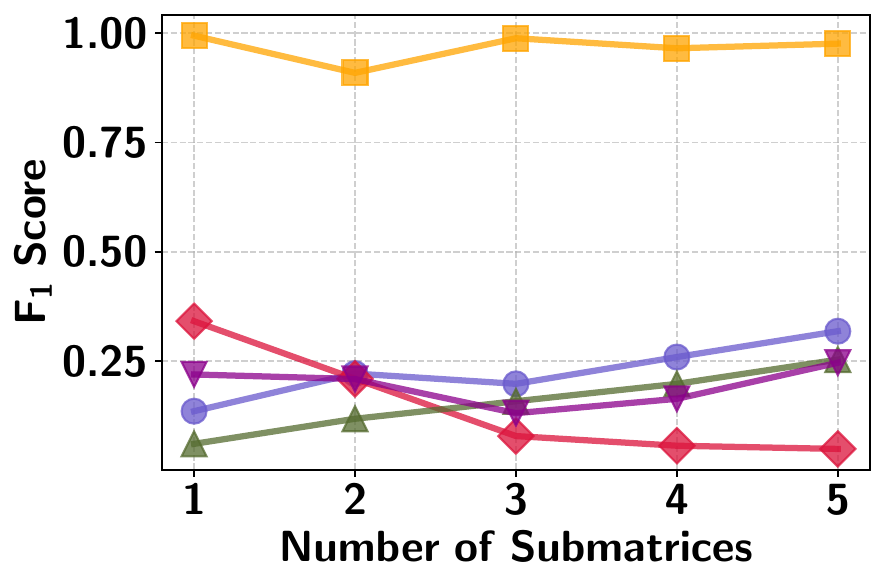} & 
        \includegraphics[width=0.3\textwidth]{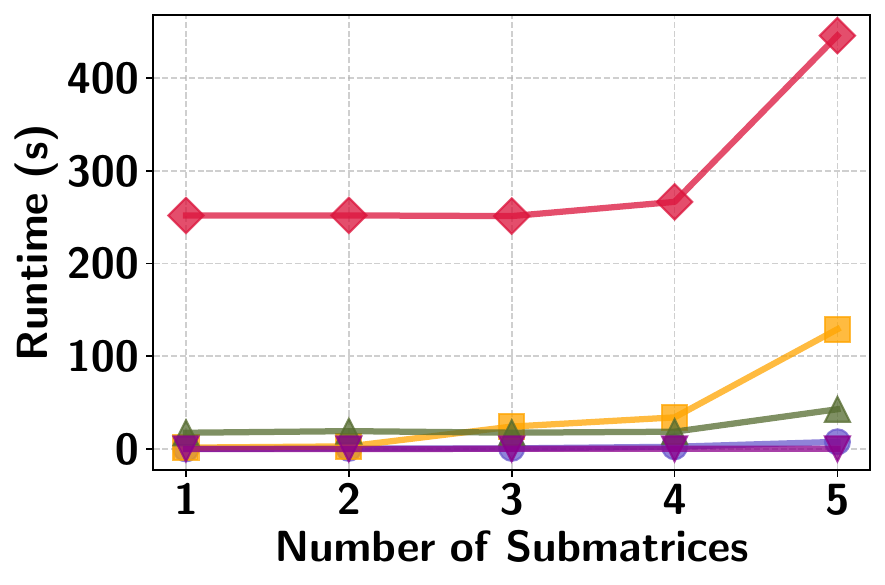} \\
        
         & \textsc{Uniform} & \\
         \includegraphics[width=0.3\textwidth]{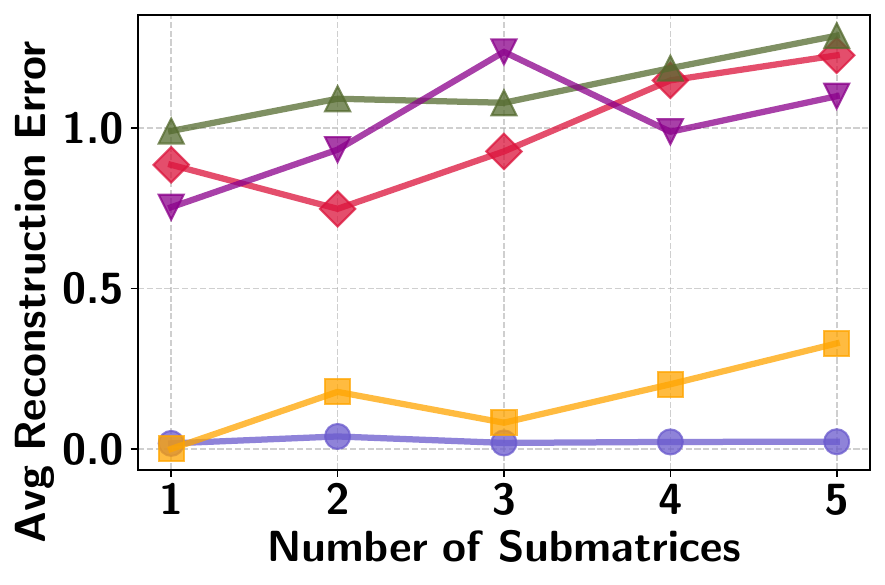} & 
        \includegraphics[width=0.3\textwidth]{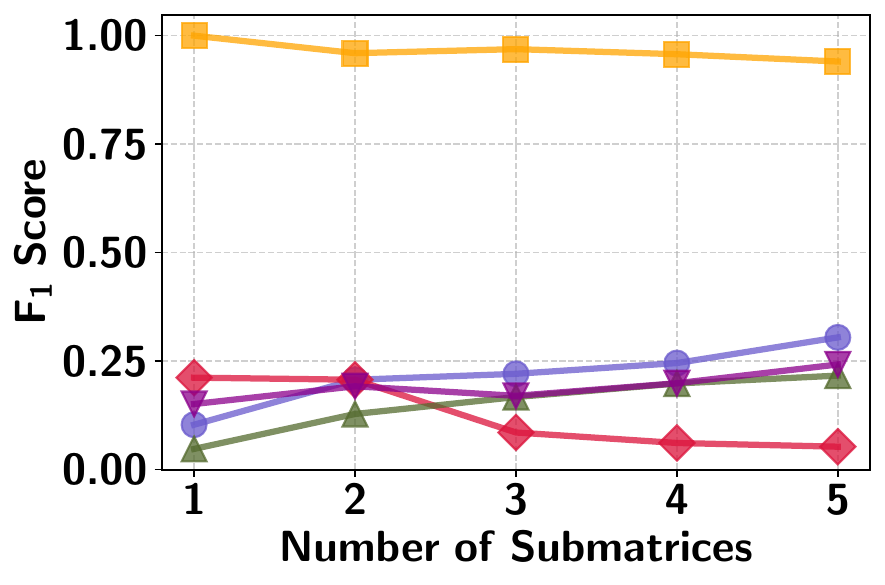} & 
        \includegraphics[width=0.3\textwidth]{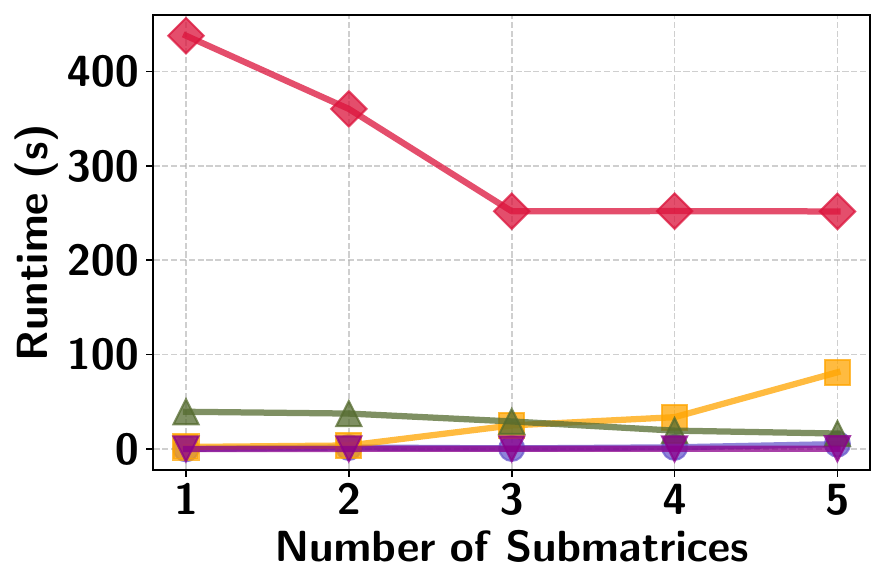} \\

         & \textsc{Exponential} & \\
        \includegraphics[width=0.3\textwidth]{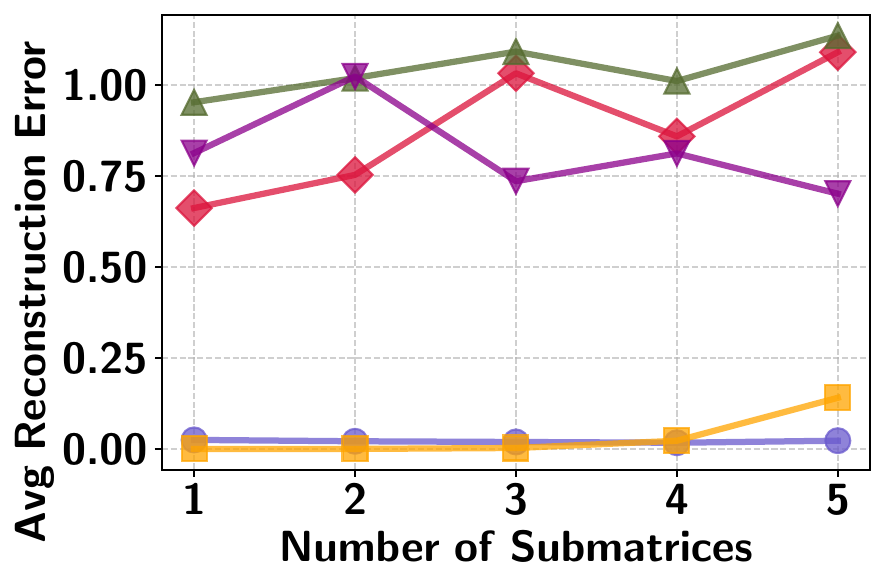} & 
        \includegraphics[width=0.3\textwidth]{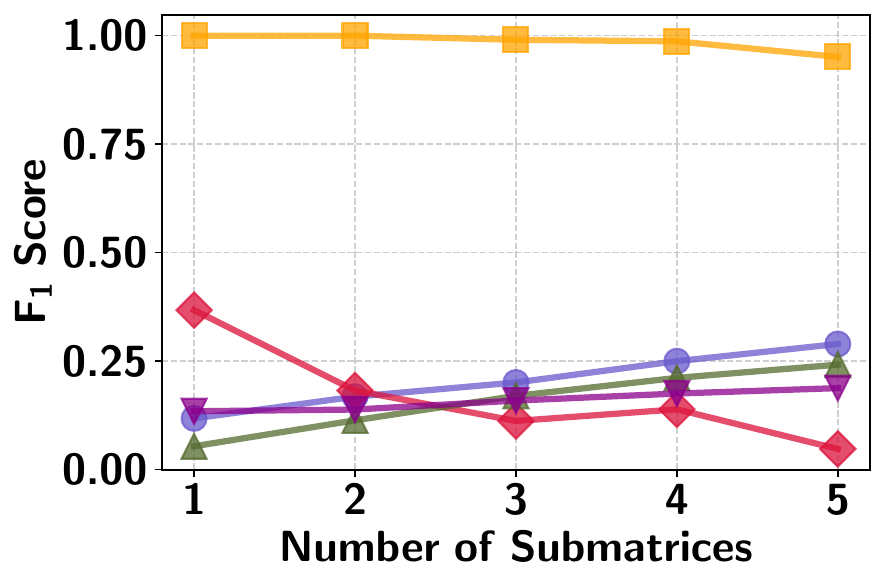} & 
        \includegraphics[width=0.3\textwidth]{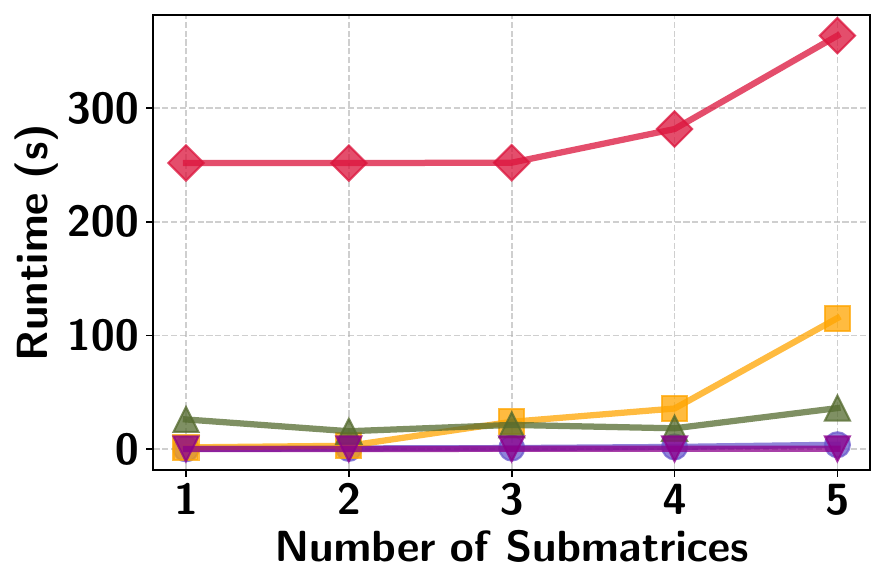} \\

         & \textsc{Beta} & \\
        \includegraphics[width=0.3\textwidth]{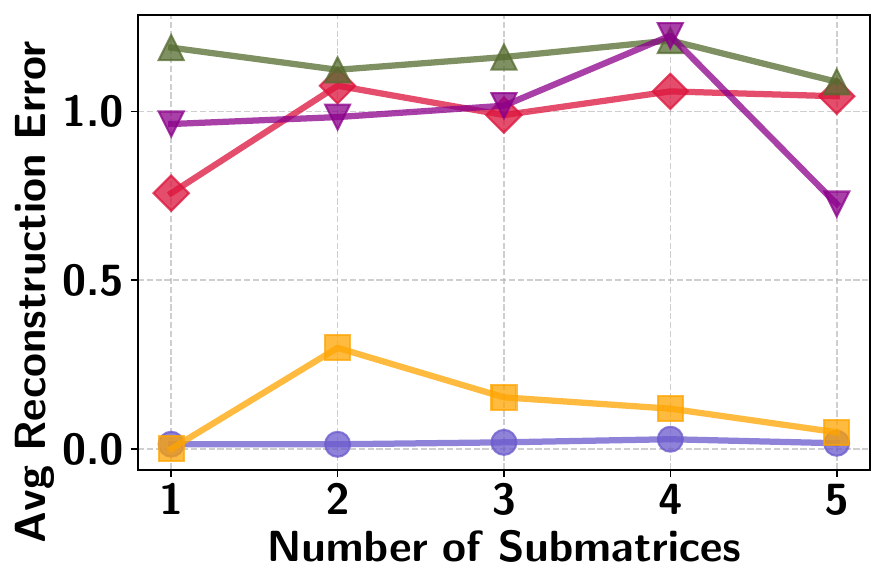} & 
        \includegraphics[width=0.3\textwidth]{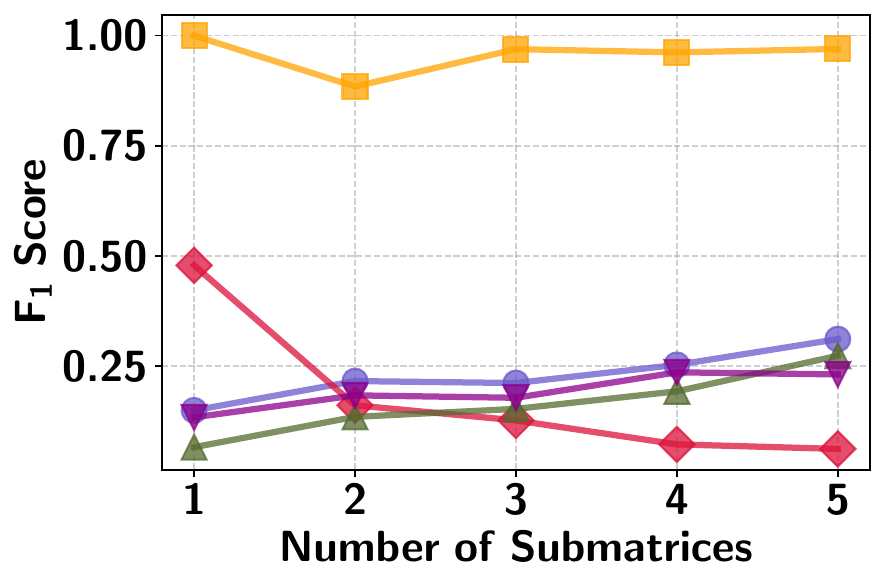} & 
        \includegraphics[width=0.3\textwidth]{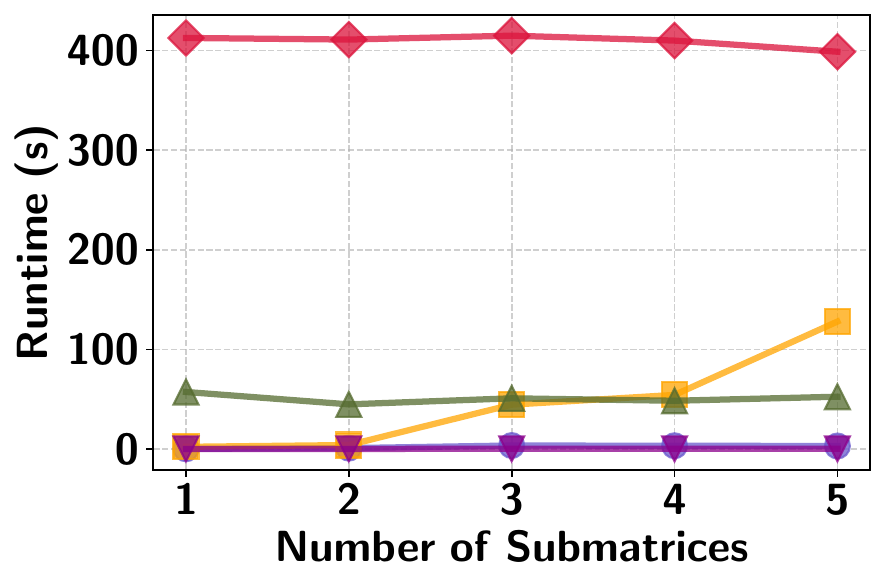} \\

         & \textsc{Gamma} & \\
        \includegraphics[width=0.3\textwidth]{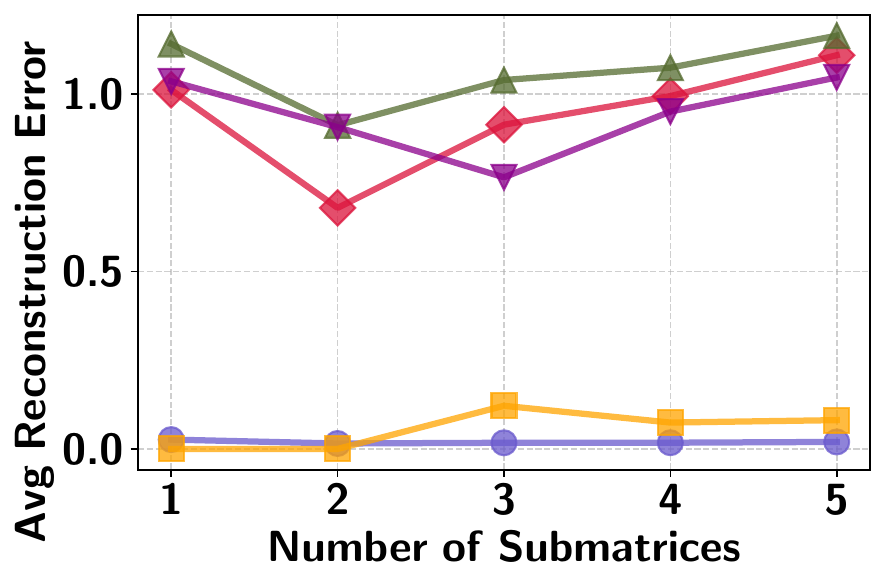} & 
        \includegraphics[width=0.3\textwidth]{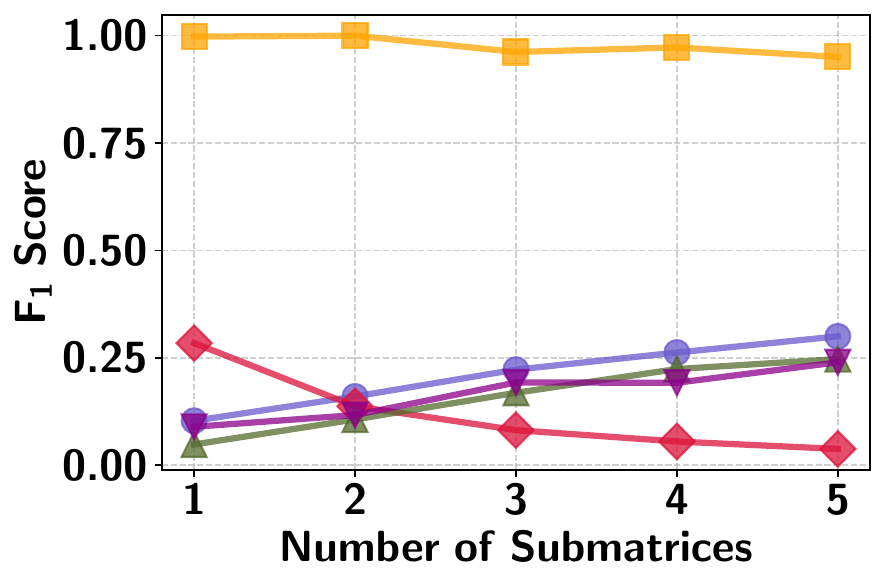} & 
        \includegraphics[width=0.3\textwidth]{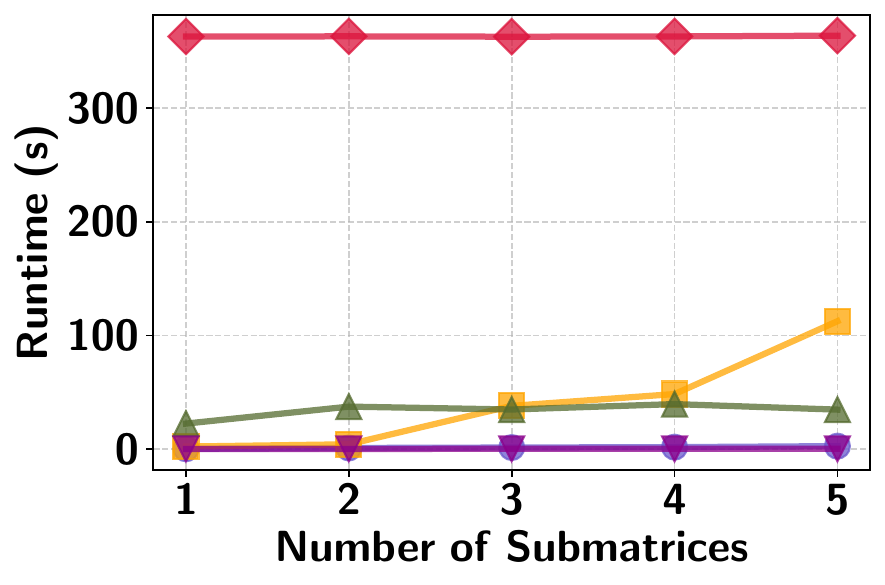} \\

         & \textsc{Poisson} & \\
         \includegraphics[width=0.3\textwidth]{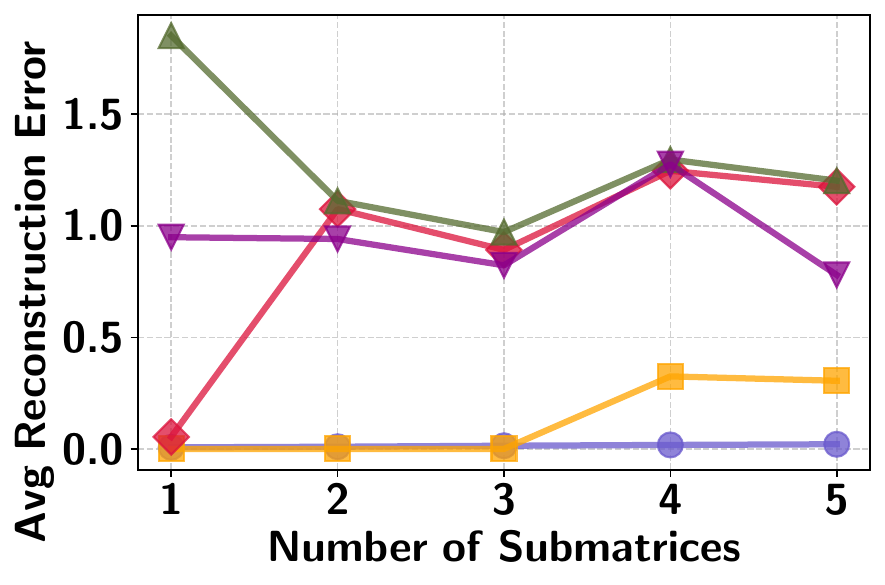} & 
        \includegraphics[width=0.3\textwidth]{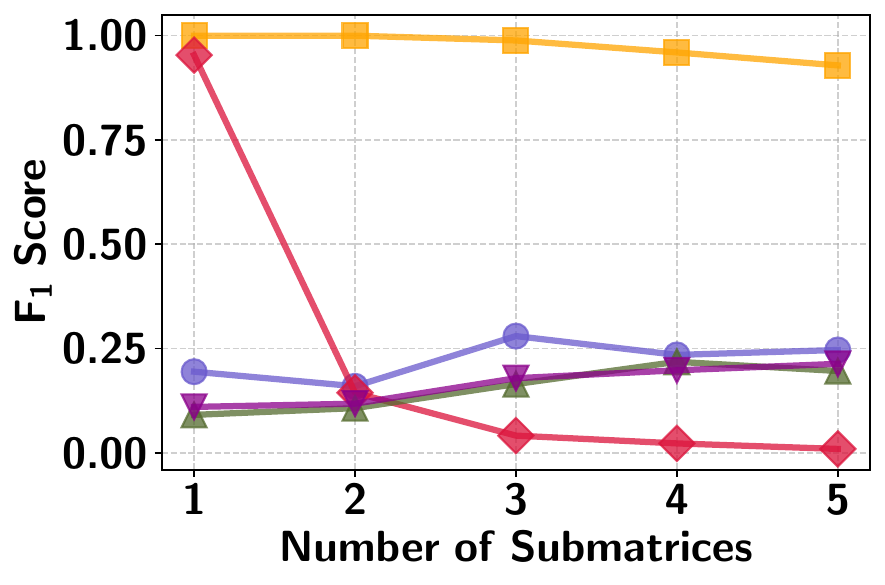} & 
        \includegraphics[width=0.3\textwidth]{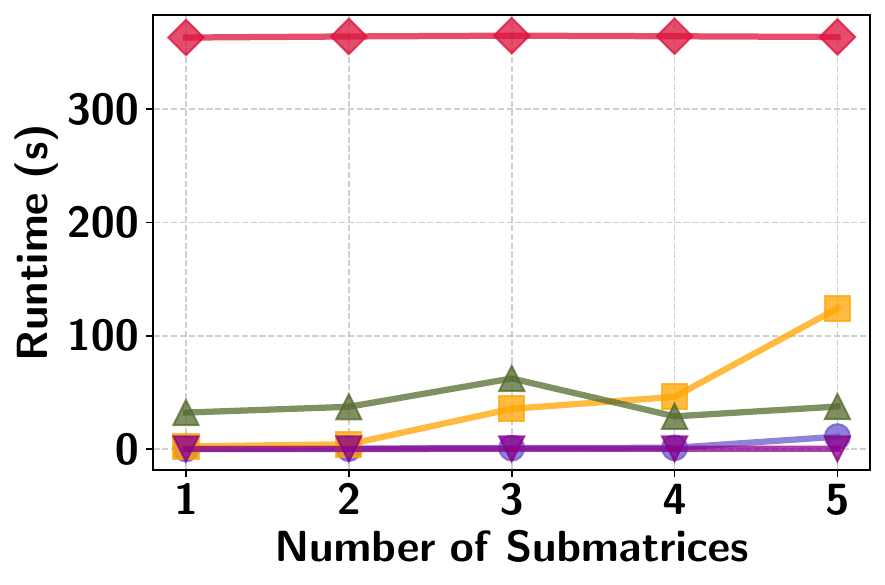} \\
        \vspace*{-0.65cm}
    \end{tabular}
\end{center}
\caption{
\label{fig:local_low_rank_approximation_by_submatrices}
Full-rank synthetic $250 \times 250$ matrices generated from different probability distributions with multiple (possibly overlapping) planted near-rank-$1$ submatrices.
Performance of different methods in discovering planted submatrices. 
We show the average (per-entry) reconstruction error (left), the $F_1$ score (center) and the runtime (right) as a function of the number of planted submatrices. 
}
\end{figure}

Figure~\ref{fig:local_low_rank_approximation_by_submatrices}  
shows the same metrics as in Figure~\ref{fig:near_rank_one_recovery}, for synthetic data with entries generated from the same probability distributions,  
but in the setting where multiple, possibly overlapping, near-rank-$1$ sub\-matrices are planted and discovered. 
The results in this more challenging setting highlight that \ourmethod is the only method 
that consistently retrieves the ground-truth sub\-matrices.
Among the baselines, \sparsePCA stands out for its accurate reconstruction. 
However, the estimate $\estimate{\inputmat}$ of the input matrix it generates quickly becomes very dense as more sub\-matrices are discovered, and hence this approach fails to identify the locations of the ground-truth sub\-matrices. 

Finally,  
Figure~\ref{fig:noise} in~\ref{ap:additional_experiments} 
demonstrates the robustness of our method with respect to noise. 

\spara{Results on real-world data.}
Table~\ref{tab:results_real_data} reports low-rankness score and size averaged over the top-$5$ 
sub\-matrices retrieved by our method, \svp and \rpsp in real-world datasets. 
To determine the top-$5$ sub\-matrices returned by each method, 
we select those that maximize the minimum between the low-rankness score and the size. 
Moreover, to offer a more complete picture, in the appendix (Figure~\ref{fig:real_data_top_five_patters}), we additionally display the low-rankness and size of the individual top-$5$ patterns. 

%
Finding sub\-matrices with high low-rankness is not an easy task. 
\svp returns large sub\-matrices. 
However, those sub\-matrices usually have smaller low-rankness 
compared to those discovered by our method and \rpsp, and, in several cases, compared to the input matrix. 
\ourmethod and \rpsp 
are more likely than \svp to return sub\-matrices with large low-rankness.
Furthermore, \ourmethod tends to discover sub\-matrices that strike a more desirable balance between low-rankness and size compared to \rpsp. 

As concerns runtime, \ourmethod is drastically faster than \rpsp in smaller datasets, but it can become slower in larger datasets. 
Nonetheless, the runtime of our method could be significantly reduced by leveraging a more efficient approach to maximum-edge-bi\-clique extraction and by reducing the number of iterations, which, however, could deteriorate the quality of the results. As mentioned in Section~\ref{sec:scalability}, future work will study the performance of different heuristic approaches to maximum-edge-bi\-clique extraction in terms of quality of the results, efficiency and scalability.


For our method, 
we also explore the trade-off between size and low-rankness by varying the value of $\tol$;
the results are presented in~\ref{ap:additional_experiments}.

\begin{table}[t]
\centering
\caption{\label{tab:results_real_data} Performance in real-world data. For the top $5$ local low-rank patterns identified by the methods, we show the average relative percentage increase (L-R) with respect to the low-rankness score of the input matrix, the size (in percentage of entries of the input matrix) and the runtime (in seconds) to obtain them. }
    \begin{tabular}{l rrr rrr rrr}
        \toprule
        \multirow{2}{*}{Dataset} & \multicolumn{3}{c}{\svp}  & \multicolumn{3}{c}{\rpsp} & \multicolumn{3}{c}{\ourmethod} \\
        \cmidrule(lr){2-4} \cmidrule(lr){5-7} \cmidrule(lr){8-10}
        & L-R & Size & Runtime & L-R & Size & Runtime & L-R & Size & Runtime \\
        \midrule
        \Hyperspectral & 4.05 & 2.88 & 47.0 & 9.84 & 2.10 & 590 & 11.09 & ~21.43 & 1\,697 \\ 
       

        \Isolet & -8.08 & 3.21 & 11 & 4.06 & 3.51 & 409 & 7.08 & 9.9 & 650 \\ 
        \Olivetti & -1.39 & 4.3 & 3 & 2.23 & 2.81 & 366 & 3.83 & 5.2 & 207 \\ 
         \MovieLens & 49.61 & ~8.59 & 1.0 & ~46.69 & ~2.88 & 373 & ~116.42 & 1.06 & 152 \\ 
        \OrlRnSp & -1.51 & 4.39 & 1 & 2.84 & 1.92 & 386 & 3.65 & 4.29 & 58 \\  
        \Golub & -0.32 & 0.29 & 15 & 0.01 & 2.39 & 161 & 0.07 & 19.43 & 61 \\ 
        \Mandrill & 3.25 & 2.3 & 1 & 6.53 & 2.37 & 404 & 6.14 & 3.16 & 43 \\ 
        \Ozone & -39.92 & 0.74 & 2 & 0.03 & 3.23 & 211 & 0.16 & 16.76 & 24 \\ 
        \BRCA & 11.39 & 5.17 & 0 & 14.19 & 2.18 & 371 & 4.97 & 39.44 & 29 \\ 
        \Google & 11.22 & 4.4 & 4 & 18.78 & 2.55 & 145 & 22.63 & 7.16 & 17 \\ 
        \NPAS & -71.29 & 15.86 & 0.2 & 8.55 & 3.93 & 246 & 17.57 & 19.52 & 23 \\ 
        \Cameraman & 1.69 & 3.64 & 0.3 & 5.67 & 3.44 & 102 & 14.44 & 24.83 & 18 \\ 
        \MovieTrust & -2.37 & 4.05 & 0.3 & -20.79 & 4.23 & 209 & 22.49 & 10.49 & 10 \\  
        \Hearth & -16.31 & 2.72 & 0 & 5.13 & 3.63 & 107 & 3.89 & 20.9 & 5 \\ 
       \Imagenet & 3.44 & 4.85 & 0.4 & 5.26 & 3.98 & 126 & 8.52 & 2.29 & 31 \\

        \bottomrule
    \end{tabular}
\end{table}

\section{Conclusion}\label{sec:conclusion}
Low-rank approximation finds applications in many data-analysis tasks. 
Typically, methods assume that the entire matrix exhibits low-rank structure, 
while in real-world data this is often true only for certain sub\-matrices.
In this work, we study the problem of finding sub\-matrices 
that are provably close to a 
rank-$\generalrank$ approximation. 
%
We introduce a novel method that finds such sub\-matrices, study the properties of the method, 
and, with a thorough experimental evaluation, we show
that our method outperforms strong~baselines.
%

There are several directions for future work. 
For instance, future work could study a more robust initialization strategy, 
develop more efficient and scalable alternative algorithms, and optimize the selection of the anchor rows and columns.
It would also be valuable to investigate more the probabilistic aspects of our method.
From a practical perspective, it would be interesting to explore further the benefits 
of our approach in different applications. 


\clearpage

\begin{credits}
\subsubsection{\ackname}
We thank the anonymous reviewers for their valuable comments, which helped us improve the paper.

Martino Ciaperoni is supported by the European Union through the ERC-2018-ADG GA 834756 (“XAI: Science and Technology for the Explanation of AI Decision-Making”) 
and the Partnership Extended PE00000013 (“FAIR: Future Artificial Intelligence Research”), Spoke 1: “Human-Centered AI”.

Aristides Gionis
is supported by ERC Advanced Grant REBOUND (834862), EC H2020 RIA project
SoBigData++ (871042), and the Wallenberg AI, Autonomous Systems and Software
Program (WASP) funded by the Knut and Alice Wallenberg Foundation.

Heikki Mannila is supported by the Technology Industries of Finland Centennial Foundation.
\end{credits}


\renewcommand{\theHsection}{Appendix\Alph{section}} 

\newcommand{\startappendix}{%
    \renewcommand{\thesection}{Appendix \Alph{section}}%
    \setcounter{section}{0}  
    \setcounter{algorithm}{0} 
}

\newcommand{\appendixsection}[2]{%
    \section{#1} 
    \label{#2} 
   \addcontentsline{toc}{section}{Appendix \Alph{section}: #1}
      \setcounter{algorithm}{0}
}

\begin{center}
    {\Large \textbf{Appendix}}
\end{center}


\startappendix

\section*{Appendix Contents}  
\begin{itemize}
    \item \textbf{Appendix A:} \hyperref[app:specialcase]{A Polynomial-time Algorithm to Find a Near-rank-$1$ Subset of Rows or Columns} \dotfill \pageref{app:specialcase}
    \item \textbf{Appendix B:} \hyperref[ap:additional_analysis]{Additional Analysis} \dotfill \pageref{ap:additional_analysis}
    \item \textbf{Appendix C:} \hyperref[ap:proofs]{Proofs} \dotfill \pageref{ap:proofs}
    \item \textbf{Appendix D:} \hyperref[ap:additional_experimental_setup]{Details of Experimental Setup} \dotfill \pageref{ap:additional_experimental_setup}
    \item \textbf{Appendix E:} \hyperref[ap:additional_experiments]{Additional Experiments} \dotfill \pageref{ap:additional_experiments}
\end{itemize}

\appendixsection{A Polynomial-time Algorithm to Find a Near-rank-$1$ Subset of Rows or Columns}{app:specialcase}
\hypertarget{app:specialcase}

In this section, we give a simple algorithm that can be used to solve Problem~\ref{prob:simplified} in polynomial time. 
The algorithm follows easily from the following result. 

\begin{lemma}\label{lemma:errortosine}
Consider two vectors $\xvec \in \real^\ncols$ and $\yvec \in \real^\ncols$ and let $\errorvec = \yvec - \proj{\xvec}{\yvec}$. 
We have: 
\[
\| \errorvec \|_2 = \| \yvec \|_2 \sin(\theta). 
\]
\end{lemma}

To solve Problem~\ref{prob:simplified} optimally in polynomial time, i.e., to find a low-rank subset of rows for a fixed set of columns \colset, we simply normalize the rows 
of \inputmat to unit norm, and then compute the matrix $\sinematrix$ of pairwise sines, defined by: 
\[
\sinematrix = \sqrt{1 - (\inputmat \inputmat^T)^2}. 
\]
The entry of indices $i$ and $j$ of \sinematrix indicates the $L_2$ norm of the error vector associated with the orthogonal projection of $\inputmat_{j,:}$ onto  $\inputmat_{i,:}$. 
Therefore, to find the largest possible set of rows,   we perform a linear scan over the rows of $\sinematrix$ and for each row we pick the largest possible subset such that all elements satisfy the input error requirements. 
The solution to Problem~\ref{prob:simplified} 
is given by the largest subset found while scanning the rows. 

The solution to the symmetric problem asking for the largest subset of columns for a fixed set of rows \rowset can be found by simply inverting the role of the rows and columns in the algorithm. 



\appendixsection{Additional Analysis}{ap:additional_analysis}
\hypertarget{ap:additional_analysis}
Theorem~\ref{th:expansion} provides approximation guarantees for the near-rank-$1$ sub\-matrices found by \ourmethod. 
In more detail, the approximation guarantees are obtained for an interpretable approximation where each row or column is collinear with a single row or column. 
The approximation given by the \svd of the sub\-matrices output by \ourmethod also incurs bounded error in terms of Frobenius and spectral norm, but not necessarily in terms of max norm. 

For the approximation given by the \svd,  it is of particular interest to study the singular values. 
More specifically, since the spectral norm of a matrix is also equal to its largest singular value, it is interesting to study the spectral norm of the matrix of differences \outputerror obtained by using the rank-$1$ \svd to approximate a sub\-matrix discovered by \ourmethod. 
The following result provides a bound on the spectral norm of such matrix of differences.


\begin{theorem}\label{th:expansion_appendix}
Let $\outputmat \in \real^{\nrowssubmatrix \times \ncolssubmatrix}$ be a near-rank-$1$ sub\-matrix output by \ourmethod, with anchor row $\anchorrow$, anchor column $\anchorcolumn$ and tolerance $\delta$. 
The matrix $\outputerror = \outputmat - \estimate{\outputmat}$ where $\estimate{\outputmat}$ is the rank-$1$ approximation of \outputmat given by \svd satisfies: 

\begin{align}
\spectralnorm{\outputerror}  \leq  
\min\Bigg\{  
\sqrt{
\gamma ( \frac{\nrowssubmatrix-1}{\nrowssubmatrix} )  - (\nrowssubmatrix-1) \sqrt{1 -  8 \max_i \| \outputmat_{i,:} \|^2_2 \tol^2 \omega_r  (1 - \tol^2 \omega_r ) } }, 
, \notag &\\
\sqrt{
\gamma ( \frac{\ncolssubmatrix-1}{\ncolssubmatrix} )  - (\ncolssubmatrix-1) \sqrt{1 -  8 \max_j \| \outputmat_{:,j} \|^2_2 \tol^2 \omega_c  (1 - \tol^2 \omega_c ) } }
\Bigg\},
\end{align}
where 
$\gamma = \frobeniusnorm{\outputmat}^2$, 
$\omega_r = \frac{\sum_{i < j} {x^r_i}^2 {x^r_j}^2 }{\| \anchorrow \|_2^2 \min_i \| \outputmat_{i,:} \|^2_2}$
and 
$\omega_c = 
\frac{\sum_{i < j} {x^c_i}^2 {x^c_j}^2 }{\| \anchorcolumn \|_2^2 \min_j \| \outputmat_{:,j} \|^2_2 }$.

\end{theorem}

Although the bound given in Theorem~\ref{th:expansion_appendix} is rather  intricate, it could be considerably simplified by normalizing data, and its interpretation is similar to that of the bounds presented in 
Theorem~\ref{th:expansion}. 
Like Theorem~\ref{th:expansion}, Theorem~\ref{th:expansion_appendix} bounds the approximation error (in this case, in terms of spectral norm) based on the tolerance parameter $\tol$. 

\appendixsection{Proofs}{ap:proofs}
\hypertarget{ap:proofs}
\subsection{Proof of Lemma~\ref{lemma:errortosine}}
Given two vectors \(\yvec \in \mathbb{R}^\ncols \) and \(\xvec \in \mathbb{R}^\ncols \), the orthogonal projection of \(\yvec\) onto \(\xvec\) captures the component of \(\yvec\) that lies in the direction of \(\xvec\). To prove Lemma~\ref{lemma:errortosine}, we need to show that the $L_2$ norm of the difference between $\yvec$ and its projection onto $\xvec$ is equal to the $L_2$ norm of \(\yvec\) multiplied by the sine of the angle \(\theta\) between \(\yvec\) and \(\xvec\).

Recall that the projection of the vector \(\yvec\) onto the vector \(\xvec\) is given by:
\[
\proj{\xvec}{\yvec} = \frac{\yvec^T \xvec}{\xvec^T \xvec} \xvec = \frac{\yvec^T \xvec}{\|\xvec\|_2^2} \xvec.  
\]

The projection error \errorvec associated with  $\proj{\xvec}{\yvec}$ is the vector given by the difference between \(\yvec\) and its projection onto \(\xvec\), i.e., 
\[
\errorvec = \yvec - \proj{\xvec}{\yvec}  = \yvec - \frac{\yvec^T \xvec}{\xvec^T \xvec} \xvec.
\]
The $L_2$ norm of the projection error $\|\errorvec \|_2$ is thus
\(
\|\errorvec\|_2 = \left\| \yvec - \frac{\yvec^T \xvec}{\xvec^T \xvec} \xvec \right\|_2. 
\)
By definition, the dot product \(\yvec^T \xvec\) can be expressed in terms of the $L_2$ norm of \(\yvec\), the $L_2$ norm of \(\xvec\), and the cosine of the angle \(\theta\) between \(\yvec\) and \(\xvec\) as follows: 
\[
\yvec^T \xvec = \|\yvec\|_2 \|\xvec\|_2 \cos(\theta). 
\]
Using this equality, we can rewrite the definition of the the projection error \errorvec as:
\[
\boldsymbol{\epsilon} = \yvec - \frac{\|\yvec\|_2 \|\xvec\|_2 \cos(\theta)}{\|\xvec\|_2^2} \xvec
= \yvec - \frac{\|\yvec\|_2 \cos(\theta)}{\|\xvec\|_2 } \xvec. 
\]
Further, the squared $L_2$ norm of \(\boldsymbol{\epsilon}\) becomes
\(
 \left\| \yvec - \frac{\|\yvec\|_2 \cos(\theta)}{\|\xvec\|_2} \xvec \right\|_2^2. 
\)
Since the projection error is orthogonal to the projection~\cite{strang2022introduction}, the Pythagorean theorem yields:
\begin{equation}\label{eq:orthogonality_projection_error_vector}
\|\yvec\|_2^2 = \|\proj{\xvec}{\yvec}\|_2^2 + \|\boldsymbol{\epsilon}\|_2^2. 
\end{equation}

Because the $L_2$ norm is equivariant under scalar multiplication,  the  $L_2$ norm of the projection \(\|\proj{\xvec}{\yvec}\|_2\) is:
\[
\|\proj{\xvec}{\yvec}\|_2 = \frac{\|\yvec\|_2 \|\xvec\|_2 \cos(\theta)}{\|\xvec\|_2}
= \|\yvec\|_2 \cos(\theta). 
\]
Therefore, using Equation~\ref{eq:orthogonality_projection_error_vector}, the squared $L_2$ norm of the projection error can be written as: 
\[
\|\boldsymbol{\epsilon}\|_2^2 = \|\yvec\|_2^2 - \|\yvec\|_2^2 \cos^2(\theta)
= \|\yvec\|_2^2 (1 - \cos^2(\theta))
= \|\yvec\|_2^2 \sin^2(\theta),  
\]
where we have employed the trigonometric identity $1 - \cos^2(\theta) = \sin^2(\theta)$. 
Finally, upon taking the square root of both sides, we get:
\[
\|\boldsymbol{\epsilon}\|_2 = \|\yvec\|_2 \sin(\theta). 
\]
Thus, we have shown that, as claimed, given the orthogonal projection of \yvec onto \xvec, the $L_2$ norm of the projection error is equal to the $L_2$ norm of \(\yvec\) multiplied by the sine of the angle \(\theta\) between \(\yvec\) and \(\xvec\). 



\subsection{Proof of Theorem~\ref{th:expansion}}
Before proceeding with the proof of Theorem~\ref{th:expansion}, it is useful to prove the following lemma, which refers to the initialization phase only.

\begin{lemma}\label{lemma:initialization}
Consider a $2 \times 2$-sub\-matrix \initialmat and 
let $\initialerror = \initialmat - \estimate{\initialmat}$, where $\estimate{\initialmat}$ is a rank-$1$ approximation of \initialmat.
If, as for the sub\-matrices expanded by \ourmethod,  $|\det(\initialmat)| \leq \toleranceinit$, 
\initialmat admits rank-$1$ approximations $\estimate{\initialmat}$ 
such that: 
\begin{equation}\label{eq:initialization_guarantee_infinity_2by2}
       \infinitynorm{ \initialerror } \leq  
       \frac{\toleranceinit 
\max_{i,j} |\inputmat_{i,j}| } { 2 \min_{i,j} \inputmat_{i,j}^2 }
\end{equation}

and

\begin{equation}\label{eq:initialization_guarantee_frobenius_2by2}
       \spectralnorm{ \initialerror } \leq \frobeniusnorm{ \initialerror } \ \leq  \frac{ \toleranceinit }{\maxnorm}, 
\end{equation} 
where $\maxnorm$ denotes the maximum $L_2$ norm of a row or column of \initialmat. 

\end{lemma}

Lemma~\ref{lemma:initialization} is interesting  not only because it provides a useful criterion to set the value of \toleranceinit, but it also formalizes the determinant-based notion of closeness to rank $1$ for $2 \times 2$ sub\-matrices on which \ourmethod builds. 
This notion is particularly simple; it only requires few basic operations, two products, a difference and possibly a change of sign, while knowledge of the (two) singular values is not required. 
Moreover,  Lemma~\ref{lemma:initialization} is helpful for understanding the main result presented in Section~\ref{sec:analysis}, and its proof, given later in this section.

\begin{proof}
Consider a $2 \times 2$ sub\-matrix: 
\[
\initialmat = \begin{pmatrix}
x_1 & x_2 \\
y_1 & y_2
\end{pmatrix}. 
\]
We will start by proving the bound on the max norm of the $2 \times 2$ matrix $\initialerror$, then we prove the bound on the Frobenius and spectral norms. 

\spara{Bound on the max norm.}
In what follows, we let $\xvec=(x_1,x_2)$ and $\yvec=(y_1, y_2)$.    
If \initialmat has exactly rank $1$, we have $x_1 y_2 = x_2 y_1$ which is equivalent to $\xvec = \alpha \yvec$ for some $\alpha$. In simple words, $\xvec$ and $\yvec$ are proportional. 
In the realistic case of data corrupted by noise, we are interested in near-rank-$1$ $2 \times 2$ sub\-matrices, which are such that $|x_1 y_2 - x_2 y_1| \leq \toleranceinit$ for some small $\toleranceinit$. 
If we approximate the rows of the matrix \initialmat as \xvec and $\alpha \xvec$ to obtain a rank-$1$ submatrix $\estimate{\initialmat}$, the only source of error is the approximation of \yvec as $\alpha \xvec$. 
Thus, 
define $\epsilon_1 = y_1 - \alpha x_1$ and similarly $\epsilon_2 = y_2 - \alpha x_2$. 
The vector $\errorvec = (\epsilon_1, \epsilon_2) $ is referred to as error vector. 

We can take $\alpha$ to be the scalar that defines the orthogonal projection of \yvec onto \xvec. 
It is known that, for any orthogonal projection of a vector $\yvec$ onto another $\xvec$, the  vector \( \errorvec = (\epsilon_1, \epsilon_2),  \) which we henceforth simply refer to as error vector, 
is orthogonal to $\xvec$. 
Therefore: 
\[
\errorvec^T \xvec = \epsilon_1 x_1 + \epsilon_2 x_2 = 0,
\]
from which we derive: 
\[ 
\epsilon_2 = - \frac{x_1}{x_2} \epsilon_1. 
\]
Furthermore, the condition we impose on the determinant means that: 
\[
| x_1 y_2 - x_2 y_1 | 
\leq \toleranceinit.
\]
Substituting the expression of $\yvec$ as a linear function of $\xvec$, we obtain: 
\[
 |x_1 (\alpha x_2 + \epsilon_2)  - x_2 (\alpha x_1 + \epsilon_1) |
\leq \toleranceinit, 
\]
from which: 
\[
| x_1 \epsilon_2  -  x_2 \epsilon_1 |
\leq \toleranceinit.
\]
Substituting $\epsilon_2 = - \frac{x_1}{x_2} \epsilon_1$ and collecting $\epsilon_1$, one gets: 
\[
\left| - \epsilon_1 \left(  \frac{x_1^2}{x_2} + x_2  \right)\right| 
    = \left|\epsilon_1 \left( \frac{x_1^2 + x_2^2}{x_2}\right)\right| \leq \toleranceinit.  
\]
Therefore, we conclude: 
\[ 
\abs{\epsilon_1} \leq \frac{\toleranceinit  |x_2| }{ x_1^2 + x_2^2 }. 
\]
As $|\epsilon_1| = \frac{|x_2|}{|x_1|} |\epsilon_2|$, we also have: 
\[
\abs{\epsilon_2} \ \leq \frac{\toleranceinit  |x_1| }{ x_1^2 + x_2^2 }.
\] 
Equation~\ref{eq:initialization_guarantee_infinity_2by2} follows by observing that we have no made any assumptions on $x_1, x_2, y_1$ and $y_2$, and every entry must thus be bounded by \(  \frac{\toleranceinit 
\max_{i,j} |\inputmat_{i,j}| } { 2 \min_{i,j} \inputmat_{i,j}^2 }  \).

\spara{Bounds on the Frobenius and spectral norms.}
To prove the bound on the Frobenius norm given in Equation~\ref{eq:initialization_guarantee_frobenius_2by2}, we again show that the bound holds for the rank-$1$  approximation $\estimate{\initialmat}$ obtained by stacking the rows $\xvec = (x_1, x_2)$ and $\yvec = \alpha \xvec$, where $\alpha$ is the orthogonal-projection scalar, i.e., $\alpha =\frac{\xvec^T \yvec}{\| \xvec \|^2_2}$.
Because \xvec can be approximated with \xvec itself, the $L_2$ norm of the projection error equals the Frobenius norm of the difference between $\initialmat$ and its rank-$1$ estimate $\estimate{\initialmat}$, i.e., $\| \boldsymbol{\epsilon} \|_2 = \| \initialmat - \estimate{\initialmat} \|_F = \| \initialerror \|_F $. 

Let $\theta$ be the angle between the two vectors $\xvec$ and $\yvec$. 
As stated in Lemma~\ref{lemma:errortosine}, we have that 
$\| \errorvec \|_2 =  \| \yvec \|_2 \sin(\theta)$.

Next, we show that we additionally have the identity: 
\begin{equation}
\label{eq:determinant_to_sine}
    |x_1 y_2 - x_2 y_1| = \| \xvec \|_2 \| \yvec \|_2 sin(\theta). 
\end{equation}
The $L_2$ norms of \( \mathbf{x} \) and \( \mathbf{y} \) are:
\begin{equation*}
\| \mathbf{x} \|_2 = \sqrt{x_1^2 + x_2^2} \text{ and } \| \mathbf{y} \|_2 = \sqrt{y_1^2 + y_2^2},
\end{equation*}
and the cosine of the angle \( \theta \) between the vectors $\xvec$ and $\yvec$ is given by:
\begin{equation*}
\cos(\theta) = \frac{x_1 y_1 + x_2 y_2}{\| \mathbf{x} \|_2 \| \mathbf{y} \|_2 }.
\end{equation*}
Using the identity \( \sin^2(\theta) + \cos^2(\theta) = 1 \), we get: 
\begin{equation*}
\sin^2(\theta) = 1 - \cos^2(\theta) =  1 - \left( \frac{x_1 y_1 + x_2 y_2}{\| \mathbf{x} \|_2 \| \mathbf{y} \|_2 } \right)^2,
\end{equation*}
that is: 
\begin{equation*}
\sin^2(\theta) = \frac{(x_1^2 + x_2^2)(y_1^2 + y_2^2) - (x_1 y_1 + x_2 y_2)^2}{(x_1^2 + x_2^2)(y_1^2 + y_2^2)}.
\end{equation*}
Expanding and simplifying, one obtains: 
\begin{equation*}
    \sin^2(\theta) = \frac{(x_1 y_2 - x_2 y_1)^2}{(x_1^2 + x_2^2)(y_1^2 + y_2^2)},
\end{equation*}
from which: 
\[
 \sin^2(\theta) \| \xvec \|_2^2 \| \yvec \|_2^2 = (x_1 y_2 - x_2 y_1)^2, 
\]
and thus Equation~\ref{eq:determinant_to_sine} follows by taking the square root of both sides. 
As $|x_1 y_2 - x_2 y_1| = \| \xvec \|_2 \| \yvec \|_2  \sin(\theta) $, the condition $|\det(\initialmat)| = |x_1 y_2 - x_2 y_1| \leq \toleranceinit$ implies: 
\[
\| \xvec \|_2 \| \yvec \|_2  \sin(\theta) \leq \toleranceinit,
\]
which holds if and only if:
\begin{equation*}
    \sin(\theta) \leq \frac{\toleranceinit}{\| \xvec \|_2 \|\yvec \|_2}. 
\end{equation*}
Further, since $\| \boldsymbol{\epsilon} \|_2 = \|\yvec\|_2 \sin(\theta)$, we conclude: 
\begin{equation*}\label{eq:forx}
    \| \boldsymbol{\epsilon} \|_2 =  \| \initialerror \|_F  \leq \frac{\toleranceinit}{\| \xvec \|_2}. 
\end{equation*}
We can swap the roles of \xvec and \yvec without changing the absolute value of the determinant, which leads to: 
\begin{equation}\label{eq:fory}
     \| \boldsymbol{\epsilon} \|_2  = \| \initialerror \|_F  \leq \frac{\toleranceinit}{\| \yvec \|_2}. 
\end{equation}
Similarly, 
swapping the columns does not change the absolute value of the determinant, which concludes the proof of the bound on the Frobenius norm. 
Finally, the spectral norm is always upper bounded by the Frobenius norm, and hence the lemma follows.

\end{proof}

We are now in a position to prove Theorem~\ref{th:expansion}, which describes the approximation-error guarantees associated with Algorithm~\ref{alg:rankone} in terms of max, Frobenius and spectral norm. 
To prove the theorem, we extend the proof of Lemma~\ref{lemma:initialization} from the initialization to the expansion phase.

\begin{proof}
Like for the proof of Lemma~\ref{lemma:initialization}, we prove bounds separately for the max norm and Frobenius and spectral norms.  
For simplicity, we present the proof for the row-wise ratios, and we denote by \xvec the anchor row, while \yvec denotes any other row of \outputmat. 
The proofs for the column-wise ratios are always symmetric. 

\spara{Bound on the max norm.}
The proof of the bound on the max norm (Equation~\ref{eq:initialization_guarantee_infinity}) is straightforward, as it is a simple extension of the proof of the bound on the max norm given in Lemma~\ref{lemma:initialization} for the case of $2 \times 2$ sub\-matrices.    


Given the anchor row \anchorrow, 
for every pair of entries $(y_1, y_2)$ of another row  $\yvec$, we consider the approximations $y_1 = \alpha x^r_1$ and $y_2 = \alpha x^r_2$ where $\alpha$ is the orthogonal-projection scalar for the projection of $\yvec$ onto $\anchorrow$. 

The well-developed theory of orthogonal projections guarantees that the error vector 
\( \errorvec = (\epsilon_1, \epsilon_2) \)
is orthogonal to $\anchorrow$~\cite{strang2022introduction}. 
Therefore, we have: 
\[
\epsilon_1 x^r_1 + \epsilon_2 x^r_2 = 0,
\]
i.e., 
\[ 
\epsilon_2 = - \frac{x^r_1}{x^r_2} \epsilon_1. 
\]
The condition \ourmethod imposes on the ratio matrices 
implies that: 
\[
| \frac{y_1}{x^r_1} - \frac{y_2}{x^r_2} | 
= 
| \frac{y_1x^r_2 -x^r_1 y_2}{x^r_1 x^r_2}| = 
 \frac{|y_1 x^r_2 - x^r_1 y_2|}{|x^r_1| |x^r_2|} 
\leq \delta.
\]
Substituting the expression of $\yvec$ as a linear function of $\xvec$: 
\[
 \frac{|(\alpha x^r_1 + \epsilon_1) x^r_2 -x^r_1 (\alpha x^r_2 + \epsilon_2) |}{|x^r_1| |x^r_2|} 
\leq \delta.
\]
Simplifying: 
\[
 \frac{| \epsilon_1 x^r_2 - \epsilon_2 x^r_1 |}{|x^r_1| |x^r_2|} 
\leq \delta \iff | \epsilon_1 x^r_2 - \epsilon_2 x^r_1 |
\leq \tol |x^r_1| |x^r_2|.
\]
Substituting $\epsilon_2 = - \frac{x^r_1}{x^r_2} \epsilon_1$ and collecting $\epsilon_1$, one gets: 
\[
\left| \epsilon_1 \left( \frac{ {x_2^r}^2 + {x_1^r}^2 }{ x_2^r } \right) \right| \leq  \tol |x^r_1| |x^r_2|,
\]
from which we conclude: 
\[
\abs{\epsilon_1} \leq \frac{\tol |x^r_1| {x^r_2}^2 }{ {x_1^r}^2 + {x_2^r}^2  }. 
\]
As $ |\epsilon_1| = \frac{|x^r_2|}{|x^r_1|} |\epsilon_2|$, we also have: 
\[
\abs{\epsilon_2} \ \leq \frac{\tol {x^r_1}^2 |x^r_2| }{{x_1^r}^2 + {x_2^r}^2 }. 
\]
We have no made any assumptions on $x^r_1, x^r_2, y_1$ and $y_2$. Therefore, the bound:
\[
\frac{\tol \max_{i} |x_i^r|^3 }{2 \min_{i} {x^r_i}^2 }
\]
necessarily hold for any entry in \inputmat. 
Furthermore, a symmetric bound can be obtained by considering the column-wise ratios instead of the row-wise ones, and then
Equation~\ref{eq:initialization_guarantee_infinity} follows.


\spara{Bounds on the Frobenius and spectral norms.}
To prove the theorem, we extend the proof of Equation~\ref{eq:initialization_guarantee_frobenius_2by2} in Lemma~\ref{lemma:initialization} to the general case of $\nrowssubmatrix \times \ncolssubmatrix$ sub\-matrices.  
First notice that, given the orthogonal projection of a row \yvec onto \anchorrow, as in the case of Lemma~\ref{lemma:errortosine}, which does not make any assumption on the dimensions of the vectors under consideration, for the norm of the (now $\ncolssubmatrix$-dimensional) error vector, it holds that $\|\boldsymbol{\epsilon}\|_2^2 = \| \yvec \|_2^2 \sin^2 (\theta)$.
In the expansion stage, we seek submatrices such that the $2 \times 2$-determinants within it are near zero. Instead of computing determinants, however, we compute ratios, which is more convenient. 
As already pointed out, 
the ratios that are computed in the expansion phase of our method are directly related to the determinants of the $2\times2$-sub\-matrices.
In particular, considering again row \anchorrow as the anchor row we divide by and another row $\yvec$, we require the following inequality on the ratios: 
\[
\left| \frac{y_i}{x^r_i} - \frac{y_j}{x^r_j} \right| \leq \tol \quad \text{for all } i \text{ and } j, 
\]
which implies: 
\[
| y_i x^r_j  - y_j x^r_i | \leq | x^r_i | | x^r_j | \tol \quad \text{for all } i \text{ and } j. 
\]
Thus, the bound on the absolute difference of ratios guarantees that the associated $2 \times 2$-sub\-matrix determinant is bounded by \tol, rescaled by the absolute values of the entries $x^r_i$ and $x^r_j$ of the anchor row. 

Keeping in mind the relationship between ratios and $2\times2$ determinants, we continue the proof by finding a bound on the
the norm of the error vector $\| \boldsymbol{\epsilon} \|_2^2 = \|\mathbf{y}\|_2^2 \sin^2(\theta)$ obtained by projecting \yvec onto \anchorrow.
In particular, to find a bound for  $\| \boldsymbol{\epsilon} \|_2^2$, we need to bound $\sin^2(\theta)$. 
First, notice that: 
\[
\sin^2(\theta) = 1 - \cos^2(\theta )= 1 - \frac{({\anchorrow}^T \yvec)^2}{ \| \anchorrow \|_2^2 \| \yvec \|_2^2 } = \frac{\| \xvec \|_2^2 \| \yvec \|_2^2  - ({\anchorrow}^T \yvec)^2}{ \| \anchorrow \|_2^2 \| \yvec \|_2^2 }. 
\]
Let us focus on the numerator, $\| \anchorrow \|_2^2 \| \yvec \|_2^2  - ({\anchorrow}^T \yvec)^2$. 
For simplicity, in what follows, we use the notation $\sum_{i \neq j}$ to denote the sum over pairs $(i,j)$ from $1$ to $\ncolssubmatrix$ such that $i\neq j$, and, similarly, the notation $\sum_{i < j}$ to indicate the sum over pairs $(i,j)$ from $1$ to $\ncolssubmatrix$  such that $i$ is lower than $j$. 
Expanding the product of the squared $L_2$ norms, one obtains: 
\[
\|\anchorrow\|_2^2 \|\yvec\|_2^2 = \left( \sum_{i=1}^{\ncolssubmatrix} {x^r_i}^2 \right) \left( \sum_{j=1}^{\ncolssubmatrix} y_j^2 \right) = \sum_{i=1}^{\ncolssubmatrix} {x^r_i}^2 y_i^2 + \sum_{i \neq j} {x^r_i}^2 y_j^2.
\]
Similarly, expanding the squared dot product, we get:
\[
({\anchorrow}^T \yvec)^2 = \sum_{i=1}^{\ncolssubmatrix} {x^r_i}^2 y_i^2 + 2 \sum_{i < j} x^r_i y_i x^r_j y_j.
\]
Thus:
\[
\|\anchorrow\|_2^2 \|\yvec\|_2^2 - ({\anchorrow}^T \yvec)^2 = \sum_{i=1}^{\ncolssubmatrix} \anchorrowentryis^2 y_i^2 + \sum_{i \neq j} \anchorrowentryis^2 y_j^2 - \left( \sum_{i=1}^{\ncolssubmatrix} \anchorrowentryis^2 y_i^2 + 2 \sum_{i < j} \anchorrowentryi y_i \anchorrowentryj y_j \right).
\]
Canceling out the  terms \( \sum_{i=1}^{\ncolssubmatrix} {x^r_i}^2 y_i^2 \), we are left with:
\[
\sum_{i \neq j} \anchorrowentryis^2 y_j^2 - 2 \sum_{i < j} \anchorrowentryi y_i \anchorrowentryj y_j,
\]
%
By symmetry, the term $\sum_{i \neq j} \anchorrowentryis^2 y_j^2 $ can be re\-written as $\sum_{i < j} (\anchorrowentryis^2 y_j^2 + \anchorrowentryjs^2 y_i^2)$, and hence: 
\begin{align*}
    \sum_{i \neq j} \anchorrowentryis^2 y_j^2 - 2 \sum_{i < j} \anchorrowentryi y_i \anchorrowentryj y_j =&\\ \sum_{i < j} (\anchorrowentryis^2 y_j^2 + \anchorrowentryjs^2 y_i^2) - 2 \sum_{i < j} \anchorrowentryi y_i \anchorrowentryj y_j = \sum_{i < j} ( \anchorrowentryis^2 y_j^2 + \anchorrowentryjs^2 y_i^2 - 2 \anchorrowentryi y_i \anchorrowentryj y_j).
\end{align*}
The expression above corresponds to a sum of squared differences. Specifically, it is equivalent to: 
\[
\sum_{i < j} (y_i \anchorrowentryj - y_j \anchorrowentryi)^2.
\]
Thus, we have shown that:
\[
\|\anchorrow\|_2^2 \|\yvec\|_2^2 - ({\anchorrow}^T \yvec)^2 = \sum_{i < j} (y_i \anchorrowentryj - y_j \anchorrowentryi )^2.
\]
As explained above, the condition we impose on the ratios leads to $ | y_i \anchorrowentryj - y_j \anchorrowentryi   | \leq |\anchorrowentryi| |\anchorrowentryj| \tol $, or, equivalently $  (y_i \anchorrowentryj - y_j \anchorrowentryi  )^2 \leq {\anchorrowentryi}^2 {\anchorrowentryj}^2 \tol^2 $. 
Therefore: 
\[
\sin^2 (\theta) = \frac{\|\anchorrow\|_2^2 \|\yvec\|_2^2 - ( {\anchorrow}^T \yvec)^2}{\|\anchorrow\|_2^2 \|\yvec\|_2^2} = \frac{ \sum_{i < j} (\anchorrowentryi y_j - \anchorrowentryj y_i)^2 } {\|\anchorrow\|_2^2 \|\yvec\|_2^2} \leq  \frac{ \tol^2 \sum_{i < j} {\anchorrowentryi}^2 {\anchorrowentryj}^2 }{\|\anchorrow\|_2^2 \|\yvec\|_2^2}.
\]
Finally, recalling that $\| \boldsymbol{\epsilon} \|_2^2 = \| \yvec \|_2^2 \sin^2 (\theta)$, we have: 
\begin{equation}\label{eq:guarantee_k}
\| \boldsymbol{\epsilon} \|_2^2 \leq \frac{ \tol^2 \sum_{i < j} {\anchorrowentryi}^2 {\anchorrowentryj}^2 }{\|\anchorrow\|_2^2}. 
\end{equation}

Equation~\ref{eq:guarantee_k} applies to every non-anchor row $\yvec$ of \outputmat.  
As usual, 
the same derivations also hold for the columns of \outputmat. 
Hence, the bound on the Frobenius stated in the theorem follows, and the spectral norm is always bounded by the Frobenius norm. 
This concludes the proof of the theorem. 
\end{proof}


\subsection{Proof of Theorem~\ref{th:expansion_appendix}}
To conclude this section, we present the proof of Theorem~\ref{th:expansion_appendix} which bounds the low-rank-approximation error incurred by the rank-$1$ \svd for the sub\-matrices detected by \ourmethod. 


\begin{proof}
To prove the bound on the spectral norm of the matrix of differences \outputerror obtained by approximating a near-rank-$1$ sub\-matrix discovered by \ourmethod through its rank-$1$ \svd,
we again consider the rows of \outputmat and we re\-use the bound on the sine of the angle $\theta$ between a row \yvec and the anchor row \anchorrow obtained in the proof of Theorem~\ref{th:expansion}. 
Specifically, while proving the bound on the Frobenius and spectral norms in Equation~\ref{eq:expansion_guarantee}, we have shown that: 
\[
\sin^2(\theta) \leq \frac{\|\anchorrow\|_2^2 \|\yvec\|_2^2 - ({\anchorrow}^T \yvec)^2}{\|\anchorrow\|_2^2 \|\yvec\|_2^2} = \frac{ \sum_{i < j} ( {x^r_i} y_j - {x^r_j} y_i)^2 } {\|\anchorrow\|_2^2 \|\yvec\|_2^2} \leq  \frac{ \tol^2 \sum_{i < j} {x^r_i}^2 {x^r_j}^2 }{\|\anchorrow\|_2^2 \|\yvec\|_2^2}.
\]
For ease of notation, we let $\yvec_{min}$ denote the non-anchor row \yvec of minimum squared $L_2$ norm in \inputmat and \( \tau = \frac{ \tol^2 \sum_{i < j} {\anchorrowentryi}^2 {\anchorrowentryj}^2 }{\|\anchorrow\|_2^2 \| \yvec_{min} \|_2^2 } \), which corresponds to the largest possible bound on $\sin^2(\theta)$ across all rows of \outputmat, and will be useful later in the proof. 
%

A widely-known result in linear algebra guarantees that the square of a singular value of $\outputmat$ is equal to the corresponding eigenvalue of the symmetric and positive semi-definite Gram matrix $\gram = \outputmat \outputmat^T \in \real^{\nrowssubmatrix \times \nrowssubmatrix}$~\cite{strang2022introduction}. 
Consider the rank-$1$ approximation $\estimate{\outputmat}$ of $\outputmat$ given by the \svd.  
Subtracting $\estimate{\outputmat}$ from $\outputmat$ cancels out the leading (i.e., largest) singular value of \outputmat. 
Thus, to prove an upper bound on the spectral norm of \outputerror, we prove an upper bound on the second largest singular value of \outputmat, which is equivalent to the square root of the largest eigenvalue $\lambda_{max}(\gram)$ of \gram. 
To find an upper bound on the the second largest eigenvalue of \gram, we start by deriving a lower bound on the largest eigenvalue. 
Such lower bound follows from the inequality \( \sin^2(\theta) \leq \tau \) which holds for all angles $\theta$ between the anchor row \anchorrow and any other row $\yvec$. 
Intuitively, this inequality suggests that all the row vectors are aligned to the same vector \xvec. 
Thus, the angle between any two non-anchor rows cannot be arbitrarily large. 
More precisely, the angle between  $\yvec_i$ and $\yvec_j$ can be at most equal to the sum of the angles between $\yvec_i$ and $\anchorrow$ and between $\yvec_j$ and $\anchorrow$. 
In this case,  we have: 
\[ 
\cos (\theta_{i,j}) = \cos (\theta_i) \cos (\theta_j) - \sin (\theta_i) \sin (\theta_j), 
\]


where $\theta_i$ and $\theta_j$ denote the angle between $\yvec_i$ and $\anchorrow$ and the angle between $\yvec_j$ and $\anchorrow$, respectively, whose squared sines are bounded by $\tau$. 
Recalling that the squared cosine equals the complement to $1$ of the squared sine, we obtain: 
\[
\cos(\theta_{i,j}) \geq ( \sqrt{1 - \tau} \sqrt{1 - \tau} ) - ( \sqrt{\tau}  \sqrt{\tau}),
\]
i.e., $\cos(\theta_{i,j}) \geq 1 - 2 \tau$, which is equivalent to: 
\[
\sin^2(\theta_{i,j}) \leq 4 \tau ( 1 - \tau). 
\]
We have obtained an upper bound on the squared sine of the angle between any pair of non-anchor rows in \outputmat, which translates into a lower bound on the inner product. 
In particular, we have that: 
\begin{align*}
    \sin^2(\theta_{i,j})  = 1 - \cos^2 (\theta_{i,j})   = 1 - \frac{ (\yvec_i^T \yvec_j)^2}{ \| \yvec_i \|^2_2 \| \yvec_j \|^2_2 } \leq 4 \tau (1 - \tau),  
\end{align*}
implying: 
\begin{align*}
 - (\yvec_i^T \yvec_j)^2 \leq 4 \tau (1 - \tau)  \| \yvec_i \|^2_2 \| \yvec_j \|^2_2 - 1,
\end{align*}
and from which: 
\[
\yvec_i^T \yvec_j \geq \sqrt{1 - 4 \tau ( 1 - \tau) \| \yvec_i \|^2_2 \| \yvec_j \|^2_2 }. 
\]
A bound that holds for all possible pairs of rows $\yvec_i$ and $\yvec_j$ is thus: 
\[
\sqrt{1 - 8 \tau (1 - \tau)  \| \yvec_{max} \|^2},  
\]
where 
$\yvec_{max}$ denotes the non-anchor row of maximum $L_2$ norm. 


Now that we have obtained a lower bound on the inner products between any pair  of rows of \outputmat, note that the inner products correspond precisely to the off-diagonal entries of \gram, while the diagonal entries of \gram are all of the form $\yvec_i^T \yvec_i = \| \yvec_i \|_2^2$. 
It is known that the largest eigenvalue of a matrix can be found by maximizing Rayleigh quotient~\cite{strang2022introduction}, i.e., it satisfies:
\begin{equation*}
    \lambda_{\max}(\gram) = \max_{\mathbf{v} \neq \mathbf{0}} \frac{\mathbf{v}^{T} \gram \mathbf{v}}{\mathbf{v}^{T} \mathbf{v}}.
\end{equation*}
Choosing $\mathbf{v} = (1,1,\dots,1)^T \in \real^{1 \times \nrowssubmatrix}$, we get:
\begin{equation*}
    \mathbf{v}^{T} \gram \mathbf{v} = \sum_{i,j} \gram_{ij} \geq \frobeniusnorm{\outputmat}^2 + (\nrowssubmatrix -1)\nrowssubmatrix  \tau', 
\end{equation*}
%
where $\tau' = \sqrt{1 - 8 \tau ( 1 - \tau) \| \yvec_{max} \|^2}$. 
Since the denominator of Rayleigh quotient $\mathbf{v}^{T} \mathbf{v}$ takes value $\nrowssubmatrix$, we obtain:
\begin{equation*}
    \lambda_{\max}(\gram) \geq \frac{\frobeniusnorm{\outputmat}^2}{\nrowssubmatrix} + (\nrowssubmatrix-1)\tau'.  
\end{equation*} 
%
%
Thus, for the largest singular value $\sigma_1(\outputmat)$ of \outputmat, we have that 
\(  \sigma_1(\outputmat) \geq  \sqrt{ \frac{\frobeniusnorm{\outputmat}^2}{\nrowssubmatrix} + (\nrowssubmatrix -1)\tau'} \). 
We also have the equality~\cite{strang2022introduction}:
\begin{equation*}
    \sum_{i=1}^{n} \lambda_i(\gram) = \text{Tr}(\gram) = \frobeniusnorm{\outputmat}^2.
\end{equation*}
Therefore, since $\lambda_{\max}(\gram) \geq \frac{\frobeniusnorm{\outputmat}^2}{\nrowssubmatrix} + (\nrowssubmatrix-1)\tau'$, and the eigenvalues after the first one cannot be larger than the first one, the remaining eigenvalues satisfy:
\begin{equation*}
    \sum_{i=2}^{\nrowssubmatrix} \lambda_i(\gram) \leq \frobeniusnorm{\outputmat}^2 - \left(\frac{\frobeniusnorm{\outputmat}^2}{\nrowssubmatrix} + (\nrowssubmatrix-1)\tau'\right) = \frobeniusnorm{\outputmat}^2 - \frac{\frobeniusnorm{\outputmat}^2}{\nrowssubmatrix} - (\nrowssubmatrix-1)\tau'.
\end{equation*}
Hence, we get:
\begin{equation*}
    \lambda_2(\gram) \leq \frobeniusnorm{\outputmat}^2 - \frac{\frobeniusnorm{\outputmat}^2}{\nrowssubmatrix} - (\nrowssubmatrix-1)\tau'.
\end{equation*}
As a consequence, recalling the definition of $\tau'$, 
we can conclude: 
\begin{align*}
\spectralnorm{\outputerror} \leq  \sigma_2( \outputmat )  \leq 
\\
\sqrt{
\frobeniusnorm{\outputmat}^2 - \frac{\frobeniusnorm{\outputmat}^2}{\nrowssubmatrix} - (\nrowssubmatrix-1) \sqrt{1 -  8 \max \| \outputmat_{i,:} \|^2_2  \frac{ \delta^2  \sum_{i < j}  { x^r_i }^2 {x^r_j}^2   }{ \| \xvec \|_2^2 \min_i \|\outputmat_{i,:}\|_2^2 } (1 - \frac{ \delta^2  \sum_{i < j}  { x^r_i }^2 {x^r_j}^2 }{ \| \xvec \|_2^2 \min_i \|\outputmat_{i,:}\|_2^2 } )   }
}.
\end{align*}

As usual, the same derivations also apply to the columns of \outputmat, which concludes the proof. 
\end{proof}

\appendixsection{Details of Experimental Setup}{ap:additional_experimental_setup}
\hypertarget{ap:additional_experimental_setup}
\spara{Datasets.}
To generate synthetic data,  we proceed as follows.
\begin{itemize}
    \item First, we generate vectors $\xvec \in \real^{\nrowssubmatrix}$ and $\yvec \in \real^{\ncolssubmatrix}$ according to a given probability distribution $\mathcal{P}$. 
    
    \item We then compute the rank-$1$ matrix $\outputmat = \xvec  \yvec^T \in \real^{\nrowssubmatrix \times \ncolssubmatrix}$, and we inject additive noise generated according to the probability distribution $\mathcal{P}$, rescaled by a factor $\epsilon_{noise}$ and re-centered to have $0$ mean. The larger the parameter $\epsilon_{noise}$, the higher the noise and the more \outputmat deviates from a rank-$1$ matrix. 

    \item We finally generate a larger matrix $\inputmat \in \real^{\nrows \times \ncols}$ by resampling uniformly at random the entries of the sub\-matrix $\outputmat$ and planting the near-rank-$1$ sub\-matrix in randomly-sampled row indices $\rowset'$ and column indices $\colset'$.
\end{itemize}

The above data-generating mechanism is designed to make the task of near-rank-$1$ sub\-matrix recovery as challenging as possible. If the distribution of the entries in the  near-rank-$1$ sub\-matrix  deviates significantly from the remaining entries (or background), then co-clustering algorithms may suffice to discover the sub\-matrix.
We consider different choices for the distribution $\mathcal{P}$, namely standard normal, uniform in $[0,1]$, exponential of rate $1$, Beta with parameters $2$ and $3$, Gamma with shape $2$ and scale $1$ and Poisson of rate $5$. 

We always consider synthetic matrices $\inputmat \in \real^{\nrows \times \ncols}$ with $\nrows = 250$ and $\ncols=250$, and the parameter $\noisescale$ is set to $1^{-5}$, unless stated otherwise. 

In the experiments carried out to evaluate the performance of \ourmethod against the baselines, we first plant near-rank-$1$ square sub\-matrices of increasing size $\nrowssubmatrix \times \nrowssubmatrix$ with $\nrowssubmatrix \in \{  
25,50,75,100, 125 \}$. 

In addition to the synthetic data with a single planted near-rank-$1$ sub\-matrix, we also consider the general setting that involves multiple planted near-rank-$1$ sub\-matrices.
More formally, we generate: 
\begin{equation*}
    \outputmat = \sum_{h = 1}^{N_{patterns}  }  \outputmat^*_h, 
\end{equation*}
where: 
\[
\outputmat^*_h =
\begin{cases}
 \outputmat_{h_{i,j}}, & \text{if } i \in \rowset_h \text{ and } j \in \colset_h \\
 0, & \text{otherwise},
\end{cases}
\]
and $\outputmat_h$ is generated as explained above and planted in row indices $\rowset_h$ and column indices $\colset_h$. 
In the experiments, we consider $N_{patterns} \in \{1,2,3,4,5\}$, and we sample $\nrowssubmatrix$ and $\ncolssubmatrix$ uniformly at random between $20$ and $75$. 
For the purposes of the experimental evaluation, it is assumed that each method is aware of the number of planted sub\-matrices. 
Thus, each method (iteratively) discovers $N_{patterns}$ 
sub\-matrices. In particular, \ourmethod follows the procedure sketched in Algorithm~\ref{alg:iterated}. 
A key point to highlight is that although all the generated matrices $\outputmat^*_h$ are close to rank $1$, they can overlap with each other, yielding intersection sub\-matrices that are close to a low-rank structure, but not necessarily close to rank $1$.  
The overlap between different planted sub\-matrices makes the task of discovering near-low-rank sub\-matrices more challenging than in the case of a single planted near-rank-$1$ sub\-matrix. 
In fact, in the absence of overlap between different planted sub\-matrices, our method can simply iteratively discover each sub\-matrix independently of the others. 

As concerns real-world datasets, we consider $15$ heterogeneous matrices coming from different domains. 
Summary characteristics and references are provided in Table~\ref{tab:real_world_datasets} in Section~\ref{sec:exp_setup}. 
We choose data matrices from different domains that are common in the low-rank-approximation literature. 
In particular, we consider benchmark  datasets for supervised-learning tasks (\Isolet, \Ozone, \Hearth), image datasets (\Olivetti, \OrlRnSp, \Mandrill and \Cameraman), 
benchmark datasets for recommender systems (\MovieLens, \Google and \MovieTrust), 
gene-expression data (\Golub and  \BRCA), neural-network weights (\Imagenet), and two miscellaneous datasets  (\Hyperspectral and \NPAS).

\spara{Metrics.}
Here, we offer more details on the performance metrics adopted to assess the performance of \ourmethod against the baselines, which are briefly discussed in Section~\ref{sec:exp_setup}. 

When the ground truth is available, i.e., in synthetic data, we can assess the performance of a method in recovering the ground truth by the (balanced) $F_1$ score, which is a widely used metric in binary classification and information-retrieval tasks. 
The $F_1$ score corresponds to the the harmonic mean of precision and recall. 
Formally, it is 
defined as
\(
F_1 = \frac{2 \cdot \text{TP}}{2 \cdot \text{TP} + \text{FP} + \text{FN}}, 
\)
where TP, FP, FN are the true positives, false positives and false negatives, respectively. 
A method assigns a nonzero value to all the entries with indices in the discovered sub\-matrices and zero to the remaining entries. 
The nonzero values, in particular, are given by the low-rank approximation of the discovered sub\-matrices. 
A true positive occurs when a method assigns a nonzero value to an entry in a planted near-rank-$1$ sub\-matrix.
Similarly, a false positive occurs when the method assigns a nonzero value to an entry that is not part of a planted near-rank-$1$ sub\-matrix. Finally, false negatives correspond to those entries for which a method predicts a zero value in one of the entries that are part of a planted near-rank-$1$ sub\-matrix.

Given the ground truth, in addition to the $F_1$ score, it is of primary importance to assess the estimate of a planted sub\-matrix that is given by a method. 
For this purpose, we compute the reconstruction error, which is given by: 
\[
\frobeniusnormshort{\outputerror}^2 = 
\frobeniusnormshort{\outputmat - \estimate{\outputmat} }^2, 
\]
where  \outputmat is a ground-truth near-rank-$1$ sub\-matrix and $\estimate{\outputmat}$ its estimate. 
The choice of the squared Frobenius norm is standard, but any other norm could be considered instead. 
To ensure that the error is comparable across sub\-matrices of varying sizes, we report the reconstruction error averaged over all entries of \outputmat. 

In real-world datasets no ground truth is available. 
In this case, there are two fundamental quantities that should be monitored; a measure of closeness to a low-rank matrix and a measure of size. 
The measure of closeness to a low-rank matrix is the crucial quantity. However, the size of the sub\-matrices is also important. 
The two quantities are typically inversely related;  it is always possible to obtain a perfectly rank-$1$ sub\-matrix by reducing its size, and vice versa, it is always possible to obtain a sub\-matrix as large as the input matrix by making it deviate more and more from a rank-$1$ sub\-matrix. 
Thus, a method is preferred over another whenever it strikes a more desirable balance between closeness to rank $1$ and size.

As regards closeness to a rank-$1$ matrix,  we monitor the low-rankness score, introduced in Section~\ref{sec:preliminaries}. 
For a sub\-matrix $\outputmat \in \real^{\nrowssubmatrix \times  \ncolssubmatrix}$ with $i$-th singular value $\sigma_i^2$,   
the low-rankness score is defined as 
\( \lowrankness{\outputmat} = 
    \frac{\sigma_1(\outputmat)^2}
         {\sum_{i=1}^{\min(\nrowssubmatrix,\ncolssubmatrix)} \sigma_i(\outputmat)^2} \). 
The low-rankness score has been adopted in previous work to monitor the performance in experiments with real-world datasets~\cite{dang2023generalized}. It mathematically equals the reciprocal of the popular stable-rank measure. 
The low-rankness scores ranges between $0$ and $1$. The closer the low-rankness score is to $1$, the closer the sub\-matrix is to the ideal scenario of rank exactly $1$. 
It is important to consider the low-rankness of a sub\-matrix in relation to the input matrix it is extracted from. 
A sub\-matrix of size $\nrowssubmatrix \times \ncolssubmatrix$ with a low-rankness score of $0.95$ is extremely desirable if the low-rankness score of the input matrix is $0.3$. However, the same sub\-matrix is not as desirable when the  
low-rankness score of the input matrix is $0.9$. 
Therefore, to facilitate the understanding of the results, in addition to the raw low-rankness scores of the top-$5$ output sub\-matrices, we also report the average relative percentage increase in low-rankness with respect to the input matrix. 
The relative percentage increase in low-rankness is given by: 
\[
\frac{ \lowrankness{\outputmat}  - \lowrankness{\inputmat}    }{\lowrankness{\inputmat}}\cdot 100, 
\]
where $\lowrankness{\outputmat}$ and  $\lowrankness{\inputmat}$ indicate the low-rankness score of an output sub\-matrix $\outputmat$ and of the input matrix \inputmat, respectively. 
An important point to highlight is that, while the rank of a sub\-matrix can never exceed that of the larger matrix it is contained in, the low-rankness score of a sub\-matrix can be smaller than that of the matrix it is contained in.
Therefore, the relative percentage increase in low-rankness can be negative.


Finally, also for the size of an output sub\-matrix, 
as for the low-rankness score, it is important to consider the size of the sub\-matrix in relation to the size of the input matrix. 
Therefore, in addition to the raw number $|\outputmat|$ of entries in the top-$5$ output sub\-matrices, we report the percentage $\frac{|\outputmat|}{|\inputmat|} \cdot 100$ of entries in the input matrix that are part of the output sub\-matrix,
averaged over the top-$5$ output sub\-matrices. 

\spara{Parameters settings.}
In this paragraph, we provide more details concerning the parameter settings for the methods considered in the experiments. 

In Section~\ref{sec:exp_setup}, we discuss the values of the parameters $\tol$ and $\toleranceinit$ used in the experiments. 
In addition to $\tol$ and $\toleranceinit$, the other  important parameter to be given in input to \ourmethod is the target rank $\generalrank$. 
If $\generalrank=1$, \ourmethod utilizes Algorithm~\ref{alg:rankone} to discover near-rank-$1$ sub\-matrices, otherwise it uses Algorithm~\ref{alg:rankk} to discover near-rank-$\generalrank$ sub\-matrices. 
In the experiments with synthetic data, 
when the number of planted patterns is $1$, we consider $\generalrank =1$, when it is between $2$ and $4$, we consider  $\generalrank=1$ and $\generalrank=2$, and finally when there are $5$ patterns, we additionally consider  $\generalrank=3$.  
In all cases, however, the total number of initialization is $\nrep = 25$.  
In the case of real-world data, we always consider~$\generalrank=1$. 

Finally, the parameter $\lambda$ in the objective function $f$ controlling the relative weight of reconstruction error and size is held constant to its default value of $1$. 

As concerns the baselines, for sparse PCA, we set the sparsity-controlling parameter $\alpha$ to $1$.
In the convex-optimization-based approach, we set the parameter $\theta$ controlling the trade-off between nuclear and $L_1$ norm to $10$.
\svp is a parameter-free method. 
Finally, \rpsp samples $10^5$ $2 \times 2$ and $4 \times 4$ sub\-matrices and $10^4$ $8 \times 8$ and $16 \times 16$ sub\-matrices and  the number of score matrices that are computed (and summed up) before finding the final solution is set to $30$.

\spara{Implementation details.}
We conclude this section by providing details regarding the implementation of \ourmethod and of the baseline approaches. 

An important choice for the implementation of \ourmethod is the approach leveraged to extract 
a maximum-edge bi\-clique in the bipartite graph $\mathcal{G}_{\ind}$. 
In all the experiments with synthetic data, we 
leverage the algorithm of Lyu et al.~\cite{lyu2020maximum} for mining maximum-edge bicliques and we rely on an implementation in \textsc{C} of the algorithm available online.\footnote{\url{https://github.com/wonghang/pymbc}}


In the experiments with real-data, to eschew possible scalability issues, we heuristically extract dense submatrices from the indicator matrix \ind via spectral biclustering~\cite{kluger2003spectral}, which, as discussed in Section~\ref{sec:algorithms}, is the heuristic of choice to approximate maximum-edge bicliques. This algorithm is not guaranteed to return   
sub\-matrices of \ind consisting of all ones (i.e., bi\-cliques). 
Subsequently, the guarantees reported in Theorem~\ref{th:expansion} may not hold when spectral bi\-clustering is used. 
Despite the lack of theoretical guarantees, the implementation of \ourmethod using spectral biclustering is effective in practice, as demonstrated by our experiments.

To approximate the discovered sub\-matrices, we use the interpretable approximation in terms of the original rows or columns of the data if the target rank is~$1$ and the \svd otherwise.

For spectral biclustering, we use the implementation available in the \textsc{scikit-learn} \textsc{Python} library.\footnote{\url{https://scikit-learn.org/stable/modules/generated/sklearn.cluster.SpectralBiclustering.html}}

To handle sparse datasets, we also implement particular constraints that avoid returning trivial rank-$1$ sub\-matrices consisting of all zeros.


As for the baselines, we implement \svp in \textsc{Python}. 
For \rpsp, we instead rely on the implementation provided by the authors.\footnote{\url{https://github.com/ptdang1001/RPSP}}
For \sparsePCA and for the convex-optimization approach (\cvx), we use the \textsc{scikit-learn}\footnote{\url{https://scikit-learn.org/stable/modules/generated/sklearn.decomposition.SparsePCA.html}} and \textsc{cvxpy}\footnote{\url{https://www.cvxpy.org}} \textsc{Python} libraries, respectively.


\appendixsection{Additional Experiments}{ap:additional_experiments}
\hypertarget{ap:additional_experiments}
In this section, we present additional experimental results. 
First, we present preliminary experiments, and then we present additional results of experiments evaluating the performance of the method we introduce in both synthetic and real data. 

\spara{Preliminary experiments.}
The goal of the preliminary experiments is to 
provide valuable insights into the synthetic data used in our experiments, and on the behavior of \ourmethod in such data. 

\begin{figure}[t]
\begin{center}
    \begin{tabular}{ccc}
        & \textsc{Normal} & \\
        \includegraphics[width=0.3\textwidth]{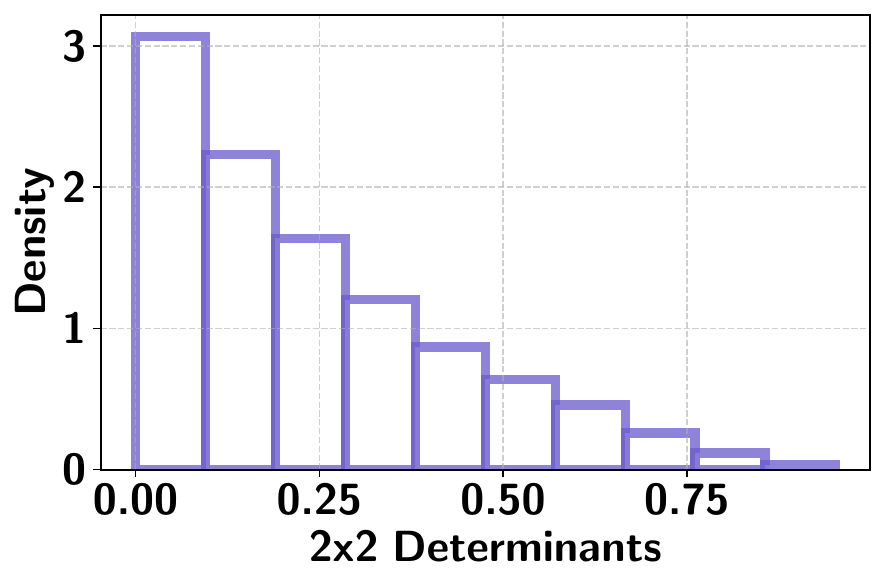} & 
        \includegraphics[width=0.3\textwidth]{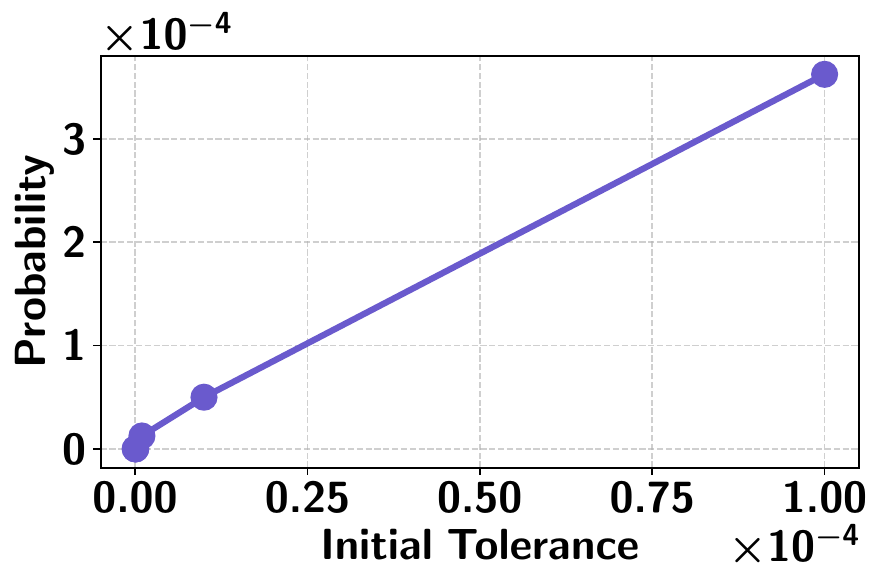} & 
        \includegraphics[width=0.3\textwidth]{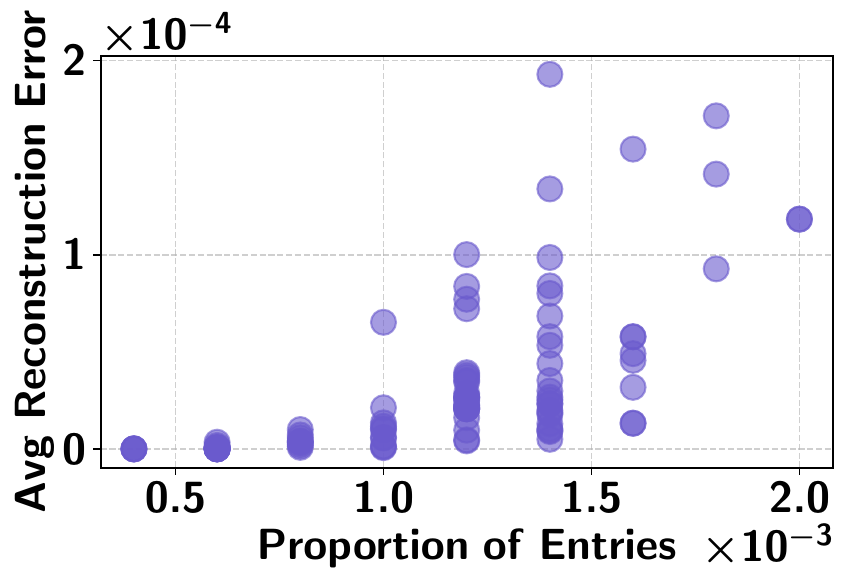} \\
        \vspace*{-1.1cm}
    \end{tabular}
\end{center}
\caption{
\label{fig:prelimary_full_rank}
Full-rank $250 \times 250$ synthetic matrix with i.i.d entries from a standard normal distribution and without planted near-rank-$1$ sub\-matrices.
We show the distribution of $2 \times 2$-sub\-matrix determinants (left), the probability of observing a $2 \times 2$-sub\-matrix with determinant in absolute value lower than the initial tolerance threshold $\toleranceinit$  as a function of $\toleranceinit$ (center) and the (per-entry) average reconstruction error as a function of the proportion of entries in the sub\-matrix for individual iterations of our method (right). 
}
\end{figure}

We first consider a full-rank $250 \times 250$ matrix with all i.i.d. entries generated from a standard normal distribution and no planted near-rank-$1$ sub\-matrix. 
For such matrix, in Figure~\ref{fig:prelimary_full_rank},  we show the distribution of the determinants of the initially-sampled $2 \times 2$ matrices as well as the probability of hitting a near-rank-$1$ $2 \times 2$ sub\-matrix with near-zero determinant as a function of the parameter $\toleranceinit$ (with $ \toleranceinit \in \{ 10^{-8}, 10^{-7}, 10^{-6}, 10^{-5}, 10^{-4} \}$). 
Finally, we show the average reconstruction error $\frobeniusnorm{\outputerror}^2$ of the sub\-matrices retrieved by \ourmethod as a function of the proportion of entries of the input matrix they include. 
As Figure~\ref{fig:prelimary_full_rank} suggests, although the input matrix is generated completely at random and with no planted near-rank-$1$ sub\-matrix, it still contain many $2 \times 2$ sub\-matrices with near-zero determinant, and the probability of \ourmethod hitting a near-rank-$1$ $2 \times 2$ sub\-matrix in the initialization phase quickly grows as the initial tolerance parameter $\toleranceinit$ is increased, 
confirming the intuition gained through the simple probabilistic analysis presented in Section~\ref{sec:analysis}. 
However, no large matrix with small reconstruction error is found at the expansion phase, since, unlike in real-world matrices, it is unlikely that large low-rank sub\-matrices occur in matrices consisting solely of i.i.d. entries. 

\begin{figure}[t]
\begin{center}
    \begin{tabular}{ccc}
        & \textsc{Normal} & \\
        \includegraphics[width=0.3\textwidth]{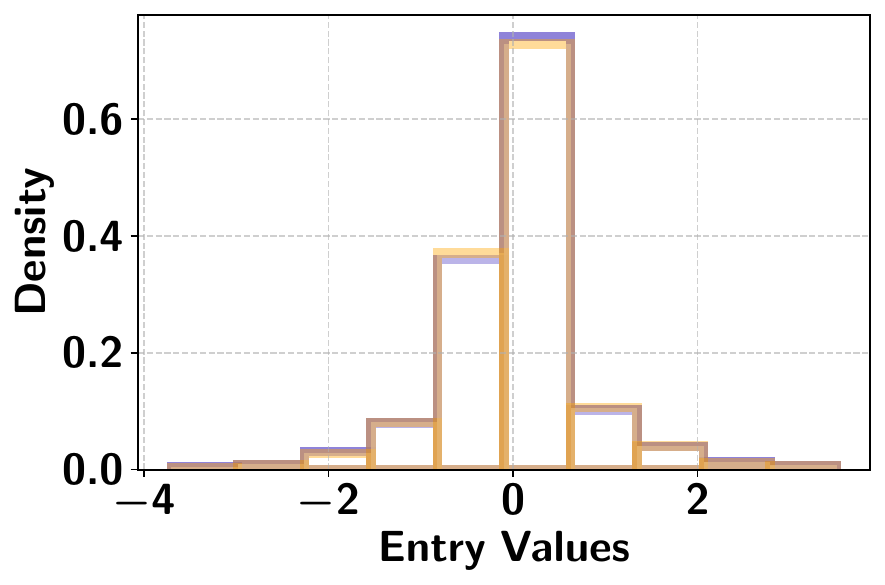} & 
        \includegraphics[width=0.3\textwidth]{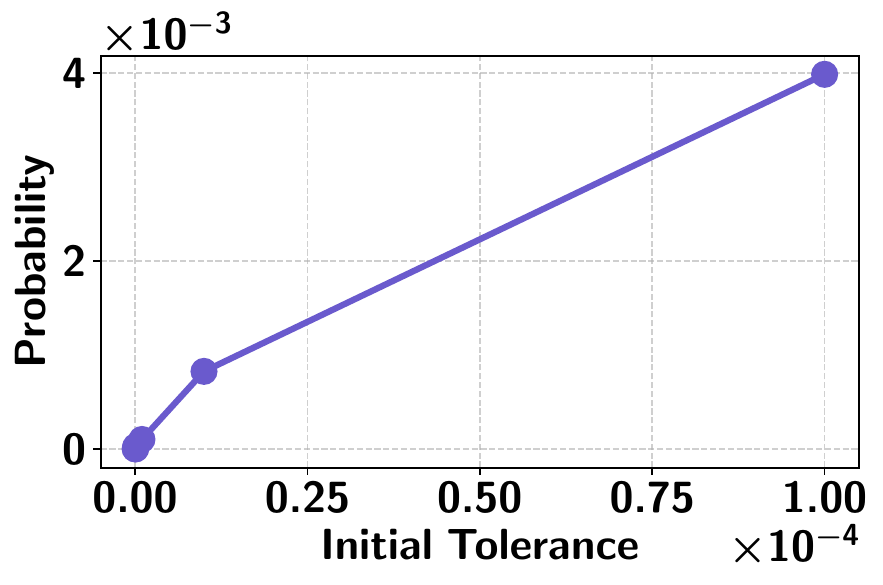} & 
        \includegraphics[width=0.3\textwidth]{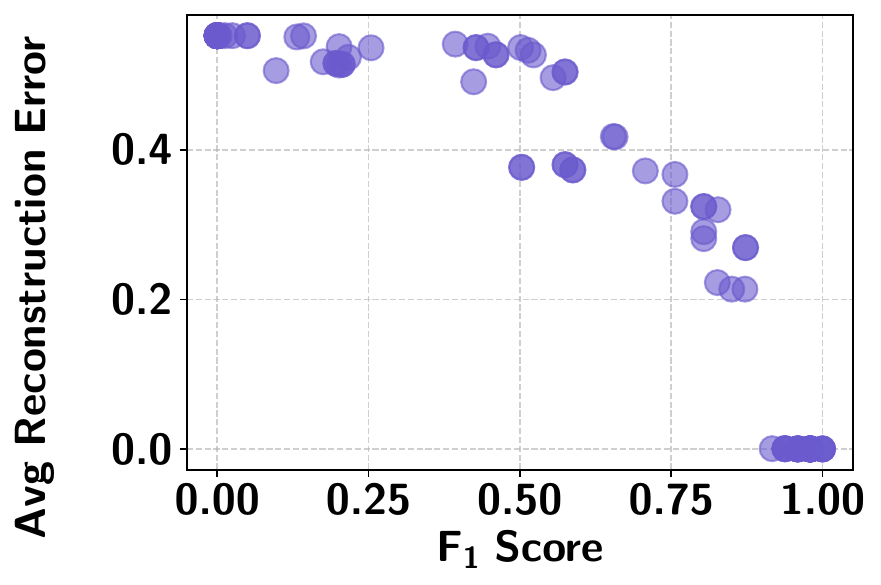} \\
        \vspace*{-1.1cm}
    \end{tabular}
\end{center}
\caption{
\label{fig:prelimary_planted}
Full-rank synthetic $250 \times 250$ matrix generated from a standard normal distribution and with a $25 \times 25$ planted near-rank-$1$ submatrix.
We show the distribution of the values in the entire matrix and sub\-matrix (left), the probability of observing a $2 \times 2$-sub\-matrix with determinant in absolute value lower than the initial tolerance threshold $\toleranceinit$ as a function $\toleranceinit$ (center) and the (per-entry) average reconstruction error as a function of the $F_1$ score for individual iterations of our method (right). 
}
\end{figure}

Figure~\ref{fig:prelimary_planted} presents similar results as Figure~\ref{fig:prelimary_full_rank}, but for a synthetic $250 \times 250$ matrix where a near-rank-$1$ sub\-matrix is generated from a standard normal distribution and planted according to the procedure illustrated in~\ref{ap:additional_experimental_setup}. 
In particular, for such matrix, the figure shows the distribution of the values in the planted near-rank-$1$ sub\-matrices and in the background, the probability that \ourmethod samples a near-rank-$1$ $2 \times 2$ sub\-matrix in the initialization phase as a function of the parameter $\toleranceinit$ (with $ \toleranceinit \in \{ 10^{-8}, 10^{-7}, 10^{-6}, 10^{-5}, 10^{-4} \}$) and the average reconstruction error as a function of the $F_1$ score for the sub\-matrices output by our method in different iterations. 
The figure demonstrates that the planted near-rank-$1$ sub\-matrices are drawn from the same distribution as the background, reflecting a scenario where standard co-clustering algorithms are generally not able to identify low-rank sub\-matrices. 
Furthermore, the figure also demonstrates that the probability of hitting a near-rank-$1$ $2 \times 2$ sub\-matrix is generally larger than in the case of Figure~\ref{fig:prelimary_full_rank} since the planted near-rank-$1$ sub\-matrices contain a large number of $2 \times 2$ rank-$1$ sub\-matrices. 
Finally, the figure confirms that \ourmethod expands many initial sub\-matrices into larger matrices with $F_1$ score close to $1$ and reconstruction error close to $0$, suggesting that the ground-truth sub\-matrix is recovered in many iterations. 

\begin{figure}[t!]
    \begin{tabular}{cccc}
        \textsc{\Hyperspectral} & 
        \textsc{\OrlRnSp} & \textsc{\Mandrill} & \textsc{Normal} \\
        \includegraphics[width=0.24\textwidth]{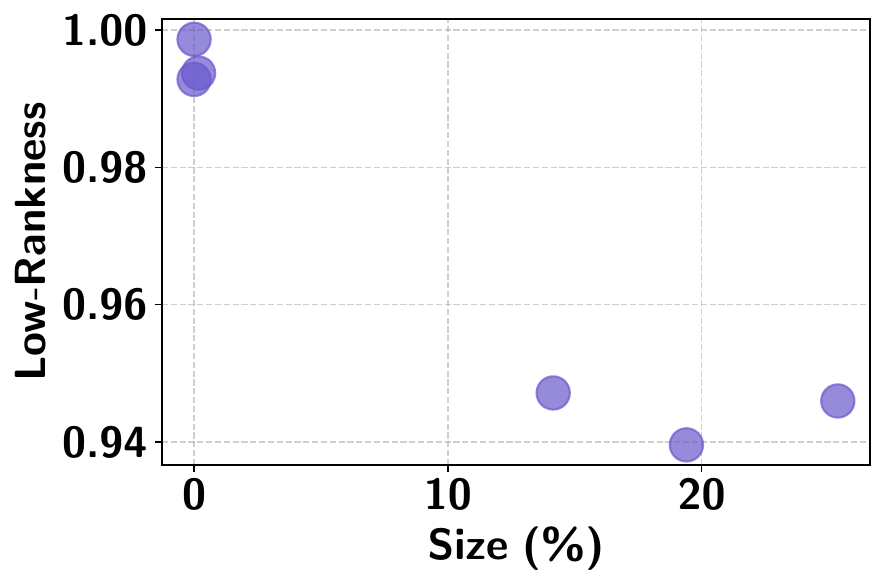} & 
        
        \includegraphics[width=0.24\textwidth]{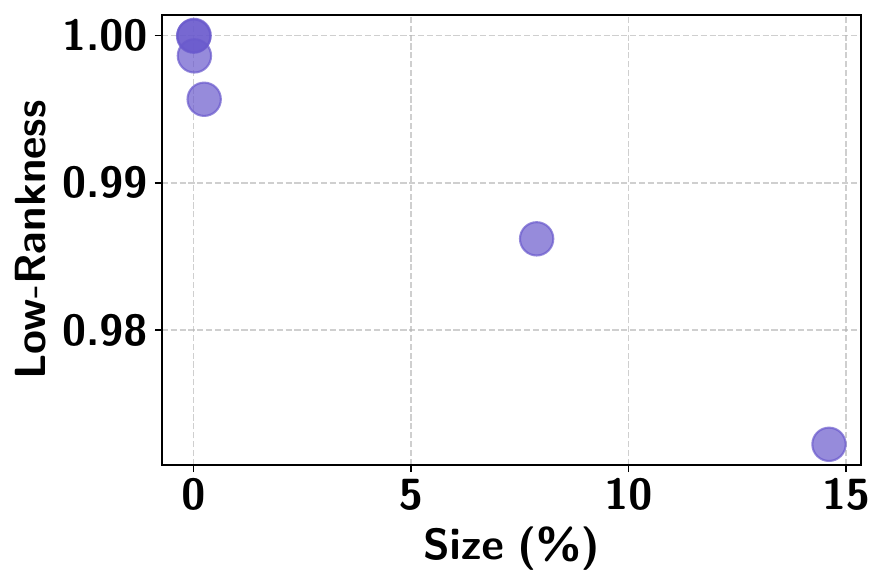} & 
        \includegraphics[width=0.24\textwidth]{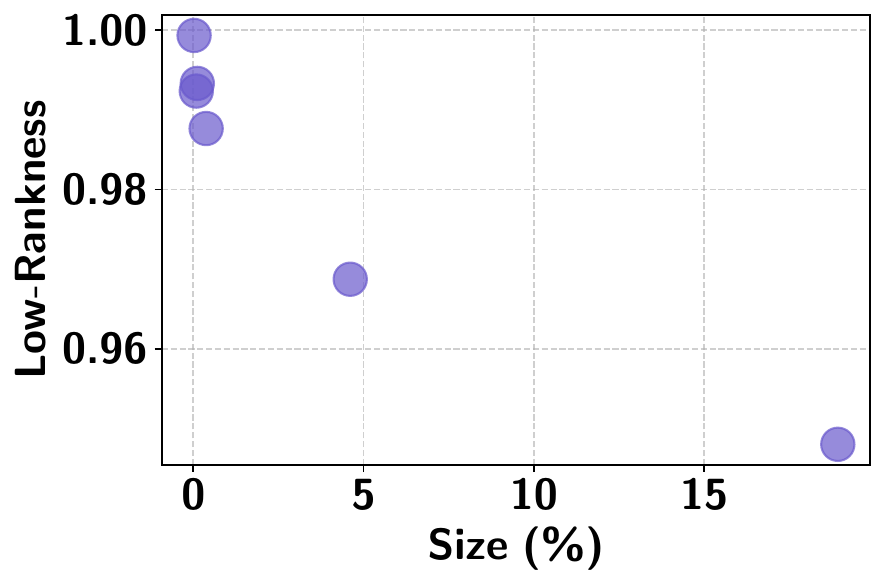} & 
        \includegraphics[width=0.24\textwidth]{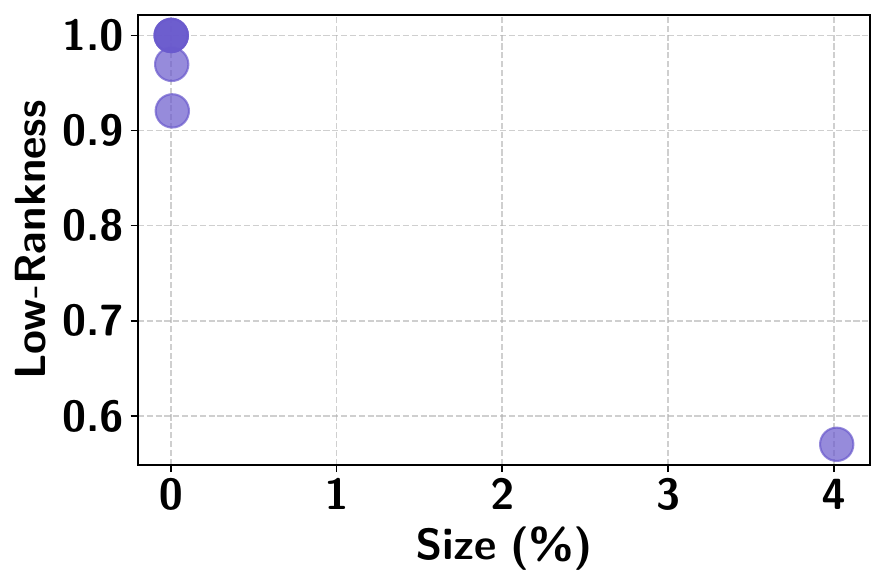} \\
        \vspace*{-0.7cm}
    \end{tabular}
\caption{\label{fig:impact_of_delta}Experiments on four example datasets (\Hyperspectral, \OrlRnSp, \Google and a $250 \times 250$ synthetic dataset with i.i.d normally-distributed entries)
to investigate the impact of $\tol$ on the output of \ourmethod.
Each point indicates the low-rankness score and the percentage of entries of the discovered sub\-matrices  
for a given value of $\tol$, with $\tol \in \{10^{-6}, 10^{-5}, 10^{-4}, 10^{-3}, 10^{-2}, 10^{-1} \}.$}
\end{figure}

\spara{Experiments on synthetic and real-world data: impact of tolerance parameter.}
The results provided by \ourmethod vary with $\tol$. 
Intuitively, this parameter indirectly bounds the approximation error and thereby it embodies the trade-off between sub\-matrix size and low-rankness. 
In more detail, a larger value of $\tol$ leads to denser indicator matrices $\ind$ and thus larger output sub\-matrices $\outputmat$. 
On the other hand, a smaller value of $\tol$ leads to generally smaller output sub\-matrices $\outputmat$ which, however, may have higher low-rankness. 
In practice, \ourmethod returns high-quality results across reasonable choices of $\tol$. 
In Figure~\ref{fig:impact_of_delta} we show, for four examples example datasets (\Hyperspectral, \OrlRnSp, \Mandrill and a synthetic $250 \times 250$ matrix with  i.i.d.  entries drawn from a standard normal distribution and no planted near-rank-$1$ sub\-matrices), the low-rankness score against the size of the sub\-matrices output by \ourmethod for different choices of the parameter $\tol$. 
Specifically, we vary $\tol$ in $\{ 10^{-5}, 10^{-4}, 10^{-3}, 10^{-2}, 10^{-1} \}$. 
As for all the experiments in real-world datasets, the results in Figure~\ref{fig:impact_of_delta} are obtained by relying on spectral bi\-clustering for approximating maximum-edge bi\-cliques. As a consequence, it is not guaranteed that larger values of $\tol$ lead to larger size and lower low-rankness.   
Nevertheless, the results confirm that, also in the case where maximum-edge bi\-cliques are approximated,  larger values of $\tol$ generally result in sub\-matrices of larger size and lower low-rankness. 
Moreover, \ourmethod outputs high-quality sub\-matrices for any appropriate choice of $\tol$.

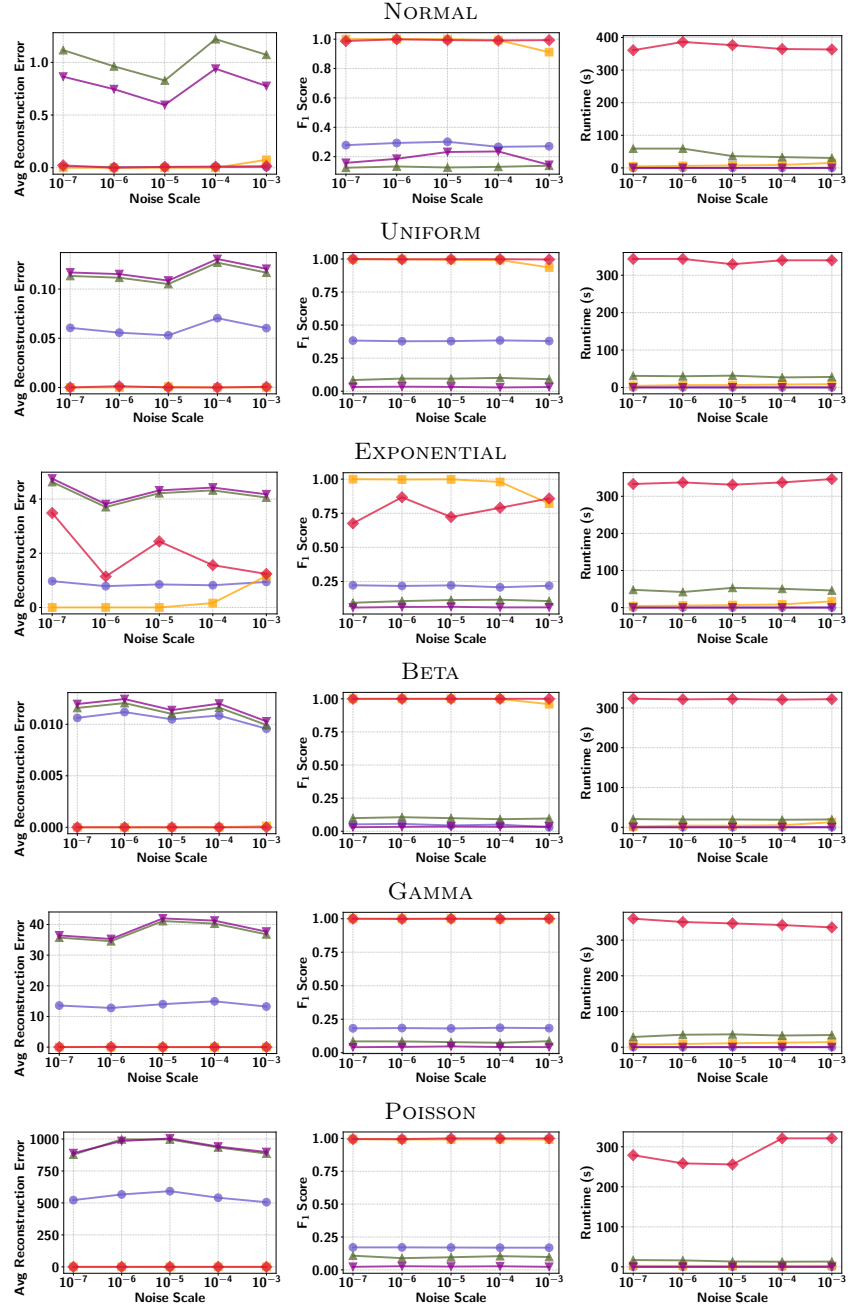
\begin{figure}[p]
\begin{center}
\begin{tikzpicture}[scale=0.1]
\matrix [matrix of nodes, 
column sep=1.0pt, 
row sep=0pt, 
nodes={anchor=center}] (m) {
\tikz \draw[slateblue, thick] (-0.18,0) -- (0.18,0) 
node[fill=slateblue, circle, inner sep=2.1pt] {};
& {\small Sparse PCA}
& 
\tikz \draw[darkolivegreen, thick] (-0.18,0) -- (0.18,0) 
node[fill=darkolivegreen, regular polygon, regular polygon sides=3, inner sep=1.4pt] {};
& {\small \cvx}
& 
\tikz \draw[darkmagenta, thick] (-0.18,0) -- (0.18,0) 
node[fill=darkmagenta, regular polygon, regular polygon sides=3, inner sep=1.7pt, rotate=180] {};
& {\small \svp}
& 
\tikz \draw[crimson, thick] (-0.18,0) -- (0.18,0) 
node[fill=crimson, diamond, inner sep=1.7pt] {};
& {\small \rpsp} 
& 
\tikz \draw[orange, thick] (-0.18,0) -- (0.18,0) 
node[fill=orange, rectangle, inner sep=2.6pt] {};
& {\small \ourmethod} \\
};
\end{tikzpicture}
    \begin{tabular}{ccc}
        & \textsc{Normal} & \\
        \includegraphics[width=0.3\textwidth]{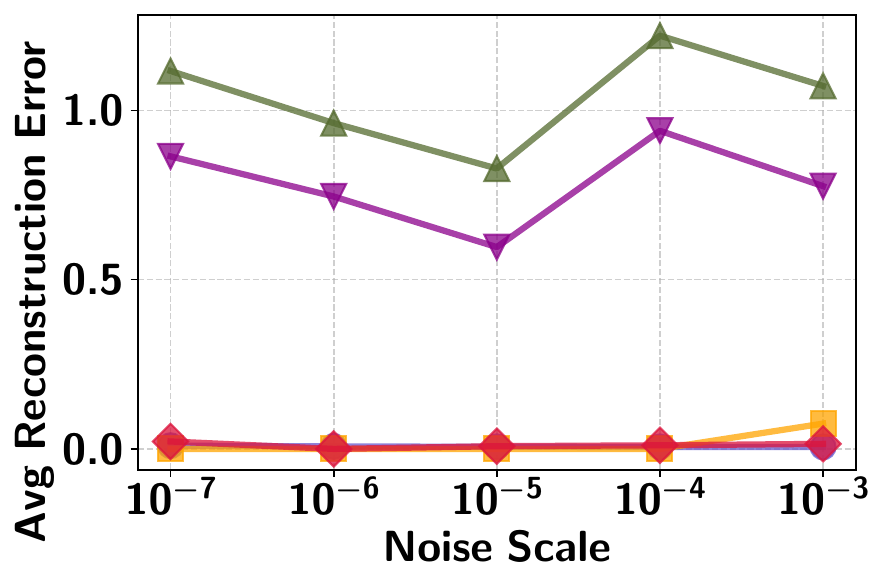} & 
        \includegraphics[width=0.3\textwidth]{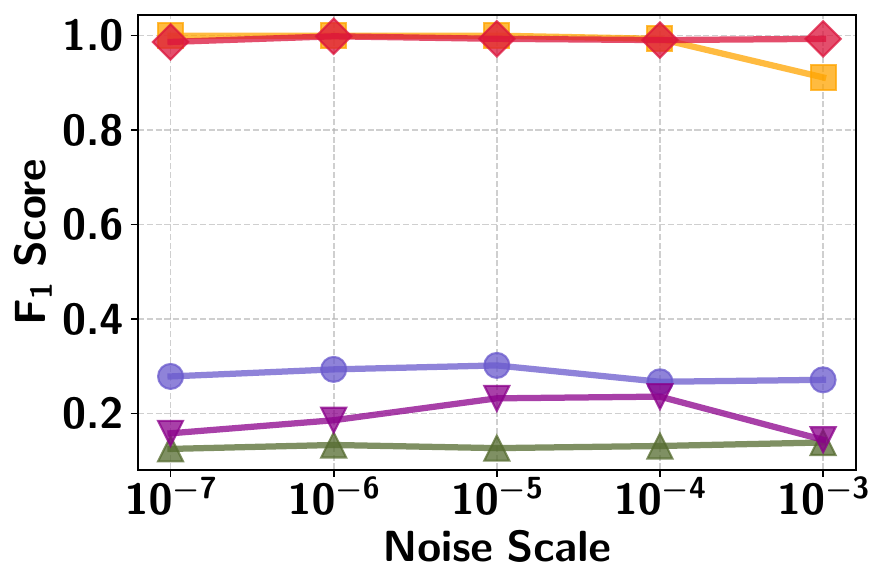} & 
        \includegraphics[width=0.3\textwidth]{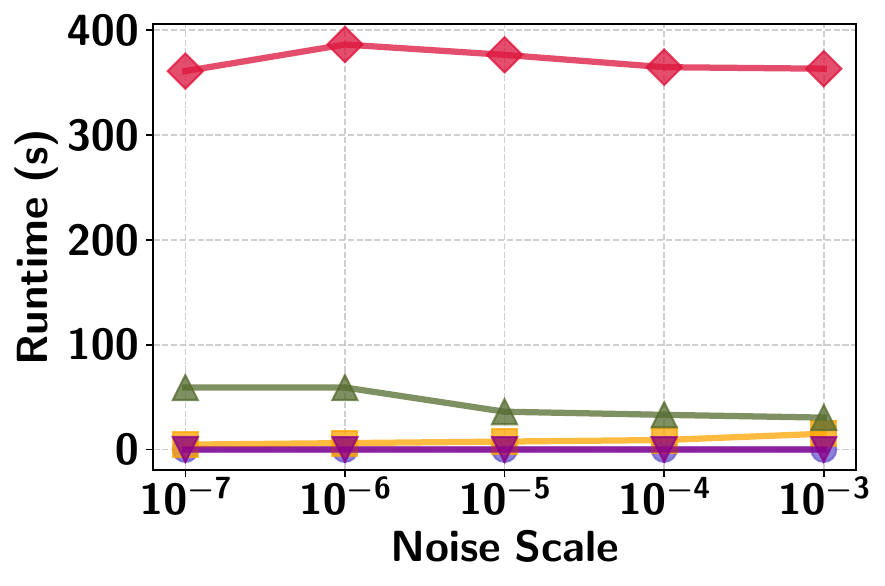} \\
        & \textsc{Uniform} & \\
         \includegraphics[width=0.3\textwidth]{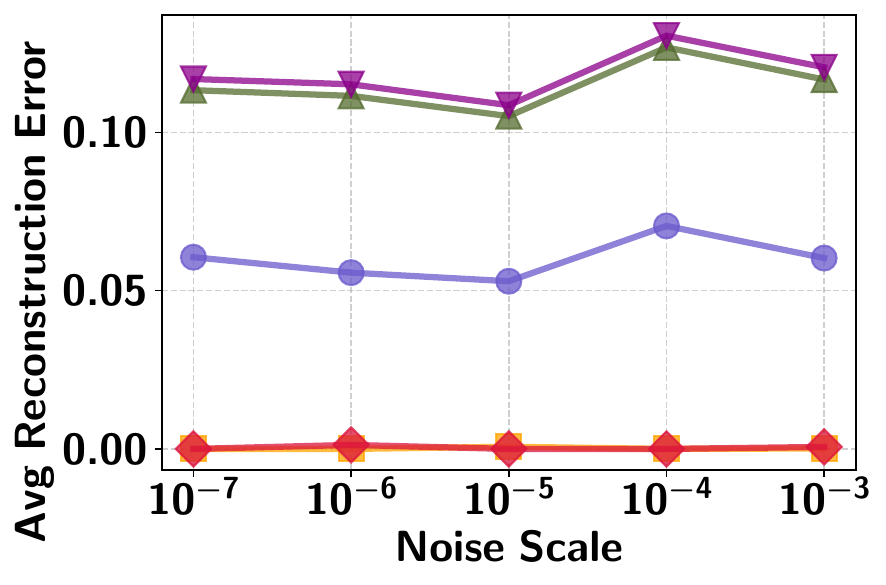} & 
        \includegraphics[width=0.3\textwidth]{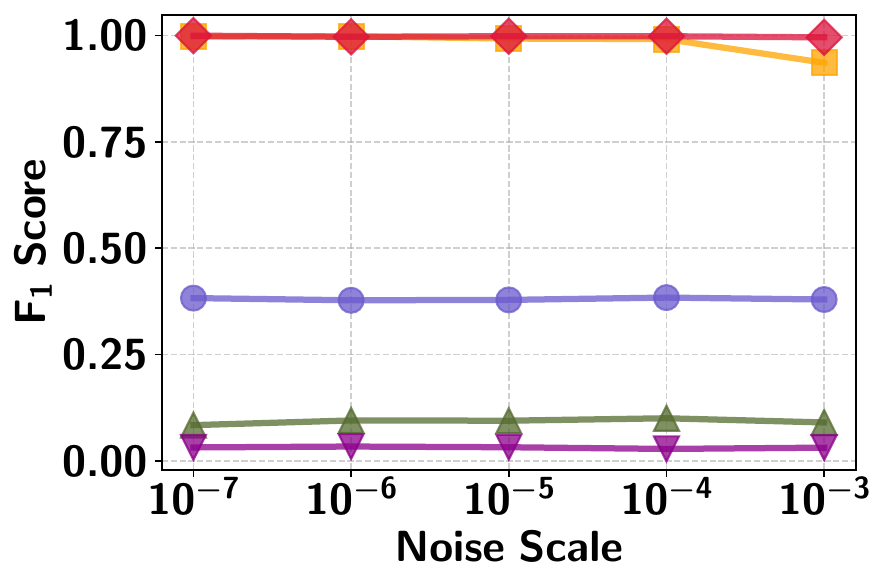} & 
        \includegraphics[width=0.3\textwidth]{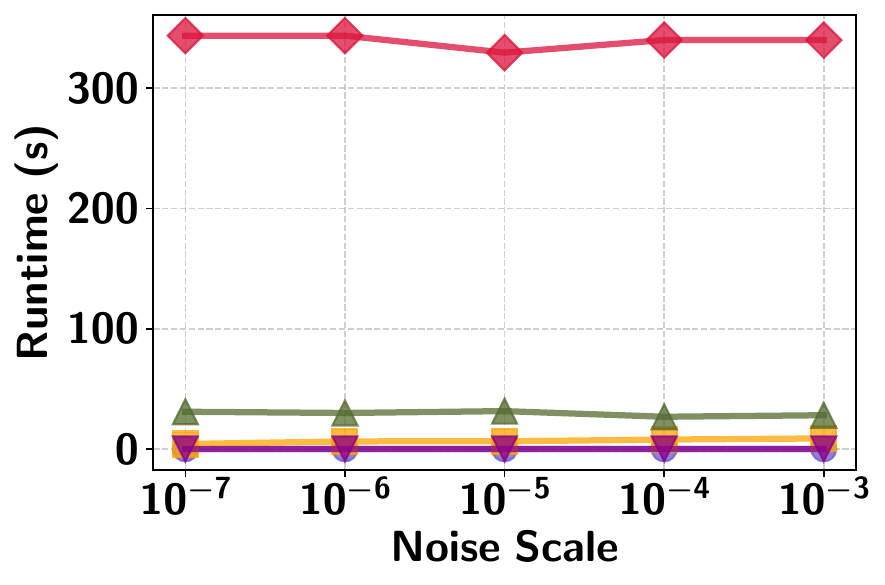} \\
        & \textsc{Exponential} & \\
        \includegraphics[width=0.3\textwidth]{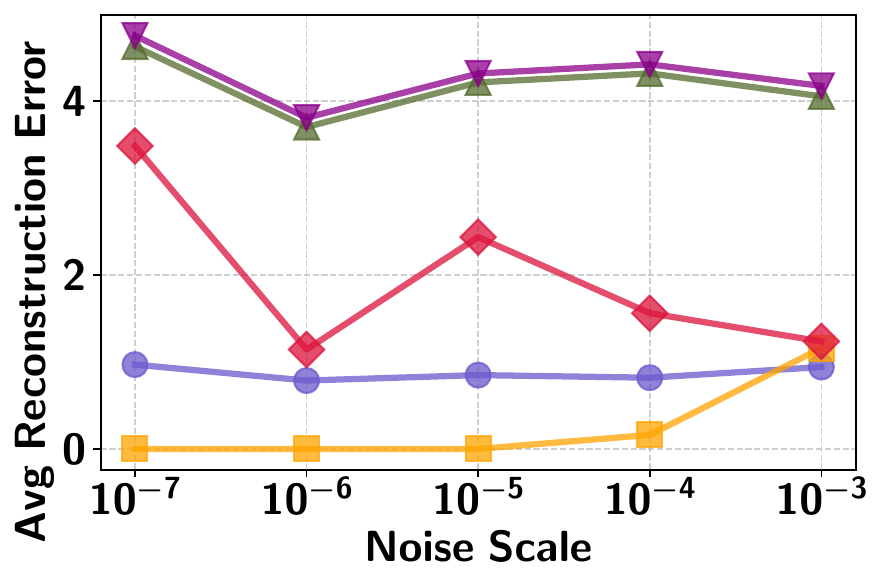} & 
        \includegraphics[width=0.3\textwidth]{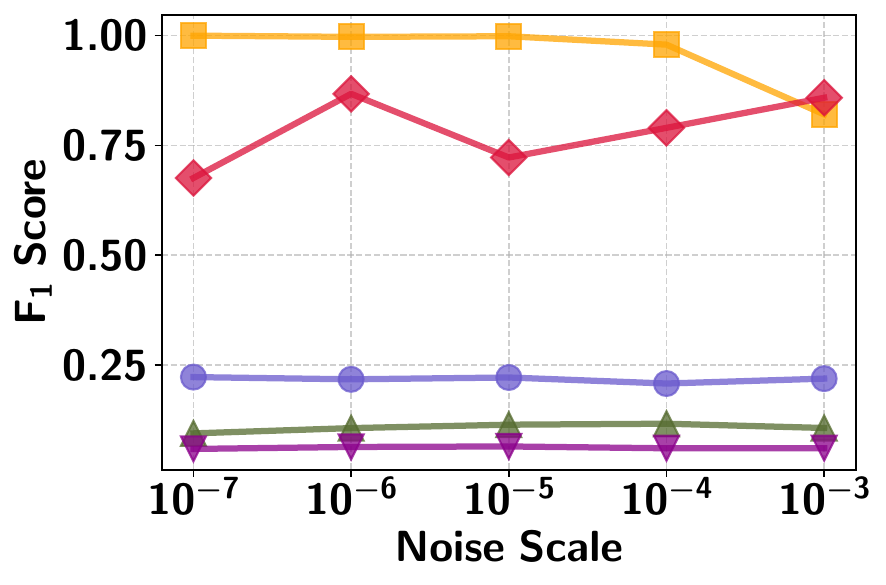} & 
        \includegraphics[width=0.3\textwidth]{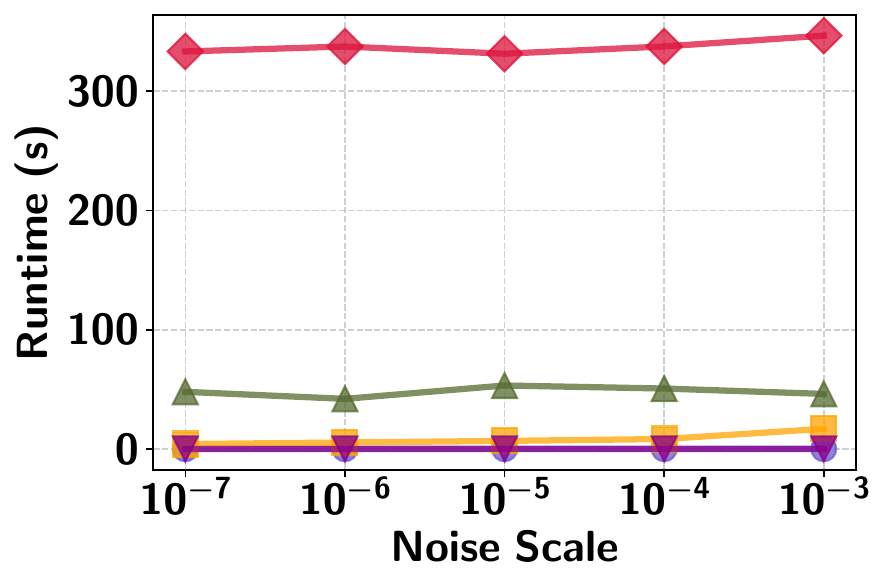} \\
        & \textsc{Beta} & \\
        \includegraphics[width=0.3\textwidth]{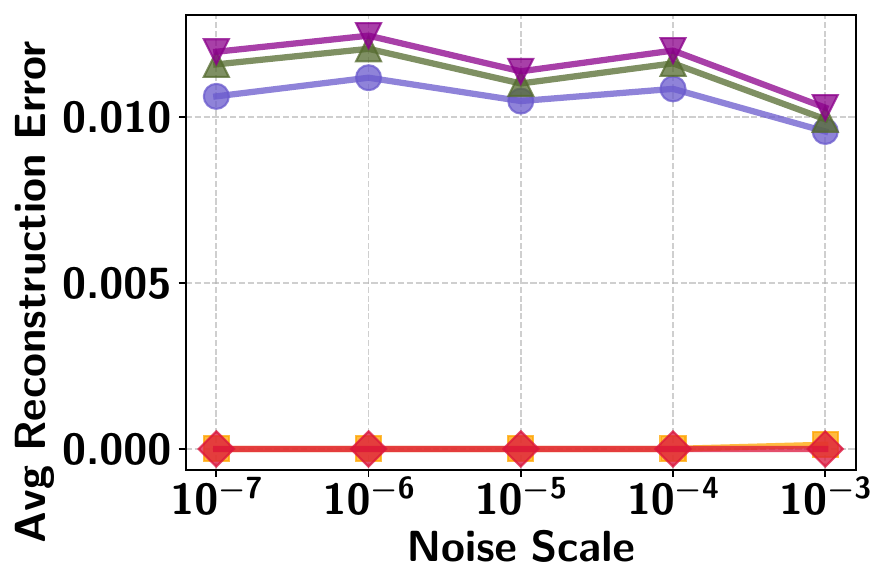} & 
        \includegraphics[width=0.3\textwidth]{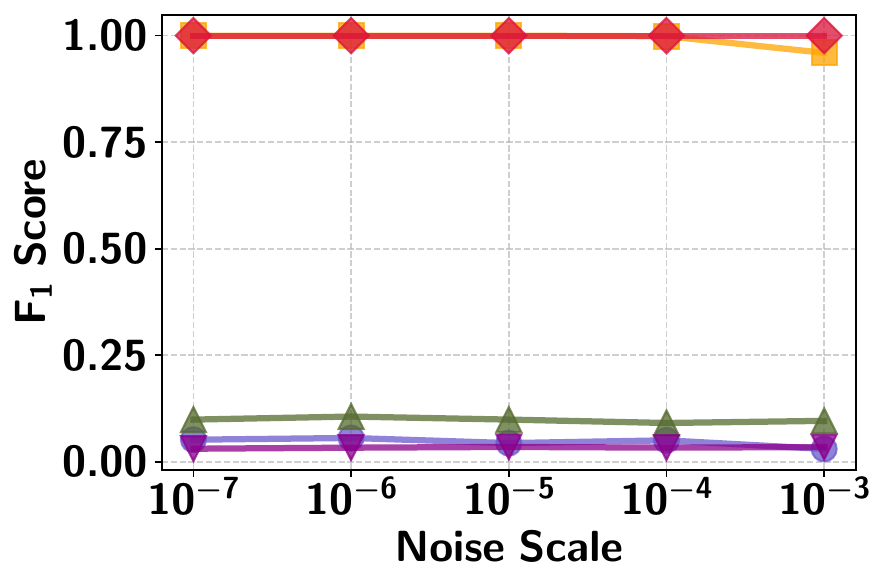} & 
        \includegraphics[width=0.3\textwidth]{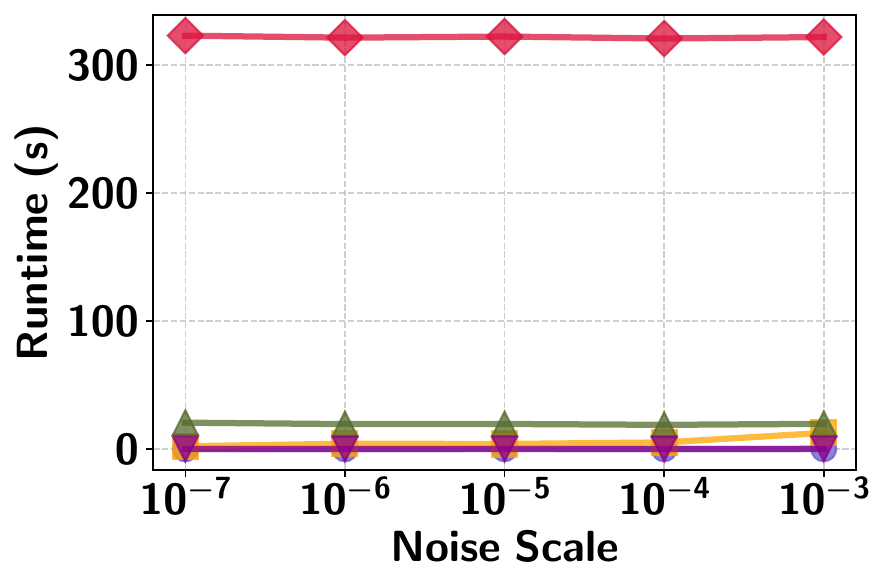} \\
        & \textsc{Gamma} & \\
        \includegraphics[width=0.3\textwidth]{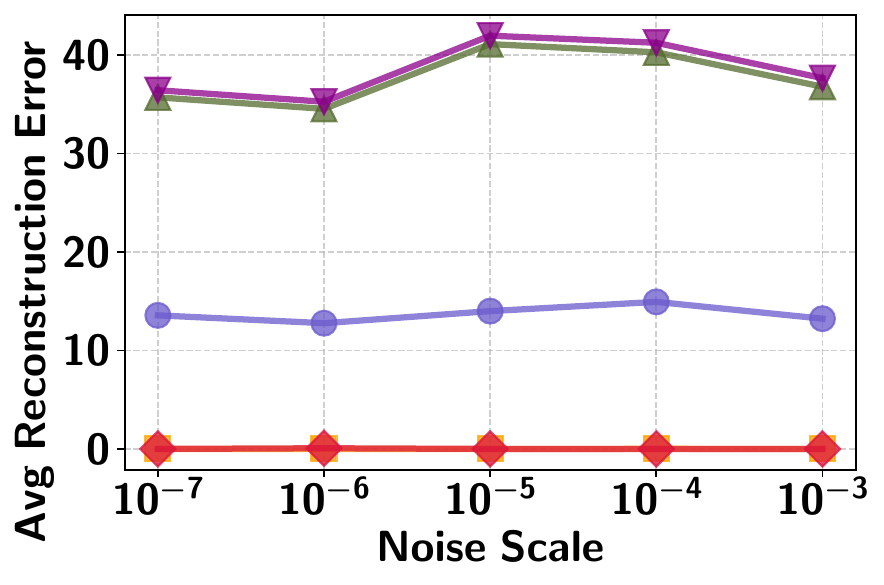} & 
        \includegraphics[width=0.3\textwidth]{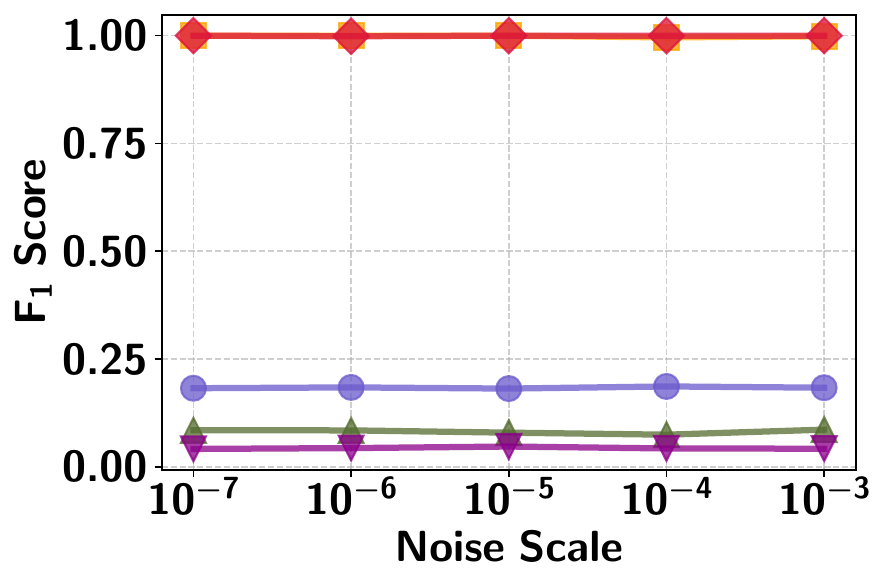} & 
        \includegraphics[width=0.3\textwidth]{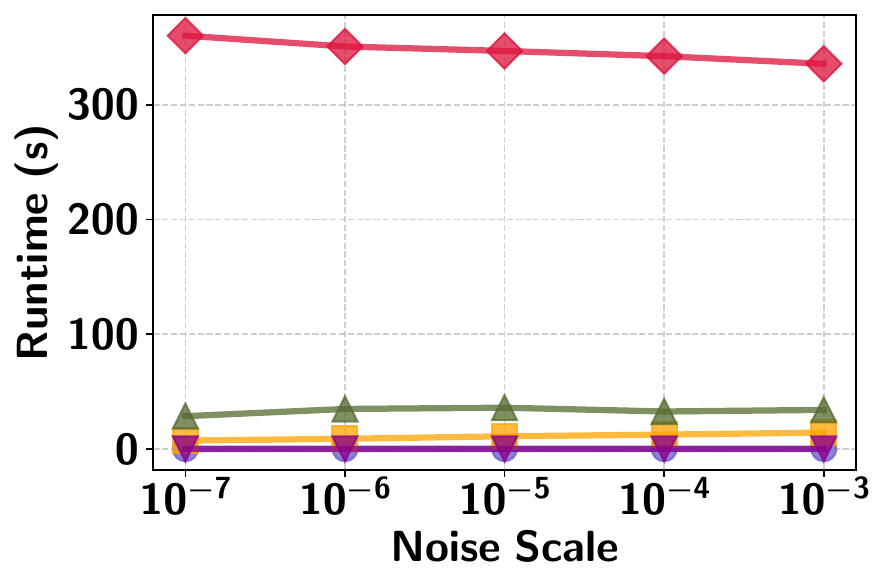} \\
        & \textsc{Poisson} & \\
         \includegraphics[width=0.3\textwidth]{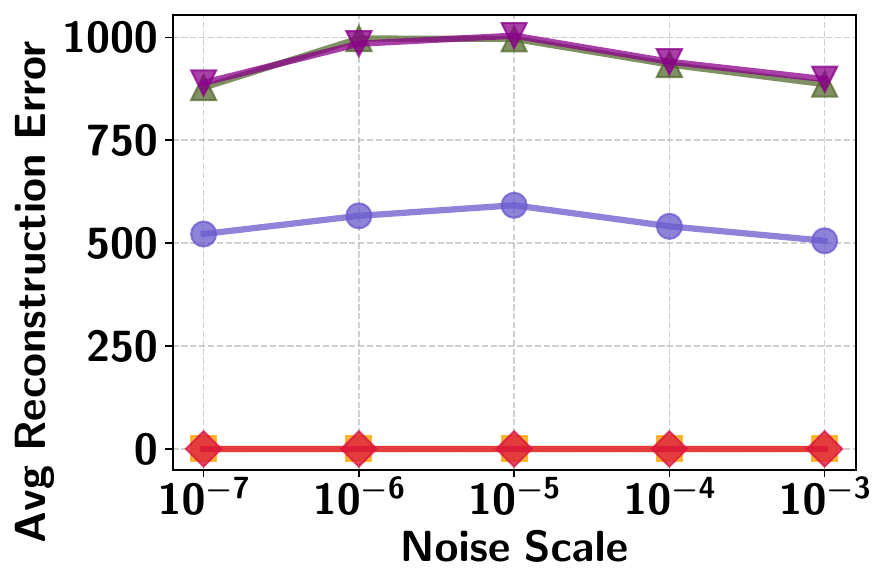} & 
        \includegraphics[width=0.3\textwidth]{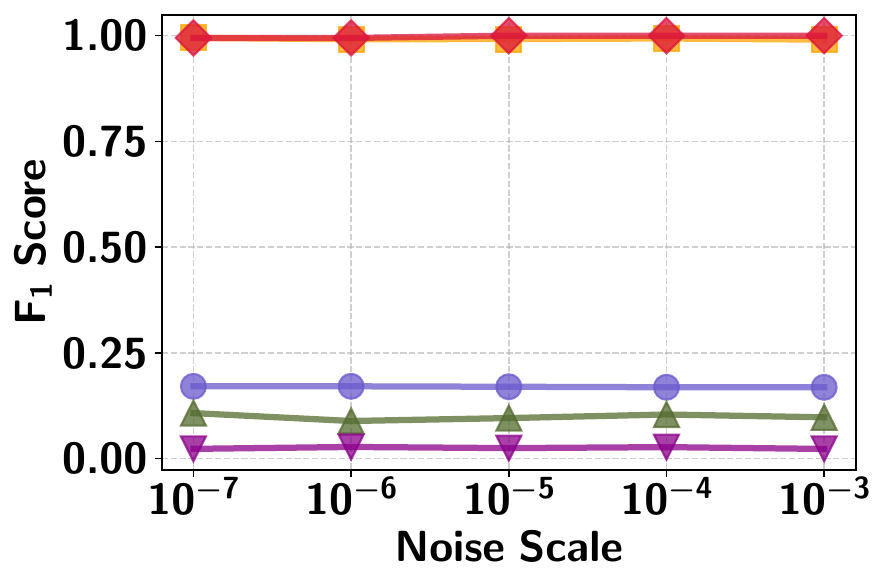} & 
        \includegraphics[width=0.3\textwidth]{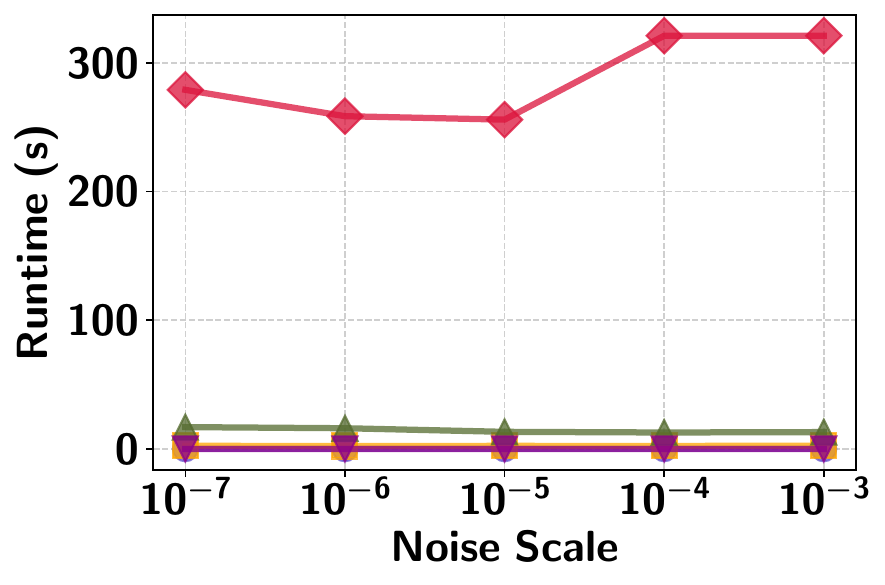} \\
        \vspace*{-0.65cm}
    \end{tabular}
\end{center}
\caption{
\label{fig:noise}
Full-rank synthetic $250 \times 250$-matrices generated from different distributions with planted near-rank-$1$ submatrices of dimensions $75 \times 75$.
Performance of different methods in the task of near-rank-$1$ submatrix discovery. 
We show the average (per-entry) reconstruction error (left), the $F_1$ score (center) and the runtime (right) of the different methods under comparison as a function of the noise scale \noisescale.
The $x$-axis is on logarithmic scale. 
}
\end{figure}

\spara{Experiments on synthetic data: impact of noise.}
To demonstrate the robustness of \ourmethod to noise, Figure~\ref{fig:noise} shows results for the performance of different methods in near-rank-$1$ sub\-matrix recovery as the level of noise increases. 
We use the same experimental setup as for the results shown in Figure~\ref{fig:near_rank_one_recovery}, except that the dimension of the planted ground-truth near-rank-$1$ sub\-matrix is held fixed to $75 \times 75$ and the scale of the noise $\noisescale$ is varied in $\{  10^{-8}, 10^{-7}, 10^{-6}, 10^{-5}, 10^{-4}  \}$. 
The figure highlights that, in this setting, \ourmethod and \rpsp outperform the other baselines. Most importantly, the results resemble those shown in Figure~\ref{fig:near_rank_one_recovery} for planted sub\-matrices of size $75 \times 75$, confirming that \ourmethod is fairly robust to noise.

\begin{figure}[p]
\begin{center}
\begin{tikzpicture}[scale=0.1]
\matrix [matrix of nodes, 
column sep=1.5pt, 
row sep=0pt, 
nodes={anchor=center}] (m) {
\node[draw=darkmagenta, fill=darkmagenta, line width=1pt, fill opacity=0.3, draw opacity=0.9, rectangle, inner sep=2.9pt] {}; 
& {\small \svp}
& 
\node[draw=crimson, fill=crimson, line width=1pt, fill opacity=0.3, draw opacity=0.9, rectangle, inner sep=2.9pt] {}; 
& {\small \rpsp} 
& 
\node[draw=orange, fill=orange, line width=1pt, fill opacity=0.3, draw opacity=0.9, rectangle, inner sep=2.9pt] {}; 
& {\small \ourmethod} \\
};
\end{tikzpicture}
    \begin{tabular}{ccc}
        \textsc{\Hyperspectral} & \textsc{\Isolet} & \textsc{\Olivetti} \\
        \includegraphics[width=0.33\textwidth]{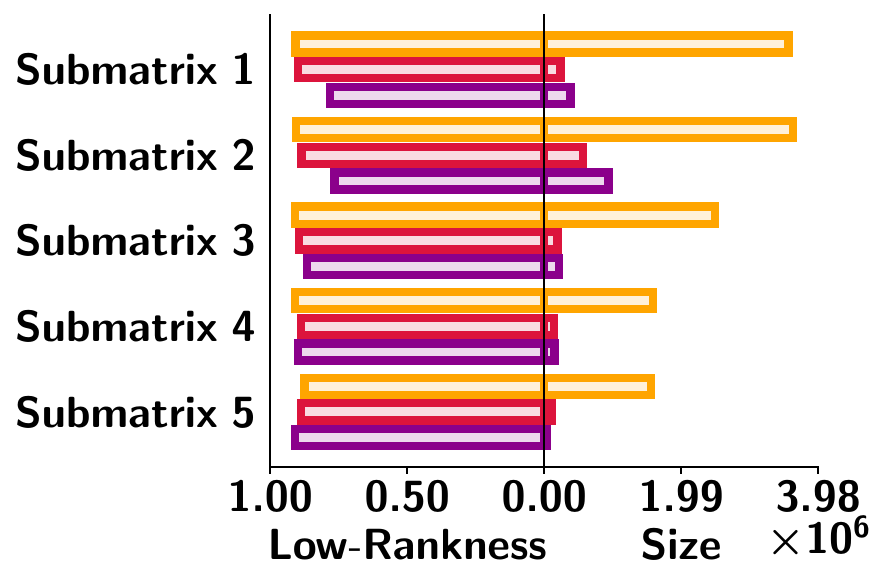} & 
        \includegraphics[width=0.33\textwidth]{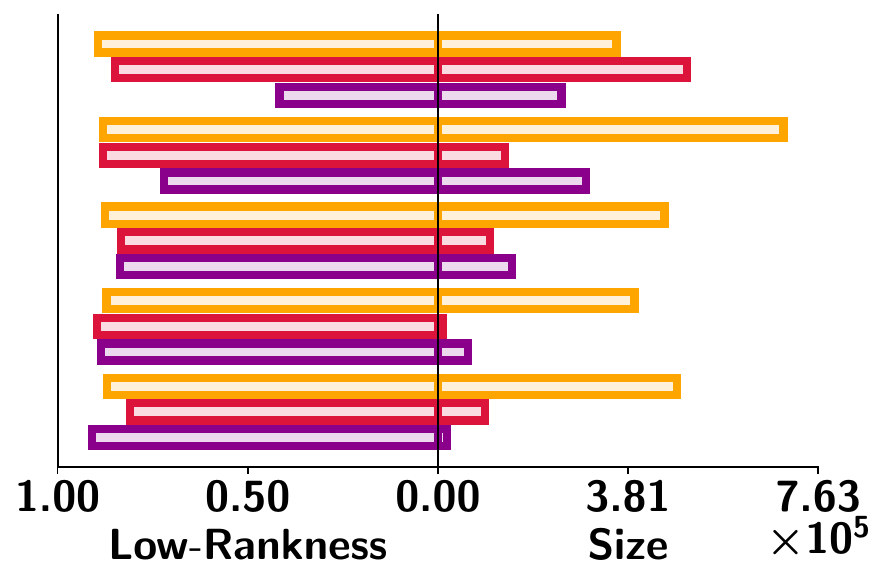} & 
        \includegraphics[width=0.33\textwidth]{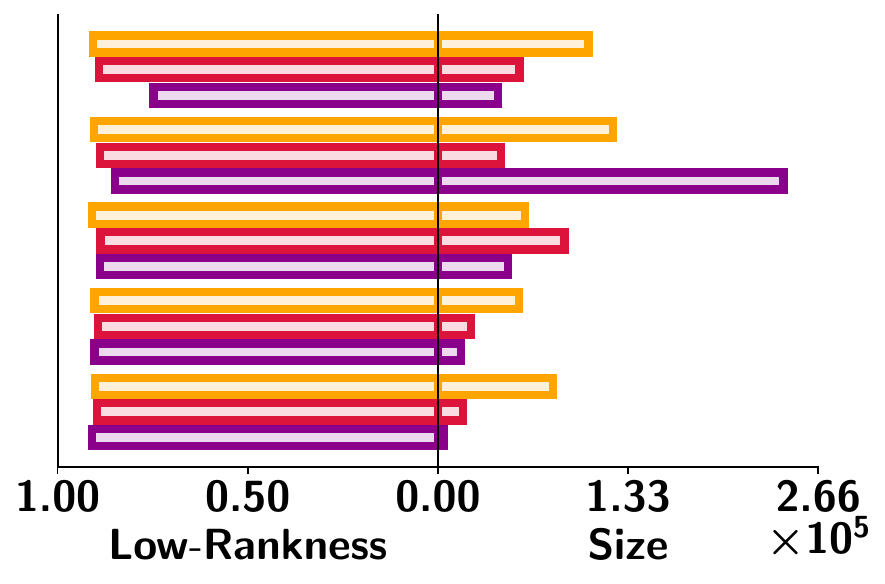} \\
        \textsc{\MovieLens} & \textsc{\OrlRnSp} &     \textsc{\Golub} \\
        \includegraphics[width=0.33\textwidth]{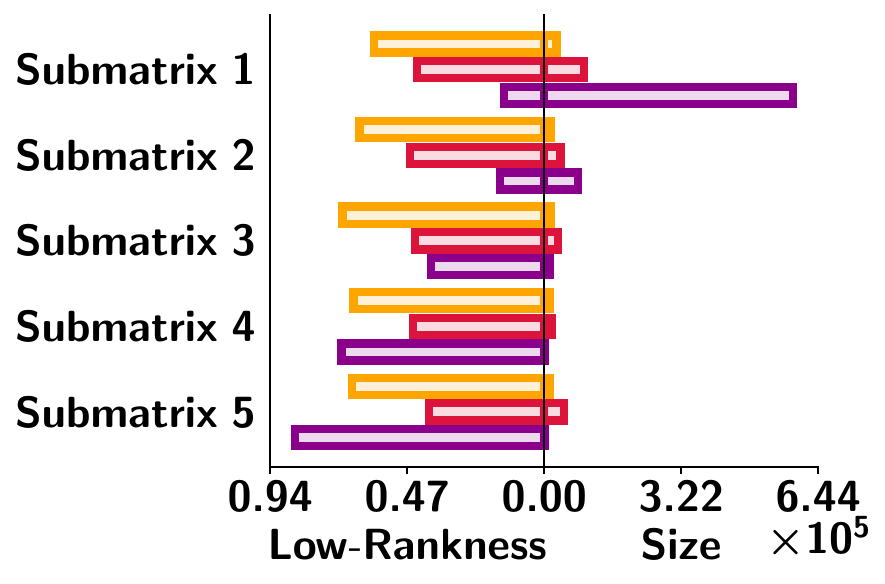} & 
        \includegraphics[width=0.33\textwidth]{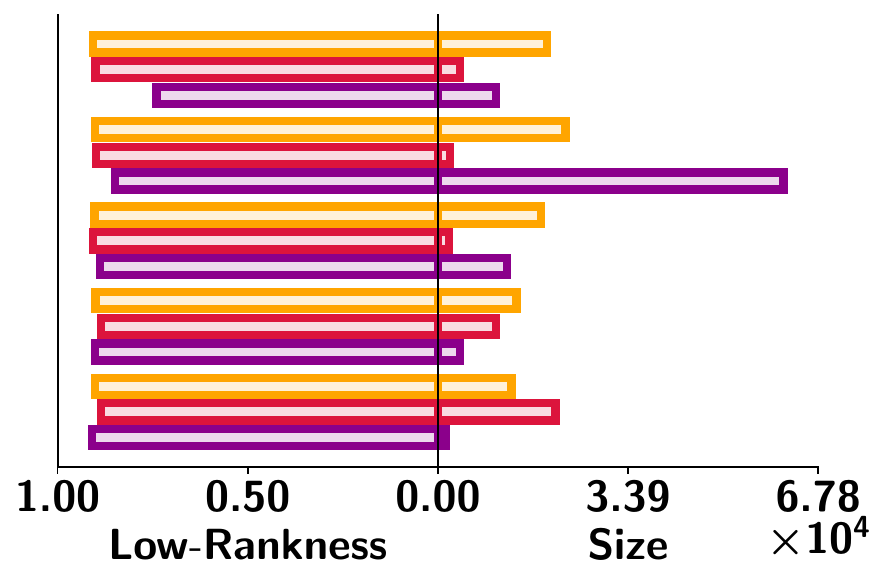} & 
        \includegraphics[width=0.33\textwidth]{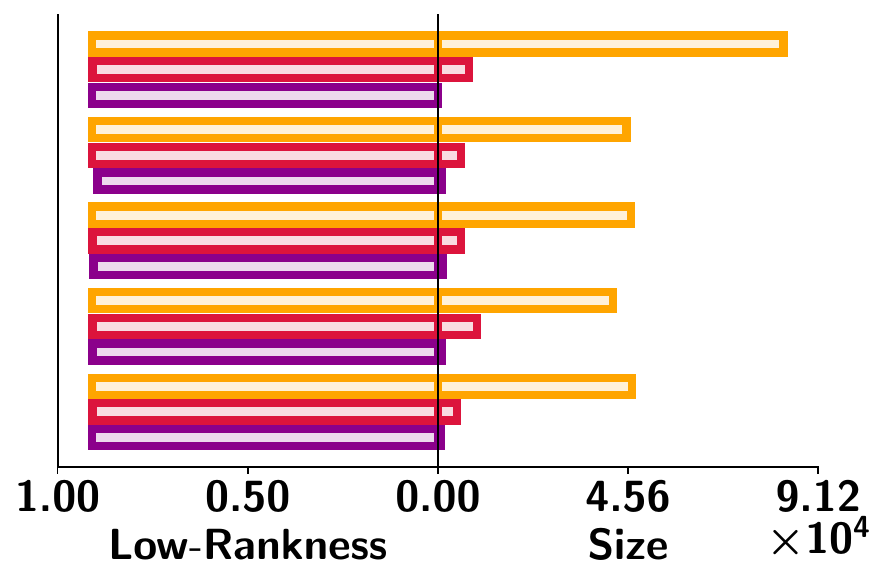} \\
        \textsc{\Mandrill} & \textsc{\Ozone} & \textsc{\BRCA} \\
        \includegraphics[width=0.33\textwidth]{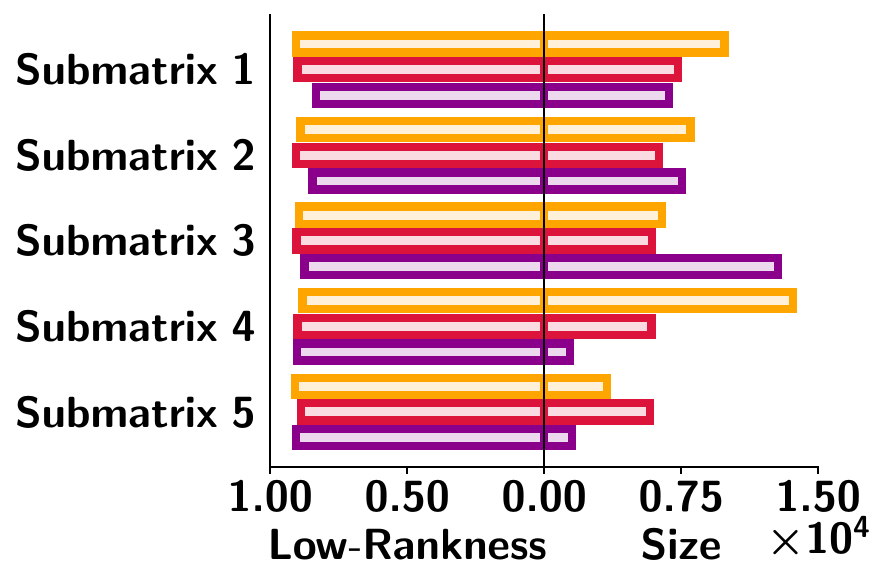} & 
        \includegraphics[width=0.33\textwidth]{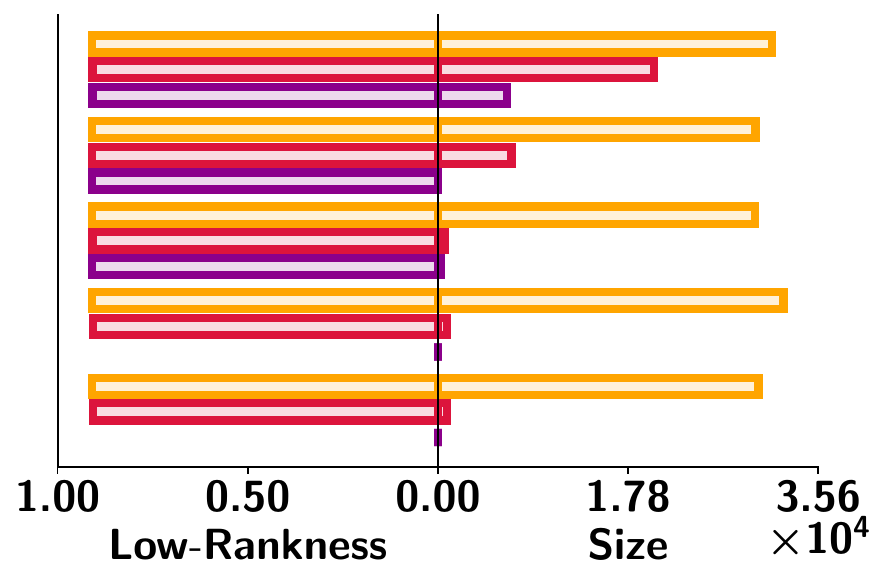} & 
        \includegraphics[width=0.33\textwidth]{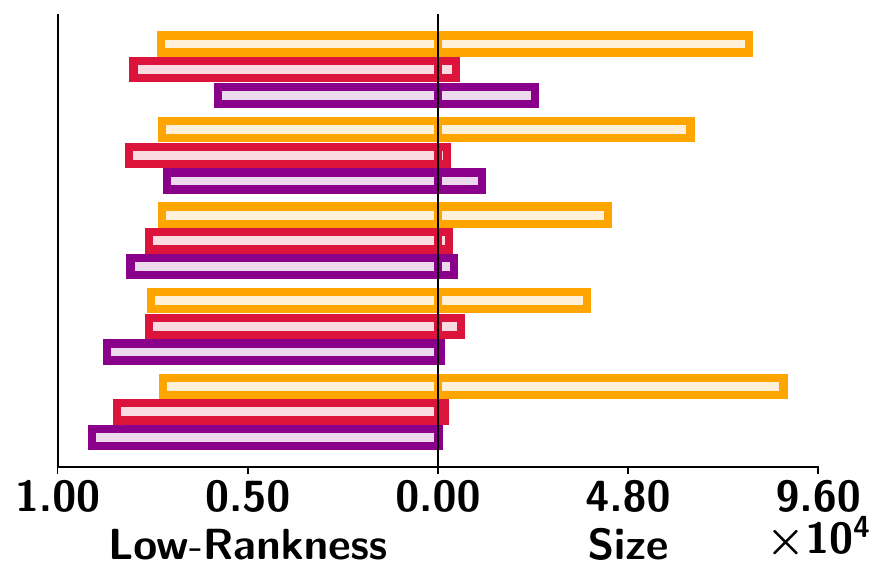} 
        \\
         \textsc{\Google} & \textsc{\NPAS} & \textsc{\Cameraman}  \\
        \includegraphics[width=0.33\textwidth]{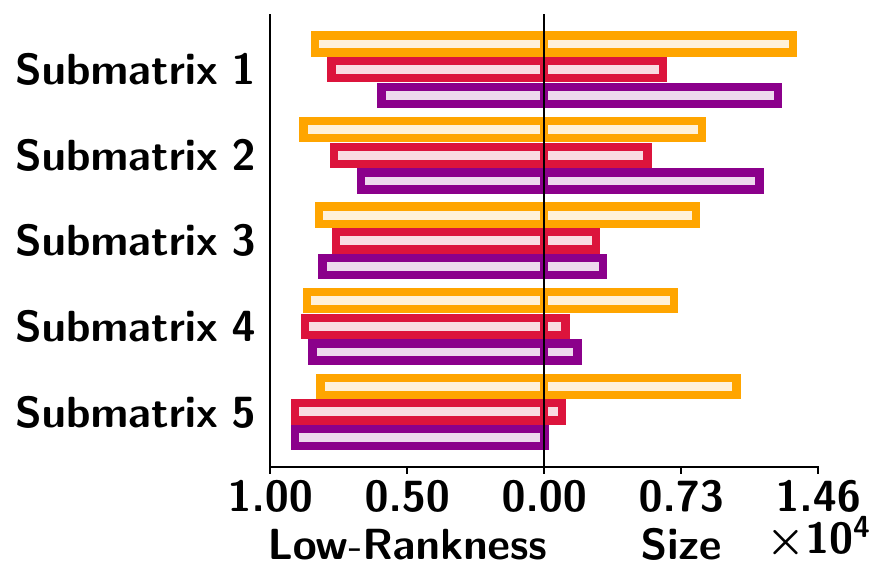} & 
        \includegraphics[width=0.33\textwidth]{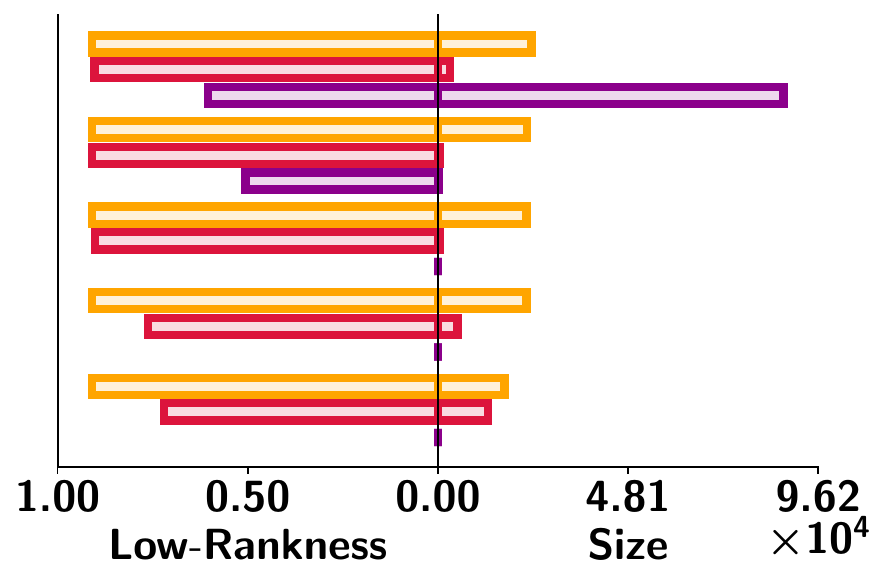} & 
        \includegraphics[width=0.33\textwidth]{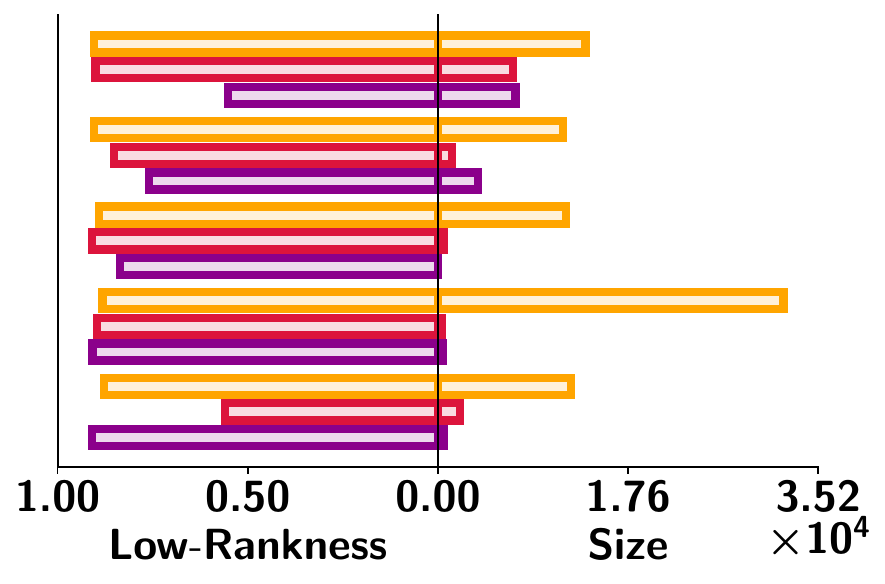} \\
        \textsc{\MovieTrust} & \textsc{\Hearth} & \textsc{\Imagenet} \\
        \includegraphics[width=0.33\textwidth]{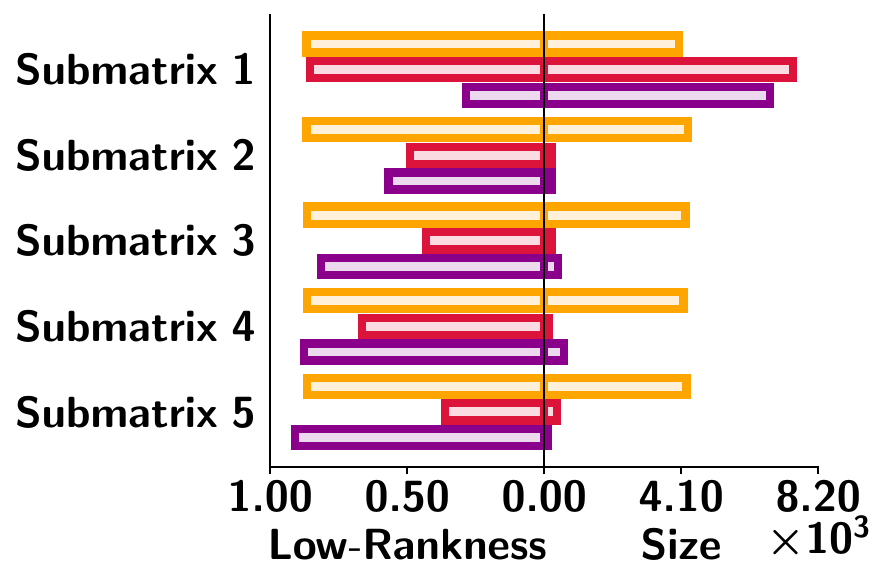} & 
        \includegraphics[width=0.33\textwidth]{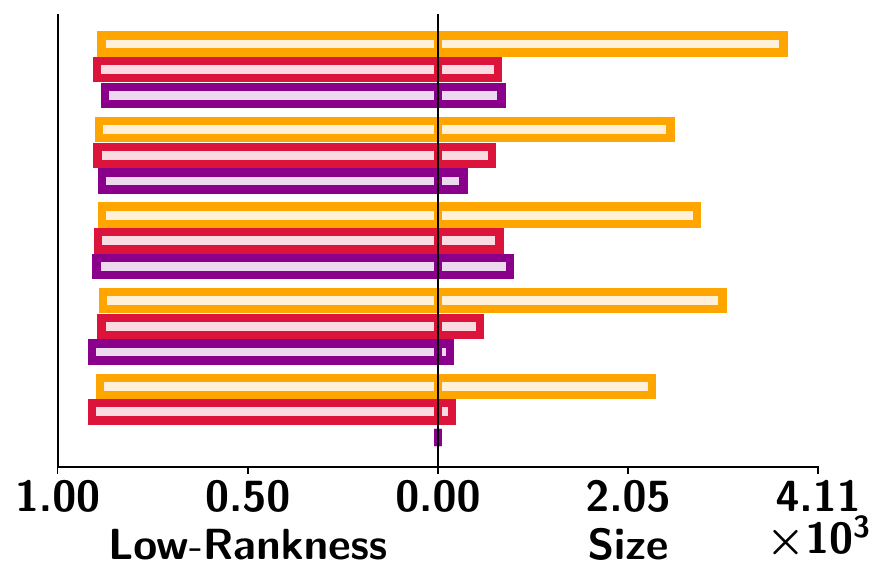} & 
        \includegraphics[width=0.33\textwidth]{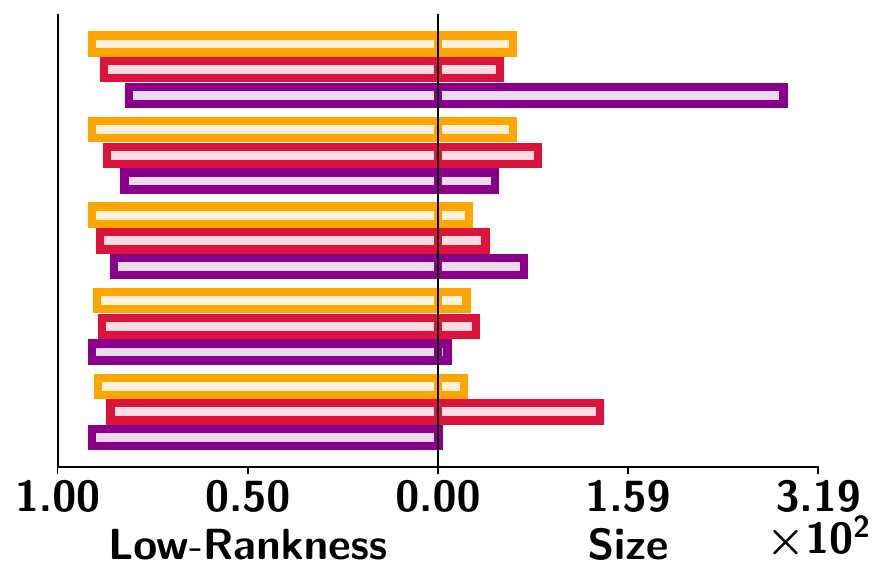} \\
        \vspace*{-0.65cm}
    \end{tabular}
\end{center}
\caption{\label{fig:real_data_top_five_patters} Real-world datasets. Low-rankness score and size for each of the top-$5$ sub\-matrices discovered by different methods.}
\end{figure}
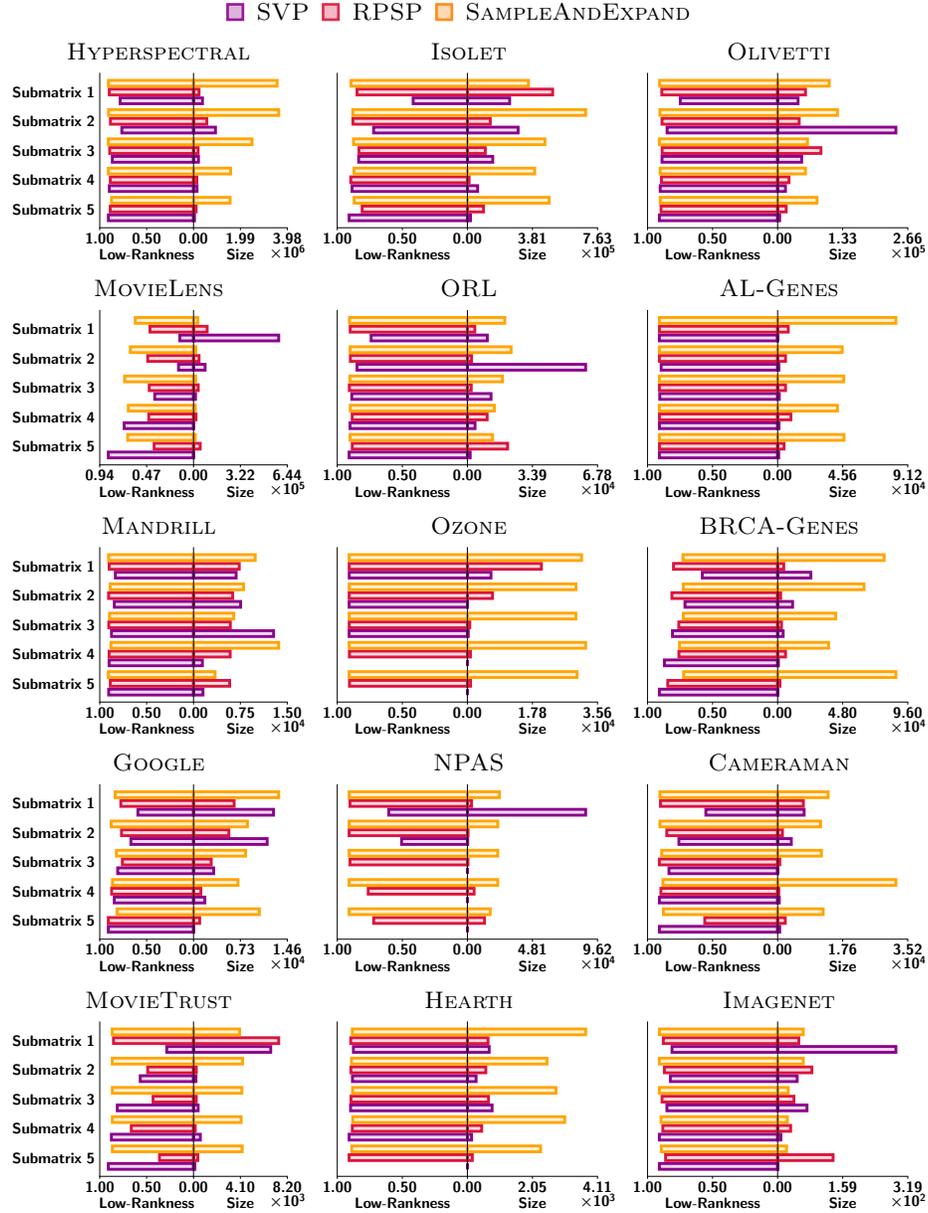

\spara{Experiments on real-world data: fine-grained results.}
Table~\ref{tab:results_real_data} shows average low-rankness and size of the top-$5$ low-rank sub\-matrices retrieved by our method, \rpsp and \svp for three datasets. 
In addition, Figure~\ref{fig:real_data_top_five_patters} provides a more fine-grained picture of the results of the experiments  for all real-world datasets, showing the raw low-rankness score and the size (i.e., the number of entries) of each of the individual top-$5$ sub\-matrices retrieved by \svp, \rpsp and \ourmethod. 
Note that, in three datasets (\NPAS, \Ozone and \Hearth), for \svp, results for less than $5$ sub\-matrices are shown because \svp is not able to discover five distinct sub\-matrices. 

\bibliographystyle{splncs04}
\bibliography{main}

%
%

%

\end{document}